\PassOptionsToPackage{final}{graphicx}
\documentclass[5p]{elsarticle}
\usepackage[monochrome]{color}

\usepackage{lineno,hyperref}
\usepackage{siunitx,fixme,xspace}
\usepackage{placeins}
\usepackage{amsmath}
\usepackage{datetime}
\usepackage{amssymb}

\newcommand{\CBE}{CBT\xspace}%CBELSA/TAPS experiment\xspace}
\newcommand{\CB}{CB\xspace}%Crystal Barrel\xspace}
\newcommand{\qtiPictureScale}{0.95}
\newcommand{\panda}{\texorpdfstring{$\overline{\mbox{P}}${ANDA}}{Panda}\xspace}

\journal{Journal of \LaTeX\ Templates}

\graphicspath{%
  {./figures/}
 }
\begin{document}

\begin{frontmatter}

\title{The new APD-Based Readout of the Crystal Barrel Calorimeter - An Overview}





%\address[fzj]{Institute for Nuclear Physics Forschungszentrum Jülich, Wilhelm-Johnen-Straße, 52428 Juelich, Germany}

%Highest Level:
\author[HISKP]{Christian Honisch\corref{correspondingauthor}}
\cortext[correspondingauthor]{Correspondence to:}
\ead{honisch@hiskp.uni-bonn.de}
\address[HISKP]{Helmholtz-Institut für Strahlen- und Kernphysik, Universität Bonn, Nussallee 14-16, 53115 Bonn, Germany}
\author[HISKP]{Peter Klassen}
\author[HISKP]{Johannes Müllers}
\author[HISKP]{Martin Urban}

\address[giessen]{II. Physikalisches Institut, Universität Gießen, Heinrich-Buff-Ring 16, 35392 Gießen, Germany}
\address[basel]{Department of Physics,
Universität Basel,Klingelbergstrasse 82, 4056 Basel, Switzerland}

%\address[gsi-det]{GSI Detector Laboratory, Planckstraße 1, 64291 Darmstadt, Germany}
\address[pi]{Physikalisches Institut, Nussallee 12, Universität Bonn, Germany}
\address[bochum]{Institut für Experimentalphysik I, Ruhr-Universität Bochum, Germany}
\address[talahassee]{Department of Physics, Florida State University, Tallahassee, FL 32306, USA}

%Primäre Beiträge
\author[HISKP]{Farah Afzal}
\author[HISKP]{John Bieling}
\author[HISKP]{Sebastian Ciupka}
\author[HISKP]{Jan Hartmann}
\author[HISKP]{Philipp Hoffmeister}
\author[HISKP]{Michael Lang}
\author[HISKP]{Dimitri Schaab}
\author[HISKP]{Christoph Schmidt}
\author[basel]{Michael Steinacher}
\author[HISKP]{Dieter Walther}

%CB Collab.
\author[HISKP]{Reinhard Beck\corref{correspondingauthor}}
\ead{beck@hiskp.uni-bonn.de}
\author[HISKP,giessen]{Kai-Thomas Brinkmann}
\author[talahassee]{Volker Crede}
\author[pi]{Hartmut Dutz}
\author[pi]{Daniel Elsner}
\author[basel]{Werner Erni}
\author[HISKP]{Eugenia Fix}
\author[pi]{Frank Frommberger}
\author[HISKP]{Marcus Grüner}
%\author[basel]{Nicolas Jermann}
\author[pi]{Tom Jude}
\author[HISKP]{Florian Kalischewski}
\author[basel]{Irakli Keshelashvili}
\author[HISKP]{Philipp Krönert}
\author[basel]{Bernd Krusche}
\author[HISKP]{Philipp Mahlberg}
\author[giessen]{Volker Metag}
\author[bochum]{Werner Meyer}
\author[basel]{Fabian Müller}
\author[giessen]{Marianna Nanova}
\author[HISKP]{Benedikt Otto}
\author[HISKP]{Lisa Richter}
\author[pi]{Stefan Runkel}
\author[HISKP]{Ben Salisbury}
\author[pi]{Hartmut Schmieden}
\author[HISKP]{Jan Schultes}
\author[HISKP]{Tobias Seifen}
\author[HISKP]{Nils Stausberg}
\author[HISKP]{Florian Taubert}
\author[HISKP]{Annika Thiel}
\author[HISKP]{Ulrike Thoma}
\author[HISKP]{Georg Urff}
\author[HISKP]{Christoph Wendel}
\author[bochum]{Ulrich Wiedner}
\author[HISKP]{Yannick Wunderlich}
\author[giessen]{Hans-Georg Zaunick}

\begin{abstract}
The Crystal Barrel is an electromagnetic calorimeter consisting of 1380 CsI(Tl) scintillators, and is currently installed at the CBELSA/TAPS experiment where it is used to detect decay products from photoproduction of mesons.\\
The readout of the Crystal Barrel has been upgraded
%was exchanged 
in order to integrate the detector into the first level of the trigger and to increase
its
%the trigger 
sensitivity for neutral final states.\\
The new readout uses avalanche photodiodes in the front-end and a dual back-end with branches optimized for energy and time measurement, respectively. An FPGA-based cluster finder processes the whole hit pattern within less than \SI{100}{\nano\second}. The important downside of APDs~--~the temperature dependence of their gain~--~is handled with a temperature stabilization and a compensating bias voltage supply. Additionally, a light pulser system allows the APDs' gains to be measured during beamtimes.

%This document describes the recently installed readout of the \CB Calorimeter, part of the CBELSA/TAPS experiment. The first section describes the mechanical properties of the detector. The second section outlines the need for an upgrade of the readout while section 2(3) presents the frontend (backend) in more detail. The advantages of Avalanche Photodiodes over PIN Photodiodes are discussed quantitatively using the physical origins of electronic noise. Also, the tradeoff between a fast response and a low detection threshold will be explained with regard to the \CB.\\
%Section 5 presents prototypes that were used to qualify the new electronics for the full upgrade of the calorimeter.\\
%The achieved performance in prototypes and in the full setup are shown in section 6. The achieved energy and time resolutions are shown.

\end{abstract}


\end{frontmatter}

%\linenumbers

\section{The CBELSA/TAPS Experiment}
The CBELSA/TAPS experiment (\CBE), which is located at the ELSA accelerator facility in Bonn, has been used to conduct baryon spectroscopy in order to obtain a better understanding of the non-perturbative regime of QCD \cite{Thiel:2022xtb}. The baryon resonances are studied by measuring polarization observables in the photoproduction of mesons \cite{sfba1}. The Crystal Barrel calorimeter (CB) is the centerpiece of the experiment and will be  introduced in Section \ref{sec:cb}.

%This section gives a very brief introduction to the CBELSA/TAPS experiment (\CBE), which the Crystal Barrel calorimeter is currently part of. The experiment is located at the ELSA accelerator facility in Bonn and measured double polarization observables in the photoproduction of mesons \cite{sfba1}.

ELSA provides electrons with energies of up to \SI{3.2}{\giga\electronvolt} for an unpolarized electron beam. In case of longitudinally polarized electrons, the electrons can have a high degree of longitudinal polarization for energies of up to \SI{2.4}{\giga\electronvolt} \cite{EPJA-WH}.

When extracted to the \CBE, the beam hits a radiator and produces real photons via bremsstrahlung. The photons can be circularly polarized \cite{CBELSATAPS:2019hhr}, when a longitudinally polarized electron beam is used, or linearly polarized \cite{CBELSA:2008nzn} when a diamond crystal is used as radiator to produce coherent bremsstrahlung.

The photon energy is tagged by measuring the secondary electron momentum using a magnet and a plastic scintillator system consisting of 576 fibers and overlapping bars in total \cite{Ponse_09_diss}.

The photon beam intensity is monitored with fully absorbing PbF$_2$ crystals, and a system to detect electron positron pairs produced on a copper converter by the photon beam \cite{CBELSATAPS:2019hhr}.
%\cite{Dielmann_08_dipl, McGehee_08_bachelor}

In the center of the CB, the photon beam hits a nucleon target \cite{Runkel_18_diss}. Available targets are, among others: (d)-butanol to provide transversely or longitudinally polarized protons or neutrons, liquid hydrogen and deuterium for nucleons that are unpolarized and a carbon foam target to study the background occurring in butanol measurements.

The focus of the physics program is put on the production of neutral pseudoscalar mesons ($\pi^0$, $\eta$, $\eta'$, $\pi^0\pi^0$, $\pi^0\eta$, ...). The \CBE is optimized for the detection of final states that are entirely composed of photons, besides the recoil nucleon. The two electromagnetic calorimeters Crystal Barrel \cite{Aker:1992ny} and MiniTAPS \cite{taps-novotny} cover almost the full solid angle.
%which is an important prerequisite to extract data precise enough to determine contributions of partial waves with high angular momentum.
This is an important prerequisite to extract data with good precision and a full coverage of even the extreme scattering angles, which can give access to contributions of partial waves with high angular momentum \cite{Wunderlich:2016imj, CBELSA/TAPS:2020yam, Wunderlich_19_diss}.

%In addition to the full coverage of the solid angle with the detector, a cutting-edge analysis code is needed which enables the detection of scattering events at extreme angles \cite{afzal_19_diss, CBELSA/TAPS:2020yam}. % using both 2- and 3 PED events 

Both calorimeters combined cover the polar angle range from \SI{1}{\degree} to \SI{156}{\degree}. The azimuthal angle is fully covered. The coverage down to $1^\circ$ in the forward region is in particular important due to the Lorentz boost of the decay products.

If the recoil nucleon is a proton, it can be detected by the fiber detector \cite{innendet-suft} installed inside the Crystal Barrel. 
%%%%%%%%
However, if the recoiling nucleon was  a neutron, the previous configuration had a very limited trigger sensitivity. The reason for this problem and its solution will be discussed in more detail in Section \ref{sec:newConcept}.
%That a recoiling neutron could be only detected with a low probability - if at all - limited the trigger sensitivity for reactions with no charged particles in the final state.
%%%%%%%%%

Expanding the measurement program to reactions off neutrons (e.g. $\gamma n \rightarrow n \pi^0$) required an improved trigger sensitivity, which was the main motivation for the readout upgrade described in this paper.

\section{The Crystal Barrel Calorimeter}
\label{sec:cb}
%This section describes the mechanical properties of the Crystal Barrel and briefly its history.

The Crystal Barrel is an electromagnetic calorimeter which was built up at LEAR, CERN to study $\bar{p}p$ and $\bar{p}d$ annihilations \cite{Aker:1992ny}.
Beginning in the year 2000 the detector was used in different configurations at the ELSA facility in Bonn for experiments on hadron spectroscopy.

%The detector consisted it its initial version 
The initial version of the detector consisted of 1380 scintillation crystals, made of Thallium doped Cesium Iodide. This material features a high light yield and a high stopping power. The scintillation signal is comparably slow with a decay time of $\tau\approx\SI{1}{\micro\second}$ \cite{ds:csi}.

Thirteen different geometric types of crystals were arranged in 20 rings of 60 crystals each and 6 rings of 30 crystals each. All crystals have shapes of truncated pyramids and point to a common center.
The configuration currently used deviates in two aspects:

The two most upstream rings are not mounted, in order to fit the target cryostat.\\
The three most downstream rings are shifted by \SI{3}{\milli\meter} downstream. The resulting gap of \SI{5}{\milli\meter} is used to feed through light guides belonging to the plastic scintillation detector in front of these rings \cite{Wendel_08_diss}. The offset results not only in a gap but also in these crystals not pointing to the common center of all other crystals.
These three rings formed the so called Forward Plug and had a faster readout than the remaining calorimeter. Fast trigger signals from this PMT based readout were available for this section of the CB.

Each crystal is wrapped into expanded PTFE foil. This material was chosen for its high reflection coefficient in order to maximize the light yield at the photodetector. The foil had to be replaced for a few crystals during the upgrade. A \SI{38}{\micro\meter} thick foil (Donaldson Tetraflex \#3101) was used in these cases.

Mechanical stability of the detector modules is achieved by a tightly fitting shell of Titanium of \SI{100}{\micro\meter} thickness \cite{Aker:1992ny}. The crystals' weight is suspended on a plastic pin attached to the front of the crystal and a bolt on the backside of the crystal.

\begin{figure}[ht]
\centering
\includegraphics[width=\columnwidth]{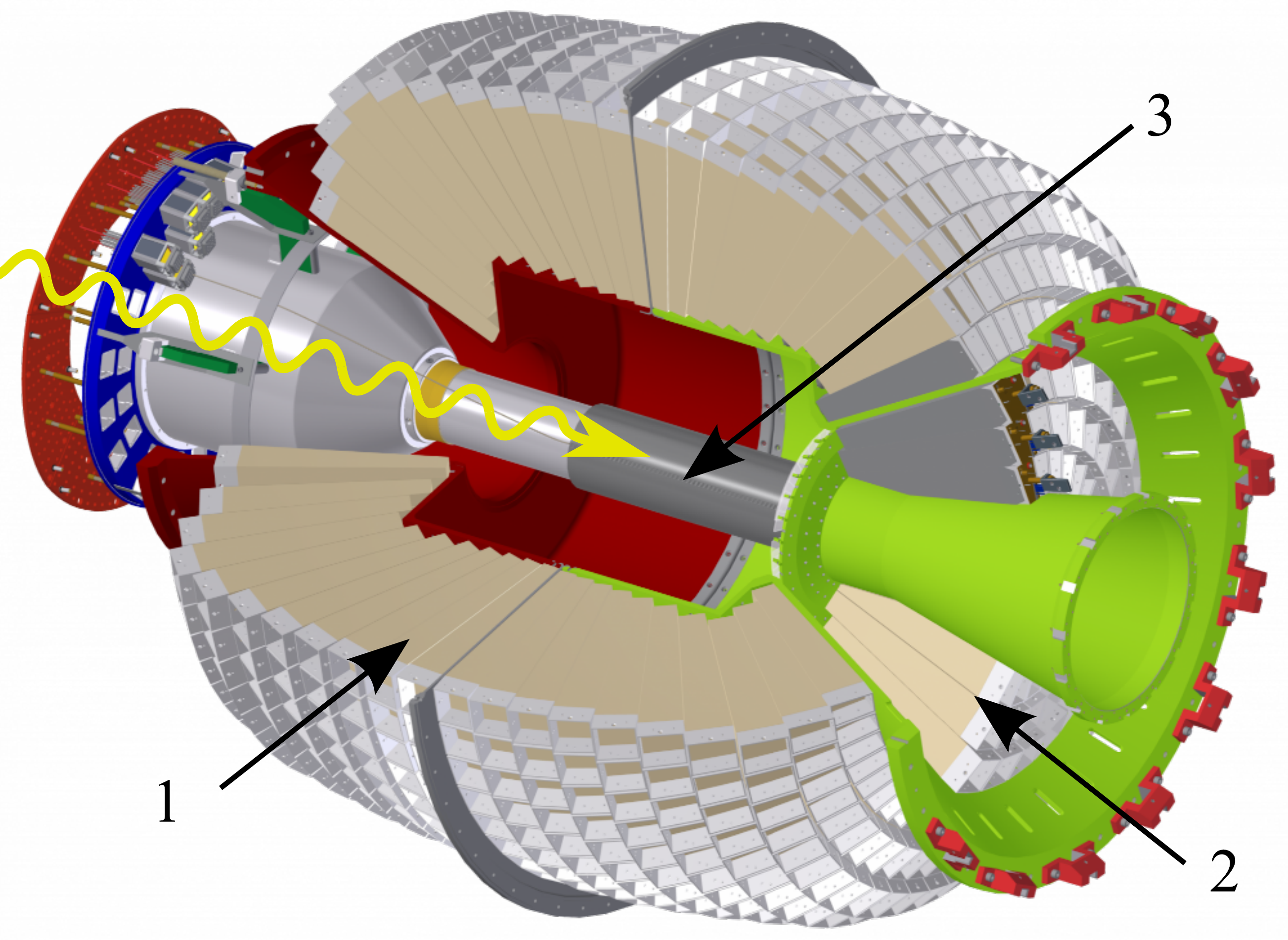}
\caption{CAD image of the Crystal Barrel calorimeter (1). The three rings in forward direction (2) had a different readout and were called Forward Plug. A detector consisting of scintillating fibers is in the center of the setup (3, one fiber shown). Photon beam enters from the left.}
\label{pic:cb-cad}
\end{figure}
\section{The new Readout Concept}
\label{sec:newConcept}
The data acquisition system of the \CBE is operated in gated mode. A trigger is needed for each event to start the digitization process.\\
Using signals from the \CB in the trigger logic turns out to be difficult. Certain other detectors of the experiment require the trigger signal latest at \SI{700}{\nano\second} past the particle detection. This time has to cover signal processing, signal detection, cluster encoding, and signal propagation. Thus, only a fraction of this time remains for the scintillation signal to form. As the scintillation light output has a decay time of roughly one microsecond, only a small fraction of the full scintillation signal is available to be used in the trigger.

%Therefore
To accommodate this, the trigger scheme of the previous version of the \CBE was split up into two stages. 
The first stage processed signals from all detectors except the \CB. The second stage used the result of the CB's cluster encoder \cite{Flemming_01_diss} to decide whether to proceed with the readout of the current event or to abort the digitization in order to save the deadtime of roughly \SI{500}{\micro\second} per event \cite{Hoffmeister_19_diss}.
An evaluation time of $t=(n+1)\cdot\SI{800}{\nano\second}$ \cite{Flemming_01_diss} with the total number of clusters $n$ was another deal-breaker for the inclusion of the \CB in the first stage of the trigger.
\subsection{Purpose of the Upgrade}
The recent progress of data on hadron spectroscopy gave rise to the need to investigate purely neutral final states, for example $\Delta^0\rightarrow n\pi^0$. Previous studies of decays at the \CBE included a proton in the final state, which could be identified with high efficiency in plastic scintillation detectors.

\begin{figure}[ht]
\centering
\includegraphics[ width=\columnwidth]{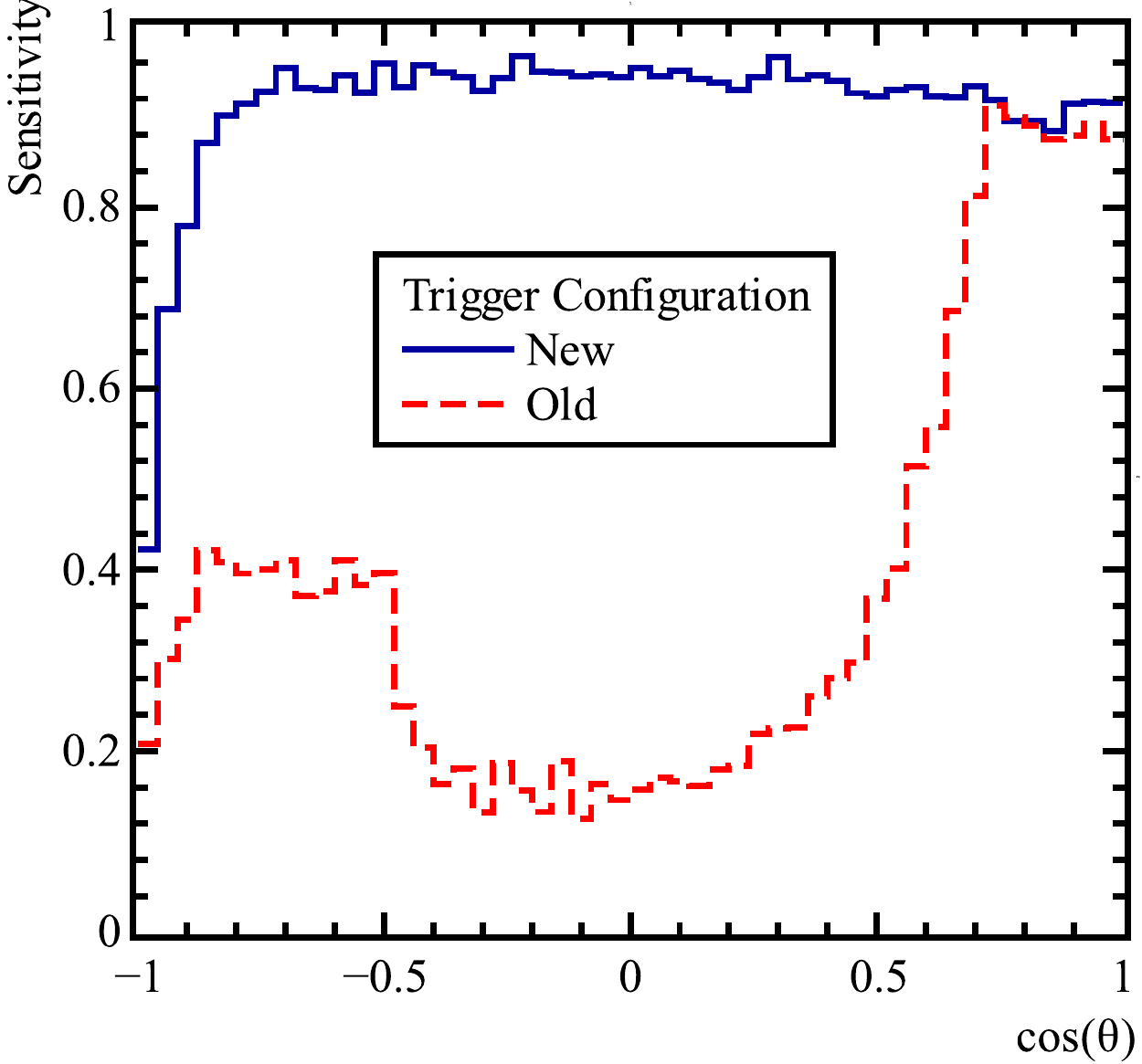}
\caption{Trigger sensitivity for $\gamma n \rightarrow n \pi^0$ and $E_\gamma=\SI{700}{\mega\electronvolt}$ as a function of $\cos(\theta)$ of the $\pi^0$. The new readout evaluates the \CB hit pattern in the first level of the trigger while the old one did not.}
\label{pic:trig_old_new}
\end{figure}

To detect neutral final states with a high sensitivity over the full kinematic range, the hit pattern of the \CB needs to be evaluated in the first stage of the trigger. Fig. \ref{pic:trig_old_new} shows the simulated trigger sensitivity of the 
%\CBE in the old form 
setup described
and for the case that the CB is included in the first level of the trigger at $E_\gamma=\SI{700}{\mega\electronvolt}$. The analyzed MC dataset is purely $\gamma n \rightarrow n \pi^0$.

The old trigger configuration results in a substantial sensitivity loss, especially in the angular region from $\text{-0.5}< \cos\theta_{\pi^0}<0.5$. Here, the fiber detector could trigger the data acquisition ($\sim\SI{15}{\percent}$) only 
%due
if one of the final state particles ($\gamma$, n) 
generated secondary charged particles, 
%being falsely marked as charged 
e.g. due to an interaction with the butanol target-holder structure. 
For $\cos\theta_{\pi^0} < \text{-}0.5$, the neutron is detected in the TAPS detector or the Forward Plug and for $\cos\theta_{\pi^0} > 0.5$ the photons are detected in TAPS or the Forward Plug.
The TAPS detector consists of BaF$_2$ scintillators which have a fast light output. Its readout uses PMTs and provides fast trigger signals. The large drop of the trigger sensitivity around $\cos\theta_{\pi^0}=-1$ is caused by photons being scattered to extreme backward angles ($\theta>156^\circ$) or for photons or neutrons going to extreme forward angles ($\theta<1^\circ$), which are not covered by the detector setup.  

 Since the detection efficiency of photons is much higher ($\sim\SI{90}{\percent}$) than for neutrons ($\sim\SI{30}{\percent}$), an asymmetric structure is observed in the trigger sensitivity. 
The positions of these features change depending on the energy of the primary photon. $E_\gamma=\SI{700}{\mega\electronvolt}$ was chosen as a representation, where all features can easily be identified.

Including the CB in the first level of the trigger leads to a noticeable improvement of the trigger efficiency of almost \SI{90}{\percent} efficiency over a large angular range.

This demand resulted in the 
%required properties 
following requirements on the new readout:
It must deliver fast signals that can be used for hit detection in a new cluster encoder, which speeds up the encoding process by two orders of magnitude.

Additionally, all changes to the current electronics need to maintain the capability to operate in a high magnetic field to allow a future tracking upgrade including a Time Projection Chamber (TPC) and a superconducting magnet \cite{Beck:2017upc}.

\subsection{Options Investigated to Obtain  Trigger Signals}
To solve the first problem (fast hit signals), two concepts were evaluated.\\
First, the existing photodiode readout could be complemented or replaced by a faster readout. The existing readout was optimized for energy measurement only. Silicon photomultipliers (SiPM) were tested as a complementary timing readout. \\
Replacing the existing photodiodes with ordinary photomultipliers, which have the advantage of a higher signal-to-noise ratio, was ruled out by the requirement to operate inside a magnetic field.

Second, the concept of photodiodes and charge sensitive preamplifiers in the front-end electronics could be maintained. The fast timing signals are achieved by signal-shaping filters, which basically add a second back-end branch, dedicated to timing and triggering.

On the basis of this evaluation the second option was chosen.
%During the evaluation phase, we decided for the second option. 
Its implementation will be discussed in the following, as well as all the needed changes in the readout which come as a consequence.\\
%Prototypes based on SiPMs were rejected due to the high dark count rates (although this is expected to improve with more recent models).\\
%, which is expected to be better with more recent SiPM models.\\
A solution based on SiPMs, would have required mechanical
%Another drawback of this solution is that it required mechanical 
modification of the existing front-end. For example, the photodiode was glued to a wavelength shifter. A part of it would have to be milled off to be able to attach the SiPM. This procedure is even more delicate as some electronic components used in the old readout were obsolete and therefore the supply of spare front-end modules was very limited.

The second problem, the fast cluster encoding, was solved by implementing a completely different clustering algorithm in FPGAs. Details will be discussed in Section \ref{ssec:clusterfinder}.

\subsection{Implementation of the Fast Branch}
\begin{figure}[h]
\centering
\includegraphics[width=\columnwidth]{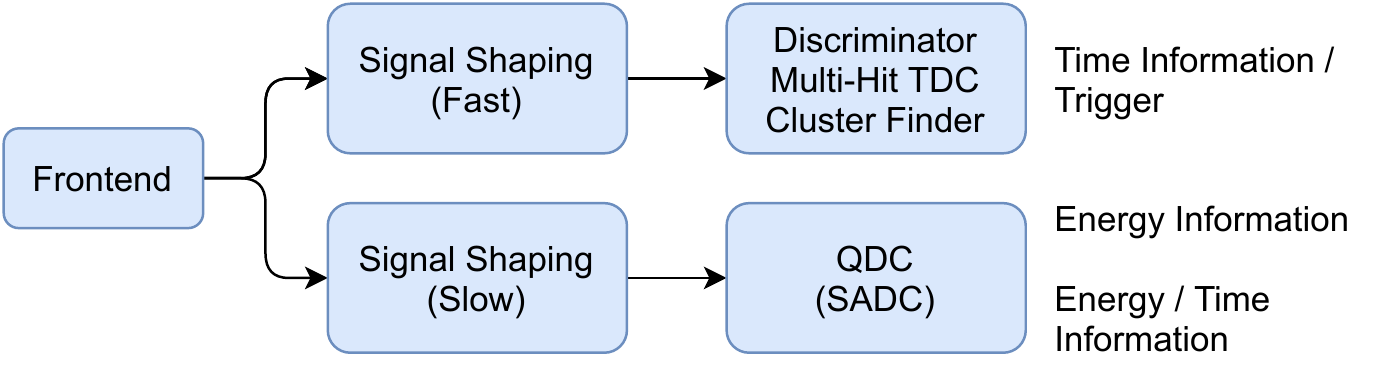}
\caption{New Readout Scheme. The existing slow branch was maintained during the first beamtimes after the upgrade and is being replaced by a modern SADC readout.}
\label{pic:readoutscheme}
\end{figure}
The new readout scheme is shown in Fig. \ref{pic:readoutscheme}. All constituents will be explained in the following sections. \\
Before that, the main challenge of the design is introduced: The optimization between a low detection threshold on the one hand and a low information latency on the other hand, which are concurring requirements.

\begin{figure}[h]
\centering
\includegraphics[width=\qtiPictureScale\columnwidth]{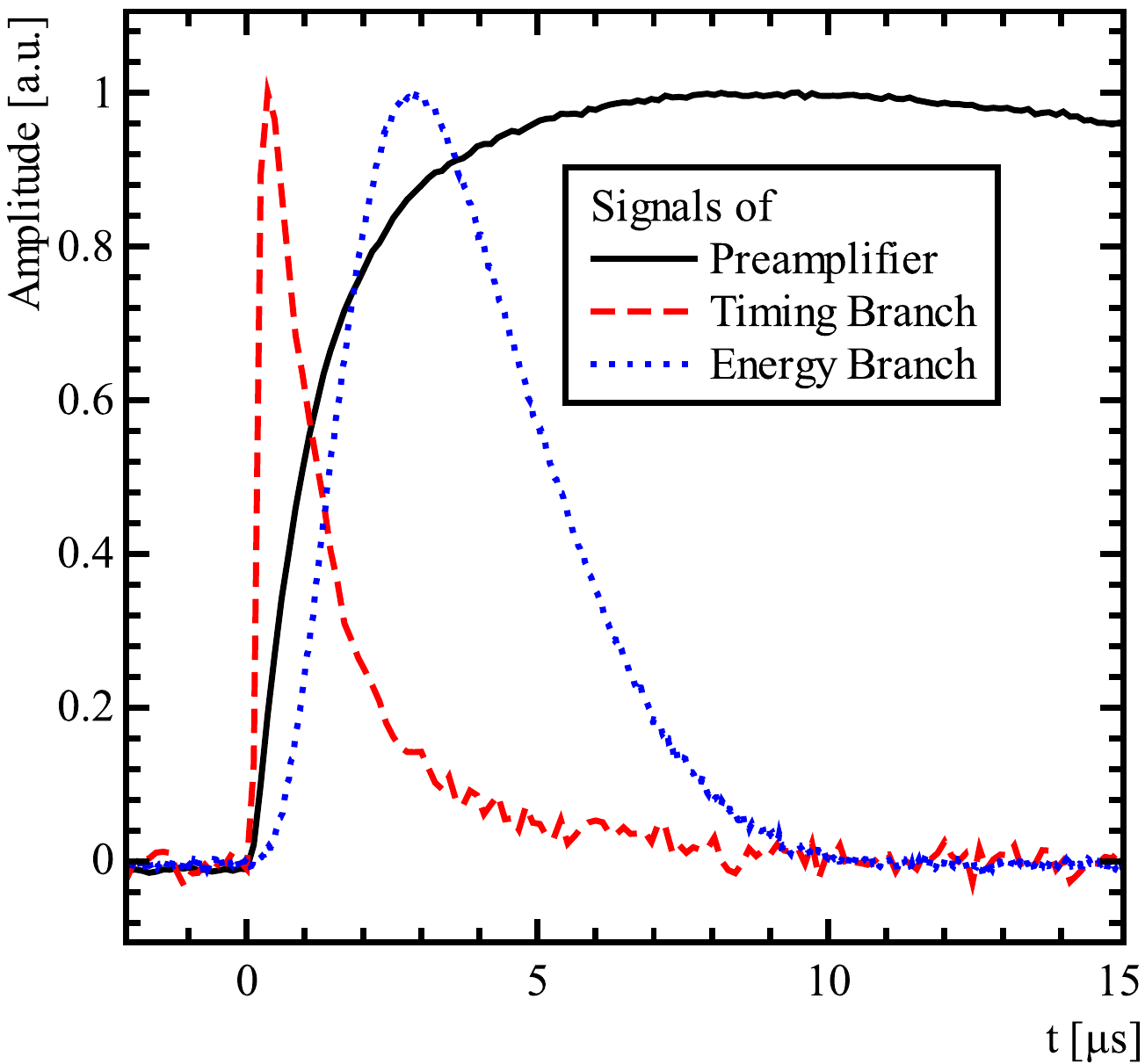}
\caption{Output signal of the preamplifier and signals in the timing and energy branches.}
\label{pic:signals-pte}
\end{figure}

%Both, fast branch and slow branch, use signal shaping filters to obtain signals for the respective purpose. 
The fast and slow branches both use signal shaping filters to obtain suitable signals.
All three signals (preamplifier, timing shaped, energy shaped) are shown in Fig. \ref{pic:signals-pte}. % together with the corresponding signals in the timing and energy branch respectively. 
The shorter rise time of the timing branch signal is clearly visible. However, the noise level is also increased, which
%limits the achievable 
increases the minimum detection threshold in each detector crystal.\\
The required value of the detection threshold results from the physics program. The entirely neutral final states of decaying $\pi^0$, $\eta$, $\eta'$ should be detected with a high efficiency and a uniform angular dependence over the whole angular range covered by the \CB (including the former Forward Plug).\\
To reach the resulting detection threshold of roughly $\SI{10}{\mega\electronvolt}$, it was also necessary to replace the existing PIN photodiodes by avalanche photodiodes which provide a better signal-to-noise-ratio due to the internal gain mechanism \cite{MOSZYNSKI2003226}.\\
The timing branch consisting of a signal shaper, a discriminator, and a cluster encoder will be discussed in Section \ref{sec:backend}. The APD as well as the newly required front-end will be presented in Section \ref{sec:apd-frontend}. It also discusses the improved SNR of the APD by analyzing the different noise contributions.

\section{The APD Front-End}
\label{sec:apd-frontend}
\begin{figure}[h]
\centering
\includegraphics[width=\columnwidth]{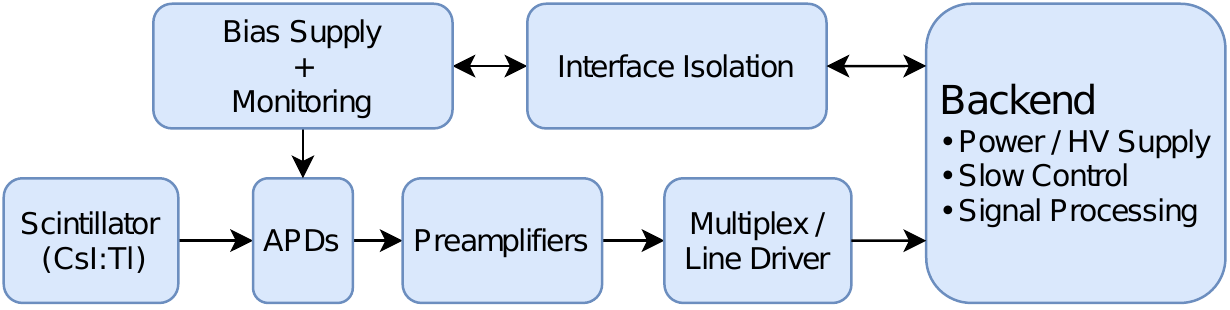}
\caption{Overview on the new detector front-end containing APDs.}
\label{pic:frontend}
\end{figure}

\begin{figure}[h]
\centering
\includegraphics[width=\columnwidth]{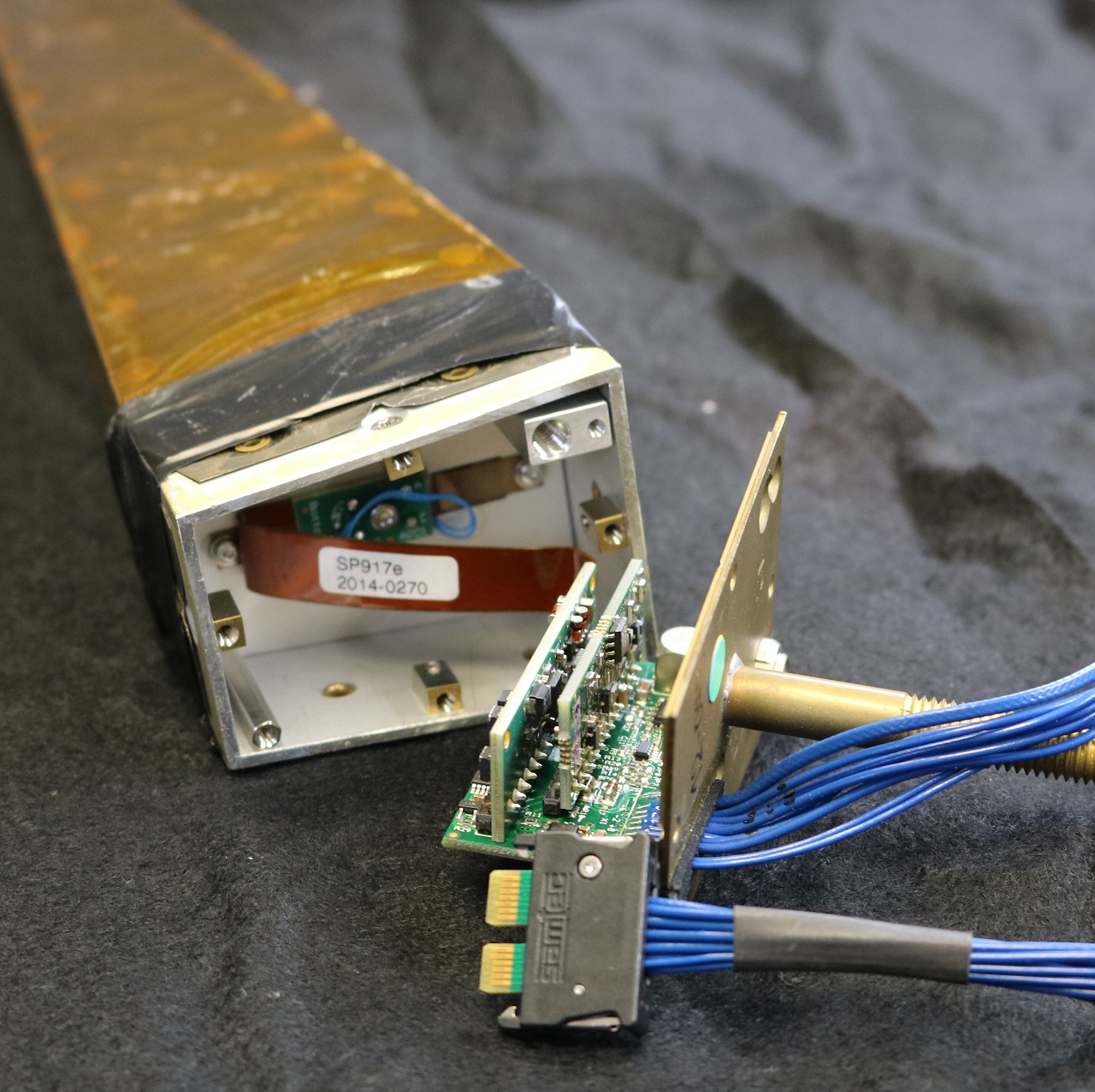}
\caption{Photograph of one detector module assembly with the new APD based front-end electronics.}
\label{pic:frontend-photo}
\end{figure}

A scheme of the new front-end is shown in Fig. \ref{pic:frontend} and a picture of one module is shown in Fig. \ref{pic:frontend-photo}. The electronics of each detector consists of two APDs, a dual-channel preamplifier, a programmable-gain line driver, and a bias supply with integrated monitoring. All components will be presented in the following sections.

\subsection{Charge Sensitive Preamplifier}
\label{ssec:preamp}
\begin{figure}[ht]
\centering
\includegraphics[width=\columnwidth]{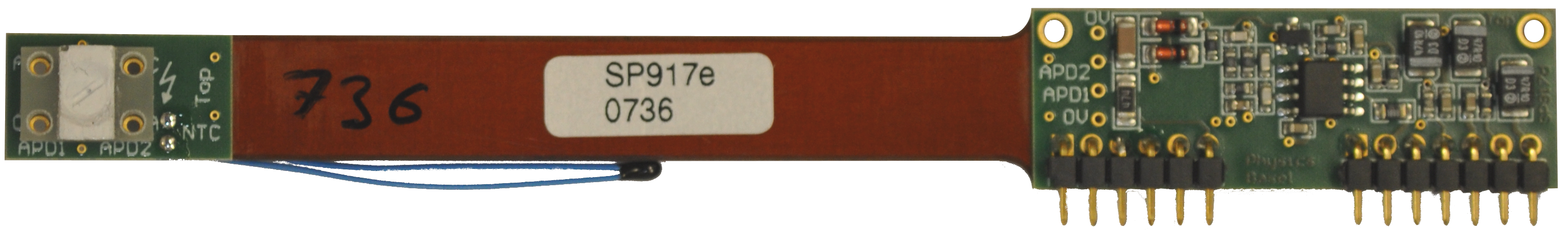}
\caption{Dual Channel Charge Sensitive Preamplifier}
\label{pic:preamp}
\end{figure}

\begin{figure}[ht]
\centering
\includegraphics[angle=270,width=\columnwidth]{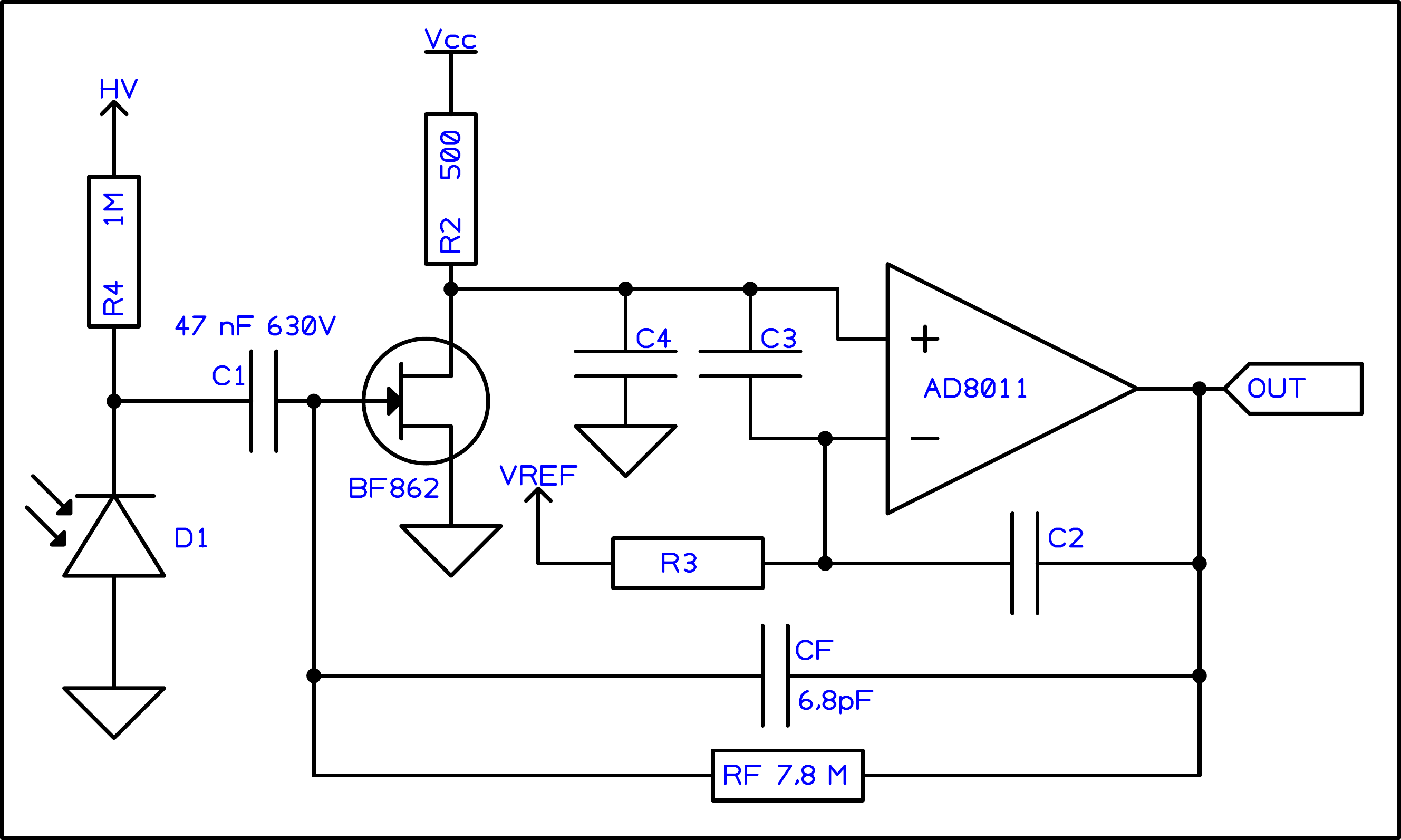}
\caption{Simplified schematic of the preamplifier used in the APD readout (one channel shown).}
\label{pic:sch-preamp}
\end{figure}

A picture of the preamplifier is shown in Fig. \ref{pic:preamp}. A simplified schematic view of one of both channels is shown in Fig. \ref{pic:sch-preamp}. It is a modified version of the charge sensitive preamplifier, developed for the electromagnetic calorimeter of the \panda experiment \cite{Keshelashvili_2015, preamp-manual}. Development and modification were done at the University of Basel.
% 2 APDs pro kristall, gewinn in t kanal nur bei zwei preamps
%%%%%%%%%%%%%% pic preamp
Two preamplifiers per detector crystal are utilized to improve the SNR and reliability.
A jFET of type BF862 is used as input transistor to achieve a high-impedance input and low noise of the amplifier. An operational amplifier (AD8011) is used for the main amplification. The gain is set to \SI{147}{\volt\per\nano\coulomb} by the value ($C_F$) of the feedback capacitor (CF). 
The resistor in parallel (RF) discharges the capacitor and therefore the components' values determine the decay time $\tau=C_F\cdot R_F$.
%The combination of $CF$ and $RF$ determines the decay time constant.
C2, C3, and C4 are required for loop stability.\\
C1 decouples the APD signal from the high bias voltage ($\sim\SI{400}{\volt}$). To allow a precise determination of the APD gain using the preamplifier,
C1 needs to have a
a comparably high capacitance.
%value is necessary.
The APD gain is measured using pulsed light (see Section \ref{ssec:apd}). To calculate the absolute value of the gain, a normalization reference point at zero bias voltage is needed, at which the avalanche gain mechanism is not present. At zero voltage, the APD has a parasitic junction capacitance of as much as $C_\text{APD}\approx\SI{4}{\nano\farad}$ \cite{Urban_18_diss}. The signal charge is distributed between the APD and C1 according to their values. Thus, $C_1\gg C_\text{APD}$ is required for negligible signal losses.\\

The origins of electronic noise in this circuit are the dark current $I_D$ of the APD and the thermal noise of the current in the jFET, which is translated to a voltage noise of the output signal using $C_\text{APD}$ and $C_F$.  The noise contributions will be discussed in more detail at the end of Section \ref{ssec:apd}.

\subsection{Hamamatsu APDs}
\label{ssec:apd}
The new readout utilizes Hamamatsu APDs of the type X11048(X3). This type is also used in the electromagnetic calorimeter of the \panda experiment and has very similar specifications as the Hamamatsu S8664-1010 \cite{apddatasheet}.\\
The X11048 has an active area of $14\times\SI{6.8}{\milli\meter}^2$, $G_\text{nominal}=50$, terminal capacitance $C_t=\SI{270}{\pico\farad (Typ.)}$, $\eta>80\%$ for $\SI{450}{\nano\meter}<\lambda<\SI{850}{\nano\meter}$.
The key disadvantage of APDs is the temperature dependence of their gain. For the X11048, it is in the order of \SI{-2.3}{\percent\per\kelvin}. We account for this property in threefold way: The temperature of the calorimeter is stabilized (see Sec.~\ref{ssec:tempStab}), the change in gain due to remaining temperature variations is counteracted by an automatic voltage adjustment (see Sec.~\ref{ssec:hv_front}), and a light pulser (see Sec.~\ref{ssec:led-pulser}) allows for a continuous determination of the APD's gain, even during production beamtimes.
The automatic voltage adjustment is part of the bias voltage supply.

To design a proper temperature dependence compensation circuit, the gain dependence on bias voltage and temperature need to be known for each APD. The manufacturer however specifies only typical values, and neither individual values nor the variation range is given. Therefore, each APD was characterized before the installation.

\subsubsection*{APD Characterization Station}

For this purpose, a characterization setup was developed \cite{Urban_18_diss}. The APD gain $M$ is measured by normalizing the signal intensity $A_i$ with present bias voltage ($V_\text{APD}=V_{M=50} \rightarrow A_\text{Bias}$) to the amplitude for zero bias voltage ($V_\text{APD}=0 \rightarrow A_0$):
$$
M=\frac{A_\text{Bias}}{A_0}
$$
To quantify the signal intensity, its amplitude can be measured as well as its integral.

\begin{figure}[ht]
\centering
\includegraphics[width=\columnwidth]{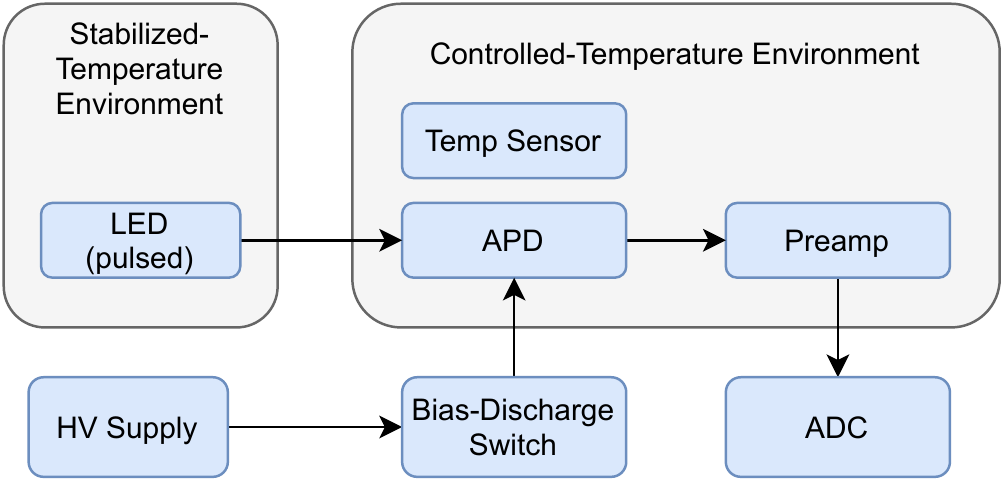}
\caption{Schematic of the setup used for APD characterization.}
\label{pic:apd_char_setup}
\end{figure}

\begin{figure}[ht]
\centering
\includegraphics[width=\columnwidth]{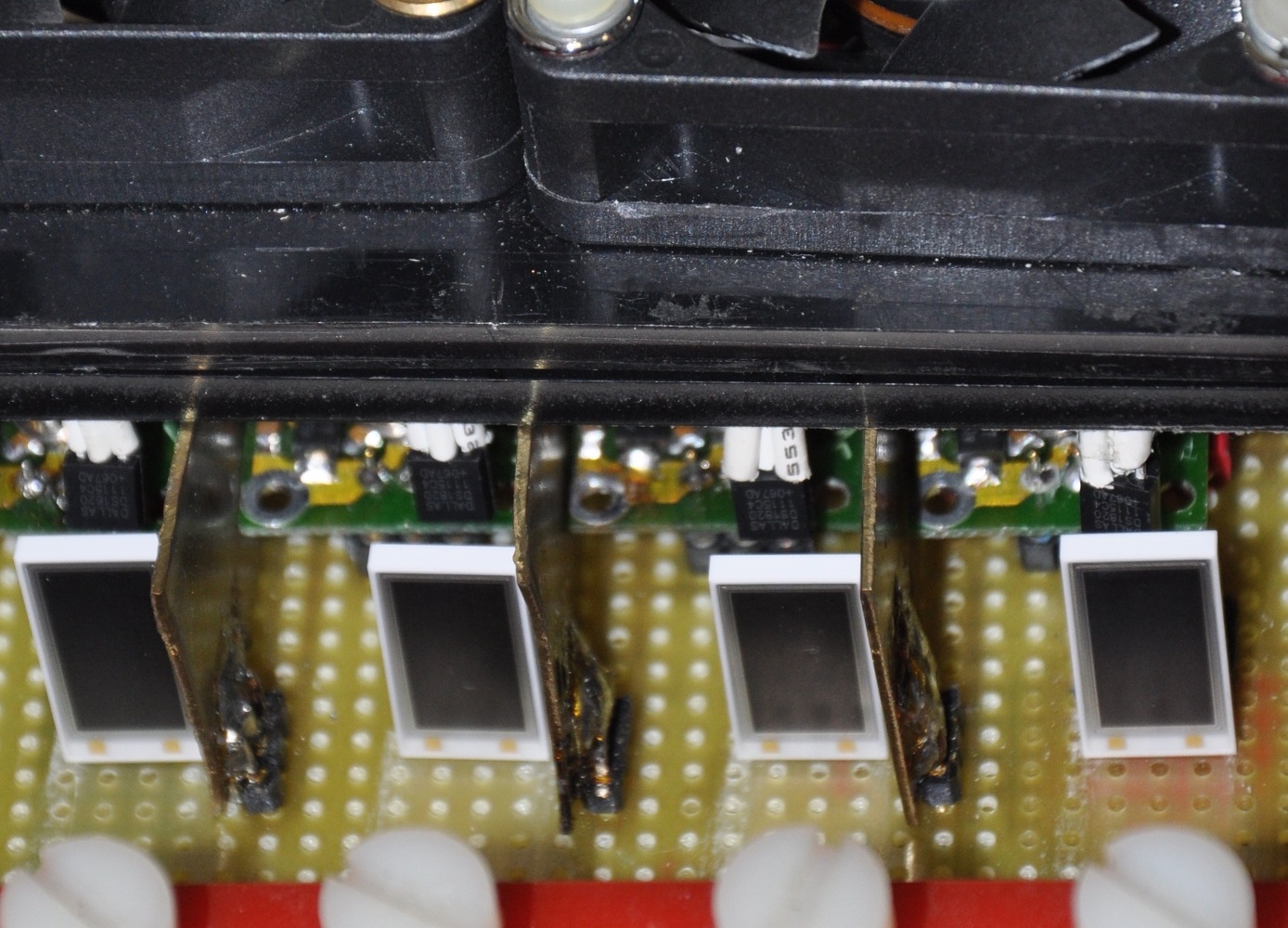}
\caption{Photograph of APDs in the  setup used for APD characterization.}
\label{pic:photo_apd_char_setup}
\end{figure}

A schematic of the setup used is shown in Fig.~\ref{pic:apd_char_setup}, a photograph of the central part of the characterization station in Fig.\ref{pic:photo_apd_char_setup}.
An LED with the same wavelength as the scintillation light is used to generate light flashes. The temperature coefficient of the LED is addressed by placing it in an environment with stable temperature. A light fiber guides the light to the APD under test. The APD is located in a separate environment in which the temperature can be set to a desired value. A temperature sensor (DS1820) directly on the backside of the APD is used to measure its temperature. The sensor's reading is used in the characterization procedure. Therefore, an absolute offset in the temperature control system is not relevant. The readout of the APD consists of a preamplifier and a sampling ADC.\\
The HV supply is programmable in order to allow for automated measurements of the APD's voltage dependent characteristics.  During reference measurements at zero bias voltage, the HV supply is disconnected with a relais. Instead, a $\SI{1}{\mega\ohm}$ resistor is used to minimize the residual bias voltage.

%An overview of this setup will be given in Section~\ref{ssec:apd-char}.

While developing the setup, a few features were found to be very important. Only a brief overview will be given here. A more detailed discussion can be found in ref. \cite{Urban_18_diss}.
%\begin{itemize}
%\item 
\subsubsection*{Increase the preamp's decoupling capacitor}
The APD's parasitic capacitance increases vastly when the bias voltage approaches zero. This causes two problems in combination with the charge sensitive preamplifiers used (see~\ref{ssec:preamp}): The noise level increases significantly and the signal level decreases. The latter is caused by charge division between the APD's parasitic capacitance and the coupling capacitor of the preamplifier (C1 in Fig.~\ref{pic:sch-preamp}).  Increasing its value allows to retain most of the signal.
%voll entladen, sonst temp abhängigkeit, oder nur großer fehler?
%\item 
\subsubsection*{Discharging the APD} It is important to minimize the residual bias voltage in order to achieve a precise reference measurement.
The increased APD capacitance at low bias voltages leads to more problems: The capacitance changes quickly at low voltages. Therefore the fraction of signal lost due to charge division changes quickly, too. An undetermined residual voltage will result in a strong systematic shift of the resulting amplitude, rendering the measurement useless.\\
We decided to permanently discharge the APD externally to zero bias voltage to achieve stable reference conditions.
%cross talk, sogar mit shield zwischen einzelnen APDs. -> einzeln messen
\subsubsection*{Cross-talk between neighboring APDs affects the amplitude}
As 3500 APDs were supposed to be characterized with this setup, accessibility of the APDs is an important feature. Switching the specimens between measurements needs to be reasonably simple. This resulted in limitations, how well neighboring APD channels in the setup can be electromagnetically shielded against each other and we were not able to fully remove cross-talk.\\
In particular, this is a problem as APDs have a spread in the bias voltage characteristics. Operating all APDs in the characterization station at the same bias would mean that one APD will most likely have a higher gain than others, introducing a systematic error in that measurement.\\
We decided to measure each of the four channels consecutively: Only one APD bias is turned on at a time, which results in constant conditions on each measurement spot for any combination of APDs.\\
As one measurement cycle (ramp bias up, measure signal amplitude, ramp bias down) is much faster (few minutes) than one full temperature cycle (several hours), APDs still can be characterized virtually in parallel.

The setup used allowed to characterize four APDs in parallel. One full characterization measurement took about five hours.
%temp sensor extrem nah an apd-> keine hysterese sichtbar
\subsubsection*{Measure the APD temperature directly at the APD}
Circulating the air inside the controlled temperature measurement chamber is important to equalize the temperature inside it. However, we still were able to see systematic variations inside the chamber.\\
%Only giving each measurement position its individual temperature sensor, located as closely to the back side of the APD as possible, allowed to get reproducible results.
To achieve reproducible results, it was necessary to allocate a separate temperature sensor to each measurement position, located as closely to the back side of the APD as possible.

A switchable attenuator improved the accuracy, practically implementing a dual range ADC.
The pins of the APDs purchased for the upgrade were made from a special non-magnetic alloy. Unfortunately the material is very soft, which causes the pins to bend easily and sometimes even break off when trying to bend them back straight.
%schaltbarer abschwächer: "dual range" adc
% luft umwälzen -> gleichmäßige temperaturverteilung
%weiche APD-Pins!!!
%\end{itemize}

A full characterization procedure consists of two parts. At first, the temperature is held constant while the voltage dependence of the gain is measured. In the second phase, the bias voltage is held constant at the value which should give a gain of $M=50$, according to the specification provided by the manufacturer. The temperature is slowly varied between \SI{20}{\celsius} and \SI{40}{\celsius}, while the gain is continuously measured. The measurement during rising and falling temperature allows to estimate a temperature differential between APD and temperature sensor. As no hysteresis is visible in the data, the differential is assumed to be negligible.

\begin{figure}[ht]
\centering
\includegraphics[width=\columnwidth]{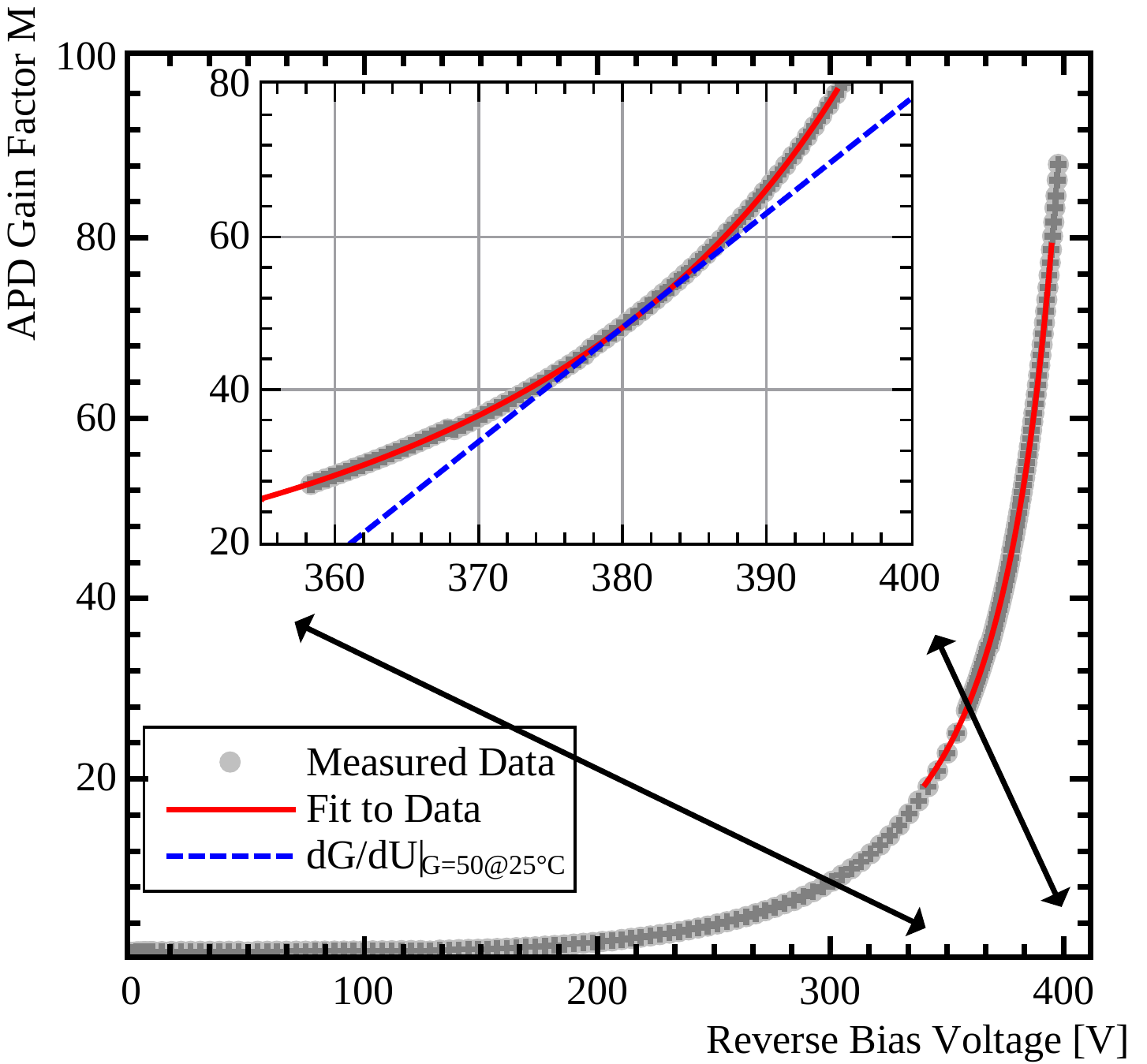}
\caption{Voltage dependence of the APD's gain \cite{Urban_18_diss}.}
\label{pic:g-vs-v}

\end{figure}

\begin{figure}[ht]
\centering
\includegraphics[width=\columnwidth]{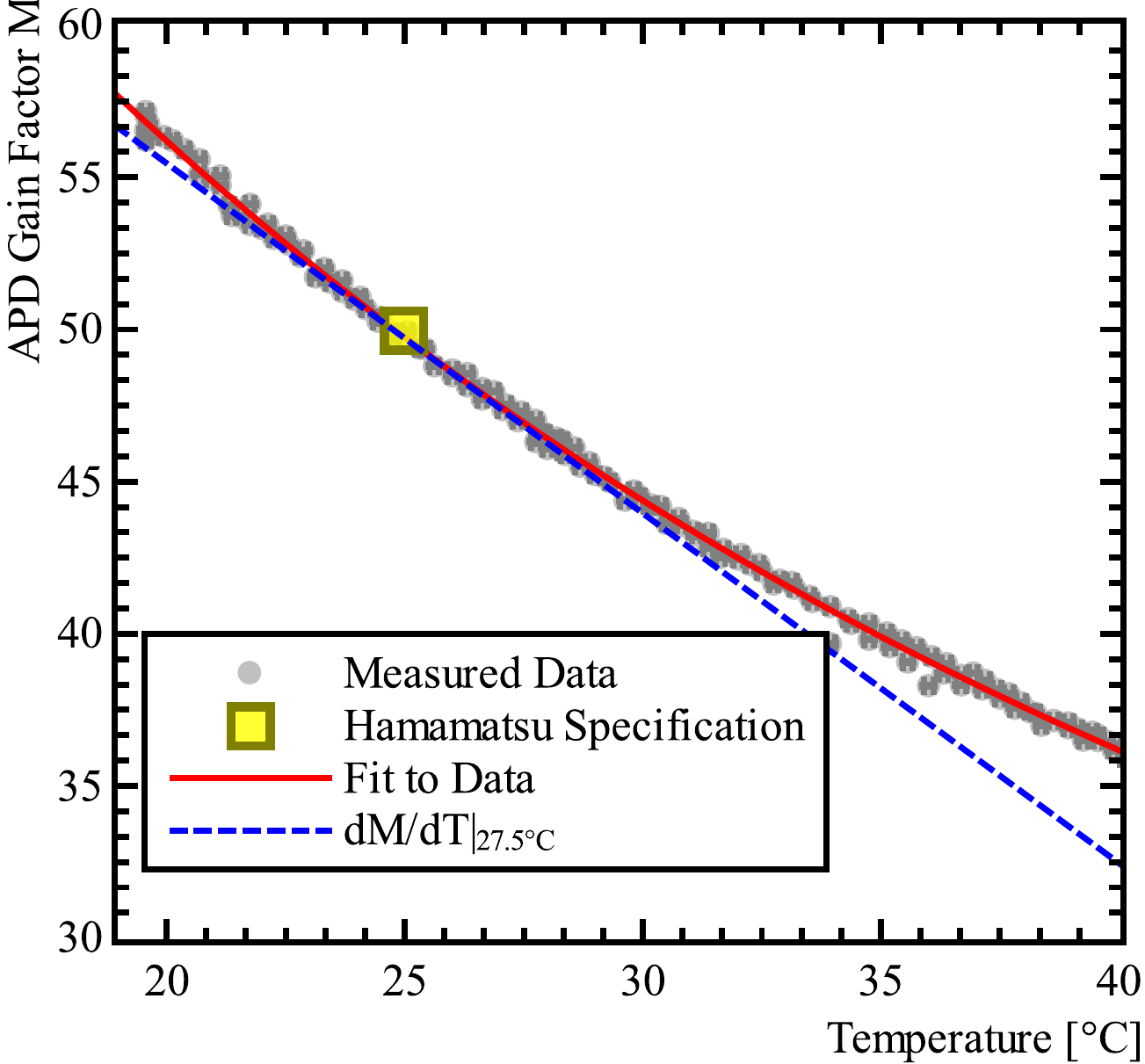}
\caption{Temperature dependence of the APD's gain \cite{Urban_18_diss}.}
\label{pic:g-vs-T}
\end{figure}

Figures~\ref{pic:g-vs-v} and~\ref{pic:g-vs-T} show typical results for the temperature dependencies on bias voltage and temperature respectively.


% x11048x3 panda apd
% gain 50, TK VK measured for all apds
% teststation, diss urban, c apd für kleine V!!! abb 3.17 diss u

The characterized temperature range is much broader than the expected operating range. The data was analyzed by fitting the following functions \cite{Urban_18_diss}. The voltage dependence was described by a modified Miller formula \cite{miller55}
\begin{equation}
M(V)=\frac{a_V}{\left(1-\frac{V}{b_V}\right)^{c_V}},
\end{equation}
and the temperature dependence by the empirical formula
\begin{equation}
M(T)=a_T+b_T\cdot\left(1+c_T\right)^T.
\end{equation}

To find the characterizing parameters, the slope of each function was determined as $\alpha_T=\frac{\text{d}M}{\text{d}T}\mid_{V=V_{50},T=\SI{27.5}{\celsius}}$ and $\alpha_V=\frac{\text{d}M}{\text{d}V}\mid_{V=V_{50},T=\SI{27.5}{\celsius}}$ respectively.

\begin{figure}[ht]
\centering
\includegraphics[width=\columnwidth]{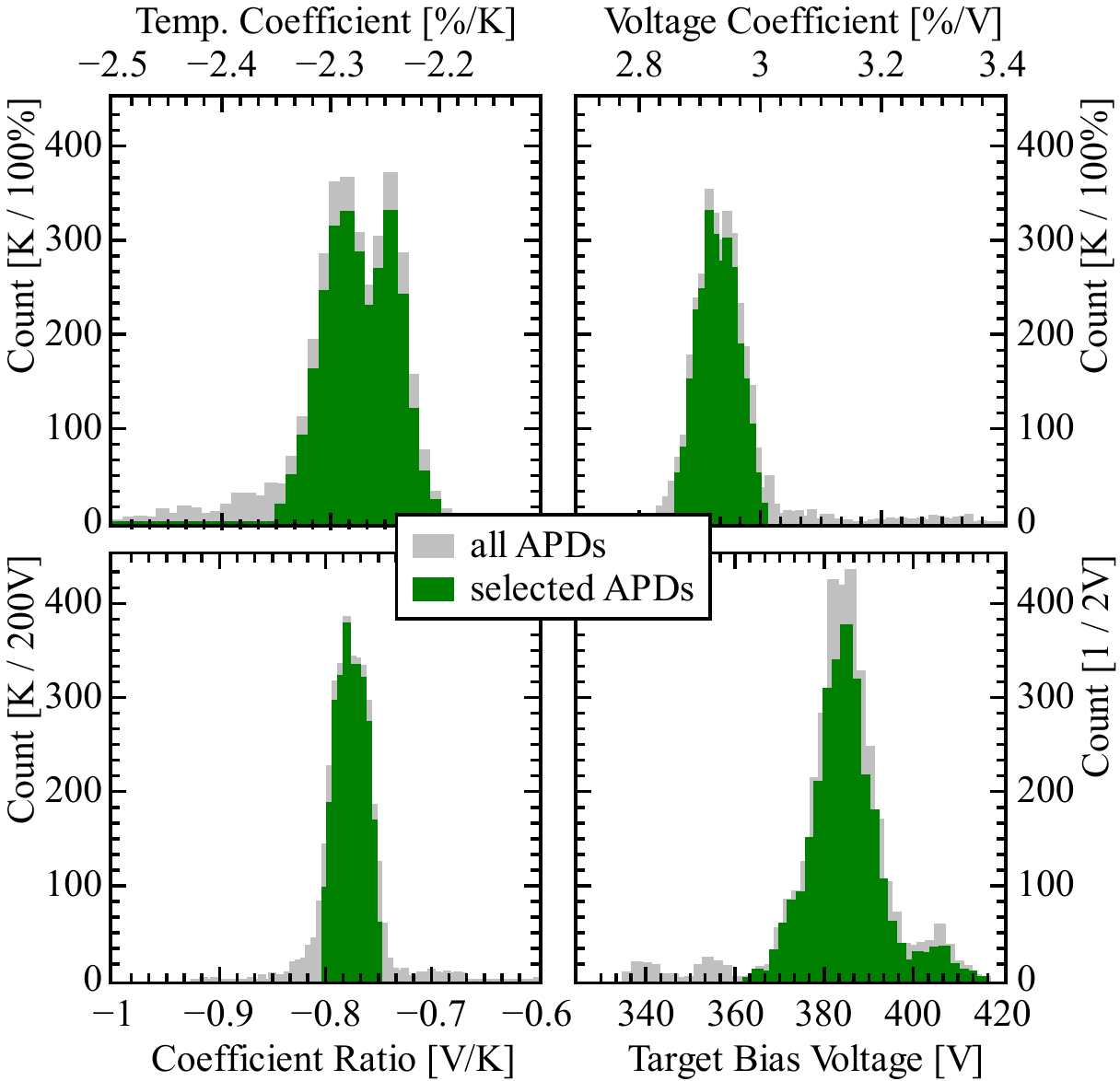}
\caption{Characteristic parameters of all (gray) and selected (green) APDs \cite{Urban_18_diss}.}
\label{pic:apd_char}
\end{figure}

Histograms of these parameters are shown in Fig.~\ref{pic:apd_char}. The variation of the gain coefficients was found to be small enough to use an identical compensation circuit for all APDs.

\begin{figure}[ht]
\centering
\includegraphics[width=\columnwidth]{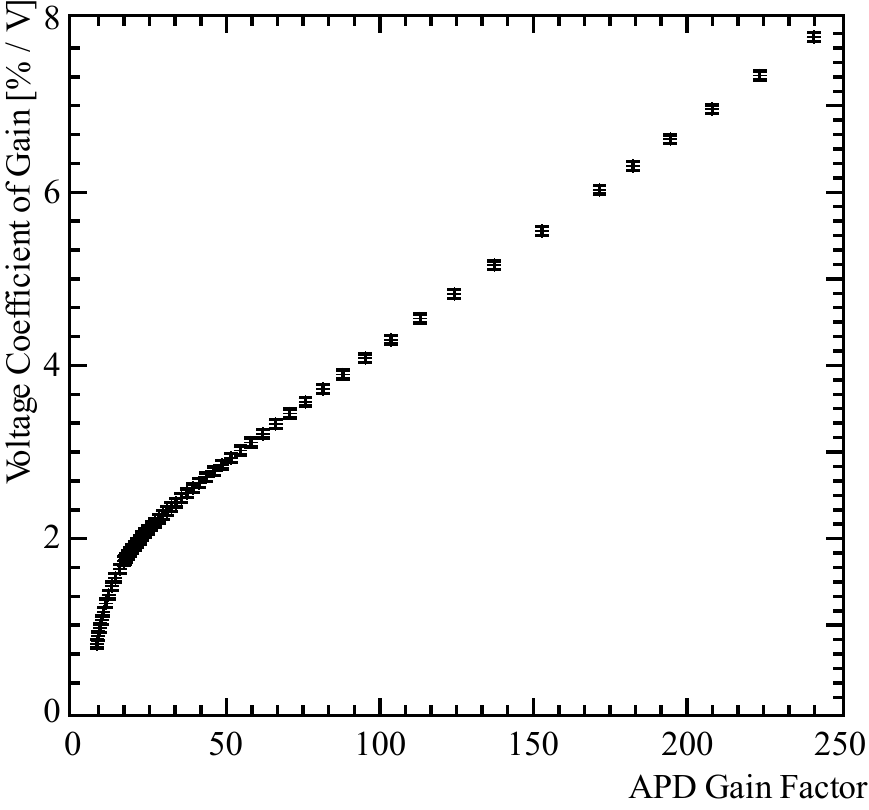}
\caption{Voltage coefficient of the APD's gain in dependence of the gain at \SI{27.5}{\celsius}. \cite{Urban_18_diss}}
\label{pic:apd_dgV_vs_g}
\end{figure}

\begin{figure}[ht]
\centering
\includegraphics[width=\columnwidth]{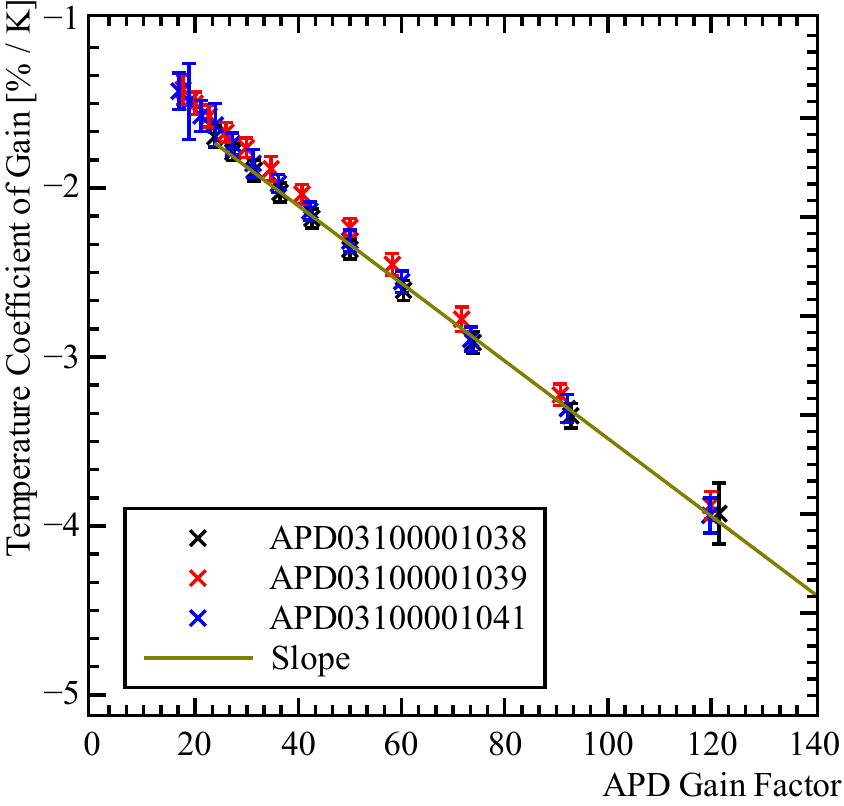}
\caption{Temperature coefficient of the APD's gain in dependence of the gain at \SI{27.5}{\celsius}. \cite{Urban_18_diss}}
\label{pic:apd_dgT_vs_g}
\end{figure}

%%%%%%%%%%%%%%
It should be noted that the parameters shown are valid for the nominal operating conditions of $M=50$ and $T=\SI{27.5}{\celsius}$. Both parameters also depend on the gain the APD is operated at. For example, increasing the gain of the APD from $M=50$ to $M=150$ increases the coefficients to $\alpha_V\approx\SI{5.6}{\percent\per\volt}$ and $\alpha_T\approx\SI{-4.5}{\percent\per\kelvin}$, respectively. The measured dependency is shown in Figs.~\ref{pic:apd_dgV_vs_g} and~\ref{pic:apd_dgT_vs_g}.\\
This increased sensitivity yields tighter constraints on the auxiliary systems. To achieve the same stability of the gain, smaller variations in temperature and voltage are allowed. For any kind of circuit that counteracts the effects of temperature variations with voltage adjustments, the difference between sensed APD temperature and true APD temperature becomes more severe as any difference leads to errors in the correction.\\
\subsubsection*{Electronic Noise}
Further APD properties depend on the gain, which have an impact on the overall detector system performance. A limiting factor for the precision of the energy and time information can be the electronic noise. In combination with a charge sensitive preamplifier two properties of the APD contribute to the noise level: The parasitic junction capacitance and the dark current.\\
The junction capacitance decreases with increasing reverse bias and reaches a plateau of \SI{254}{\pico\farad} at the nominal bias voltage. In a sample of 147 units, a variation of \SI{\pm4}{\pico\farad} was found \cite{Urban_18_diss}.\\
The dark current exhibits the opposite behavior: It increases with increasing reverse voltage. The increase is particularly pronounced when the bias voltage is large enough to enable the internal amplification mechanism.\\
APDs can be classified according to the implemented doping profile. The X11048 is a reverse type APD. Its key feature is the position of the avalanche multiplication region being close to the photo sensitive surface. However, the multiplication region is still deep enough in the semiconductor so that light is absorbed in front of it.

This results in different gain factors for photo current and dark current. The photo current electrons traverse the whole multiplication region, which results in a high multiplication gain. On the other hand, most of the dark current is generated in the bulk material of the semiconductor. 

Thus, only holes traverse the multiplication region, which have a much smaller gain than electrons.\\
More details about different APD types can be found in \cite{kataoka2005recent}.

Before the noise contributions of an APD on the preamplifier signal are discussed quantitatively, the more simple case of a PIN photodiode is mentioned. A derivation of the noise of a system consisting of photodiode and a charge sensitive preamplifier can be found in many textbooks (e.g. ref. \cite{spieler2005semiconductor}). Here, only key features will be summarized.\\
Three noise sources are commonly considered. The APD's dark current appears as shot noise, the conductive channel of the jFET exhibits thermal noise, and defects in semiconductors can lead to $1/f$ noise.

\begin{figure}[ht]
\centering
\includegraphics[width=\columnwidth]{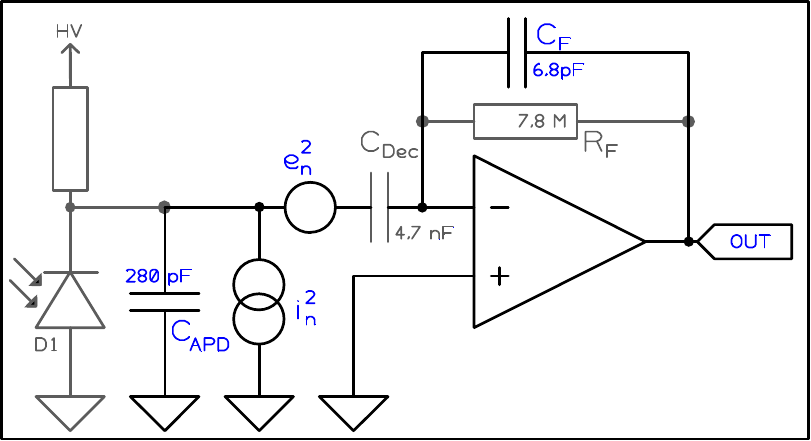}
\caption{Simplified schematic to derive the noise level of a system consisting of APD and charge sensitive preamplifier.}
\label{pic:noise-sch}
\end{figure}

The latter is mostly relevant for semiconductors with high radiation damage and ignored in the following. An equivalent circuit which is valid for both, APD and PIN photodiode, can be found in Fig.~\ref{pic:noise-sch}. The diode is implemented as a current source $i_n$, modeling the dark current, and a capacitor $C_D$, modeling the parasitic capacitance. The thermal noise of the jFET is implemented as a voltage source $e_n$.
The value of the feedback capacitor $C_F$ defines the overall gain, which applies not only for the signal but also for the noise.\\
Minor corrections arise from the values of the decoupling capacitor $C_\text{Dec}$, and the feedback resistor $R_F$, which will be neglected in the basic formulas.\\
As the noise amplitude strongly depends on the bandwidth it is measured at, it makes sense to specify the noise voltage density instead. The most simple case yields \cite{spieler2005semiconductor}
\begin{equation}
v_\text{total}^2\left(f\right)= 2 e I_D \frac{1}{\left(2\pi f C_F\right)^2} + 4k_BT\frac{2}{3 g_m}\left(\frac{C_\text{APD}}{C_F}\right)^2,
  \label{eqn:entw:rausch_csa_gesamt}
\end{equation}
where $g_m$ is the forward  transconductance of the jFET, $T$ its temperature and $I_D$ the dark current of the diode. It should be noted that this formula assumes that the first amplification stage has a sufficiently high gain to make the noise contributions of all following amplification and processing stages negligible.\\
The first term represents the shot noise originating from the dark current of the photodiode, the second term represents the thermal noise of the jFET.\\
Although both noise sources are white (i.e. not frequency dependent), the shot noise appears with a $1/f$ frequency dependence on the output. The characteristics of the charge sensitive amplifier cause this change. A full derivation can be found e.g. in ref. \cite{spieler2005semiconductor}.\\

Before the impact on time and energy resolution will be discussed, Equation \ref{eqn:entw:rausch_csa_gesamt} will be modified to cover the APD specific characteristics.
In particular, the contribution by the dark current needs to be modified. While the dark current of the PIN diode was characterized by a single value, the amplification process of the APD demands further characteristics, even for a most fundamental treatment.\\
The total dark current $I_D$ of an APD consists of two parts, one that undergoes amplification $I_B$ and one that does not $I_S$ \cite{Hamamatsu:useAPD, PerkinElmer:APD}:
\begin{equation}
I_D=I_S + M \cdot I_B,
  \label{eqn:entw:dunkelstrom_APD}
\end{equation}
with $M$ being the gain factor. The non-amplified part can be treated like the dark current of the classical photodiode as shot noise
\begin{equation}
i_{n,S}^2=2eI_S,
  \label{eqn:entw:schrotrausch_APD_IS}
\end{equation}
where $i_{n,S}$ is the noise current density. To properly describe the noise of the part that undergoes avalanche multiplication, two aspects have to be taken into consideration. First, the noise emerges when the dark current is created. Therefore, the shot noise of a $M$ fold amplified current $I_B$ is $M$ times larger than the noise of the unamplified current \cite{Hamamatsu:useAPD}. In contrast, the noise of a current that is $M$ times larger would result in an increase of the noise level by $\sqrt{M}$.\\
Second, the statistical nature of the amplification process yields a further fluctuation and therefore increase of the noise. This contribution is covered by the excess noise factor $F$ \cite{Hamamatsu:useAPD}.\\
Considering both effects, the shot noise contribution of the amplified part of the dark current is
\begin{equation}
i_{n,B}^2=2eI_B\cdot G^2\cdot F.
  \label{eqn:entw:stromrausch_apd}
\end{equation}
Considering also the 
%effect of the decoupling capacitor (leading to charge division between $C_{EK}$ and $C_P$) as well as the
feedback resistor $R_F$ (leading to a limitation for low frequencies) one gets
\begin{eqnarray}
v_\text{total}^2&=& 2 e (I_B\cdot F \cdot M^2 +I_S) \frac{1}{\left(\omega C_F\right)^2} \cdot g\left( \omega\right) \notag \\
%& & \cdot \left( \frac{\omega \tau}{\sqrt{1+\left(\omega \tau\right)^2}}\right)^2
%\left(\frac{C_\text{EK} }{C_\text{D}+C_\text{EK}}\right)^2 \notag  \\
& & + 4k_BT\frac{2}{3 g_m}\left(\frac{C_D}{C_F}\right)^2,
  \label{eqn:entw:gesamtrausch_apd}
\end{eqnarray}
%\begin{eqnarray}
%v_\text{total}^2&=& 2 e (I_B\cdot F \cdot M^2 +I_S) \frac{1}{\left(2\pi f C_F\right)^2}\cdot c_E^2 \notag \\
%& & \cdot \left( \frac{2 \pi f \tau}{\sqrt{1+\left(2 \pi f \tau\right)^2}}\right)^2\left(\frac{C_\text{EK} %}{C_\text{D}+C_\text{EK}}\right)^2 \notag  \\
%& &+4k_BT\frac{2}{3 g_m}\left(\frac{C_D}{C_F}\right)^2\cdot c_t^2,
%  \label{eqn:entw:gesamtrausch_apd_endformel_mittel}
%\end{eqnarray}
where $\tau=R_F C_F$, $\omega=2 \pi f$, $g\left( \omega\right)=\left( \frac{\omega \tau}{\sqrt{1+\left(\omega \tau\right)^2}}\right)^2$.

\begin{figure}[ht]
\centering
\includegraphics[width=\columnwidth]{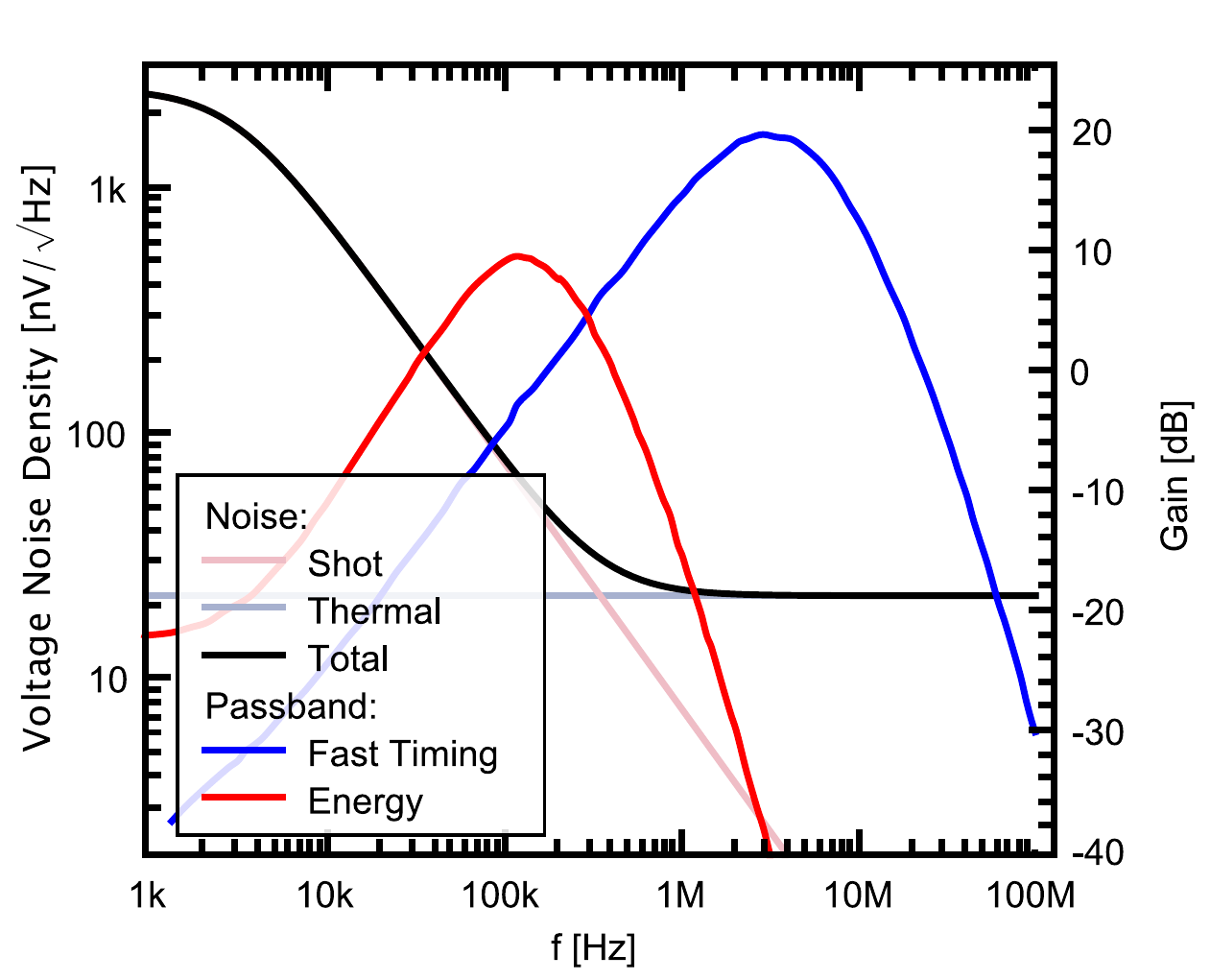}
\caption{Voltage noise density of the preamplifier output in comparison to the passbands of energy- and fast timing branch.}
\label{pic:noise_density}
\end{figure}

Fig.~\ref{pic:noise_density} shows the noise voltage density predicted by Equation~\ref{eqn:entw:gesamtrausch_apd}, as well as the individual contributions by thermal noise and shot noise.

To finish the discussion about the noise of the front-end, certain properties of the back-end need to be considered which will be introduced in Section~\ref{sec:backend}: The timing branch uses higher frequency components than the energy branch. Fig.~\ref{pic:noise_density} also shows the characteristics of the according signal-shaping filters. It demonstrates that the dominantly contributing noise sources differ for timing and energy branch. In the given setup, the passband of the timing branch is dominated by thermal noise while the energy branch has highest contributions from shot noise.\\

Considering Equation~\ref{eqn:entw:gesamtrausch_apd}, the impact of specific physical properties on the SNR can be investigated. Table~\ref{tab:entw:apd_pin} summarizes the overall relations.
For the ease of comprehension, it will be assumed in the following that the noise in the fast timing branch is entirely thermal noise and in the energy branch only shot noise.\\
In order to maximize the SNR in the energy branch, a high gain $M$ should be chosen. At a certain gain, $I_S/M^2$ becomes negligible compared to $I_BF$ and any further increase of the gain $M$ will not improve the SNR. In fact, $F$ depends on various factors and further increasing $M$ will therefore likely degrade the SNR.

More parameters are present which influence the SNR in the fast timing branch. The jFET property $g_m$ can be optimized within certain bounds by picking an appropriate component and its operation parameters. The smaller the parasitic detector capacitance is, the better the SNR will be. While the influence on this detector parameter is limited once a certain APD model is chosen, two aspects remain important: First, care needs to be taken to add as low capacitance as possible when designing the electrical connection between APD and preamplifier. Second, for zero bias voltage the APD's capacitance increases by one order of magnitude. This does not only lead to a vastly increased noise level in the fast timing branch but even becomes the dominating noise source in the energy branch.\\
Increasing the gain $M$ can improve the SNR. Limitations arise from increased shot noise for very high gain factors. It can reach levels at which it becomes the dominating noise source also in the fast timing channel. Another limitation is the avalanche breakdown for very high reverse bias voltages.\\
Cooling the jFET can decrease the noise level. As the scintillator material CsI is hygroscopic, cooling was implemented in the \CB only to a very limited degree which rules out water condensation.

As already discussed, an increased APD gain $M$ can bring another limiting factor for the resolution of the calorimeter in normal operation mode: the gain dependence on temperature and bias voltage.\\
At $M=50$, typical coefficients are $\alpha_T=\SI{-2.27}{\percent\per\kelvin}$, and $\alpha_V=\SI{2.92}{\percent\per\volt}$ \cite{Urban_18_diss}. At $M=100$ the coefficients are $\alpha_T\approx\SI{-3.4}{\percent\per\kelvin}$, and $\alpha_V\approx\SI{4.2}{\percent\per\volt}$ respectively.\\
This sets tighter limits on the stability of the APD's temperature and bias voltage.

One important factor for the stability of the bias voltage is the voltage drop over the biasing resistor (see Section~\ref{ssec:hv_front}). The signal current can yield a significant contribution for high rate conditions (total energy deposit in the scintillator per unit time).

Another limiting factor can arise for gain change compensation via bias voltage. The larger the temperature coefficient is, the more important is the necessity to have a small difference between measured temperature of the APD and its true temperature.

%%
%%
%%
%%
%%
%%Weiter mit: 
%%rauschen ändert sich unterschiedlich mit M, hoher gain hier eher für timing sinnvoll, nicht so für E. allgemeine aussagen aber schwierig: frequenzbereich des signals spielt auch eine rolle.
%%
%%

\begin{table}[ht]
  \centering
\begin{tabular}{l|l|l|}
\cline{2-3}
 & \parbox[c][2.6em][c]{0pt}{~}APD                                                                         & PIN                                          \\ \hline
\multicolumn{1}{|l|}{\parbox[c][2.6em][c]{0pt}{~}Signal Amplitude}      & $\propto\frac{1}{C_F}M$                                                     & $\propto\frac{1}{C_F}$                       \\ \hline
\multicolumn{1}{|l|}{\parbox[c][2.6em][c]{0pt}{~}Noise Amplitude (LF)} & $\propto\frac{\sqrt{I_BM^2F+I_S}}{C_F}$                                     & $\propto\frac{\sqrt {I_D}}{C_F}$             \\ \hline
\multicolumn{1}{|l|}{\parbox[c][2.6em][c]{0pt}{~}Noise Amplitude (HF)} & $\propto\frac{T}{g_m}\cdot\frac{C_D}{C_F}$                                  & $\propto\frac{T}{g_m}\cdot\frac{C_D}{C_F}$   \\ \hline
\multicolumn{1}{|l|}{\parbox[c][2.6em][c]{0pt}{~}SNR (LF)}           & $\propto\frac{1}{\sqrt{I_BF+\frac{I_S}{M^2}}}$ & $\propto\frac{1}{\sqrt{I_D}}$              \\ \hline
\multicolumn{1}{|l|}{\parbox[c][2.6em][c]{0pt}{~}SNR (HF)}             & $\propto\frac{M}{C_D}\cdot\frac{g_m}{T}$                                    & $\propto\frac{1}{C_D}\cdot\frac{g_m}{T}$     \\ \hline
\end{tabular}
  \caption{Comparison between APD and PIN photodiode: Signal amplitude and noise for low and high frequency ranges. Relations only valid close to the operating point $M=50$, $\vartheta=\SI{25}{\celsius}$}
\label{tab:entw:apd_pin}
\end{table}

\begin{figure}[ht]
\centering
\includegraphics[width=\columnwidth]{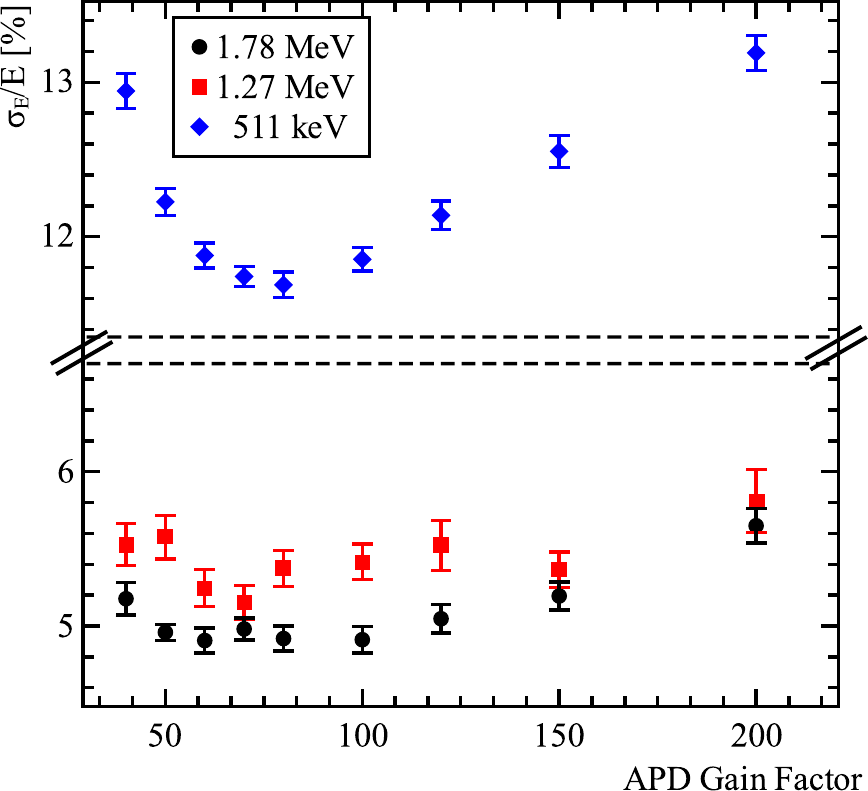}
\caption{Width of photopeaks of $^{22}$Na in one calorimeter crystal. The large size of the crystal and close position of the source allow to absorb both an annihilation photon and the decay photon of one event, resulting in a peak corresponding to the sum of both energies.}
\label{pic:sigmaE_G}
\end{figure}

Time and energy resolution were measured in a laboratory setup under specific conditions.\\
To get an idea of the gain dependence of the energy resolution, the $\gamma$ emissions from $^{22}$Na were used.
The width of the photo peaks is shown in Fig.~\ref{pic:sigmaE_G}.

The resolution values obtained for the positron annihilation photon is the data set with the smallest uncertainty and shows a minimum at $M\approx 80$. The gain was determined before each measurement.\\
As each measurement took only a few ($\lesssim 5 $) minutes, no major temperature changes are expected. Also, any voltage drop at the bias resistor is assumed to be constant. %, if present at all in relevant level.
Therefore, we assume the resolution degradation for higher gains to originate from an increased excess noise factor $F$.

As sources that limit the signal resolution ref. \cite{PANSART1997186} lists:
The number of detected photons, the electronic noise, the fluctuations in the avalanche process (expressed via $F$), and charged particles going through the APD. 
For an Hamamatsu APD ref. \cite{PANSART1997186} also shows that its excess noise factor $F$ increases by a factor of 1.5, when the gain increases from $M=50$ to $M=200$. This is compatible with the observed resolution degradation towards high gains in Fig~\ref{pic:sigmaE_G}.

It should also be mentioned that there are more factors that limit the resolution of the calorimeter (see Section~\ref{ssec:E_res}). The electronic noise is the limiting factor only for low energy deposits.

%Der Dunkelstrom einer APD setzt sich aus zwei Teilen zusammen, einem, der den Verstärkungsprozess durchläuft, und einem, der nicht verstärkt wird \cite{Hamamatsu:useAPD}. Für den Gesamt-Dunkelstrom $I_D$ gilt

%\cite{PerkinElmer:APD}, wobei $I_S$ der Anteil ist, der nicht verstärkt wird, und $I_B$ Anteil, der verstärkt wird.

\subsection{High Voltage Supply and Monitoring}
\label{ssec:hv_front}
This section describes the requirements on the bias supply electronics and its implementation. At the end of this section a few possible improvements are listed.

\subparagraph*{Requirements}
As seen in Fig.~\ref{pic:apd_char}, the APDs require individual bias voltages in the range of $\SI{320}{\volt} \lesssim V_B \lesssim \SI{420}{\volt}$. This range should be accessible with the supply. Aging of the APDs and irradiation damage might not only change the required bias voltage, but also change it by a different amount for different APDs. 
Therefore, it is favorable in the long run
%makes sense 
to be able to set each APD's bias individually, instead of only grouping APDs with similar voltages.\\
To allow monitoring, the actual bias voltage should be measurable.\\
Additionally, in order to account for the gain temperature coefficient, a compensation circuit should be implemented.

\begin{figure}[ht]
\centering
\includegraphics[width=.5\columnwidth]{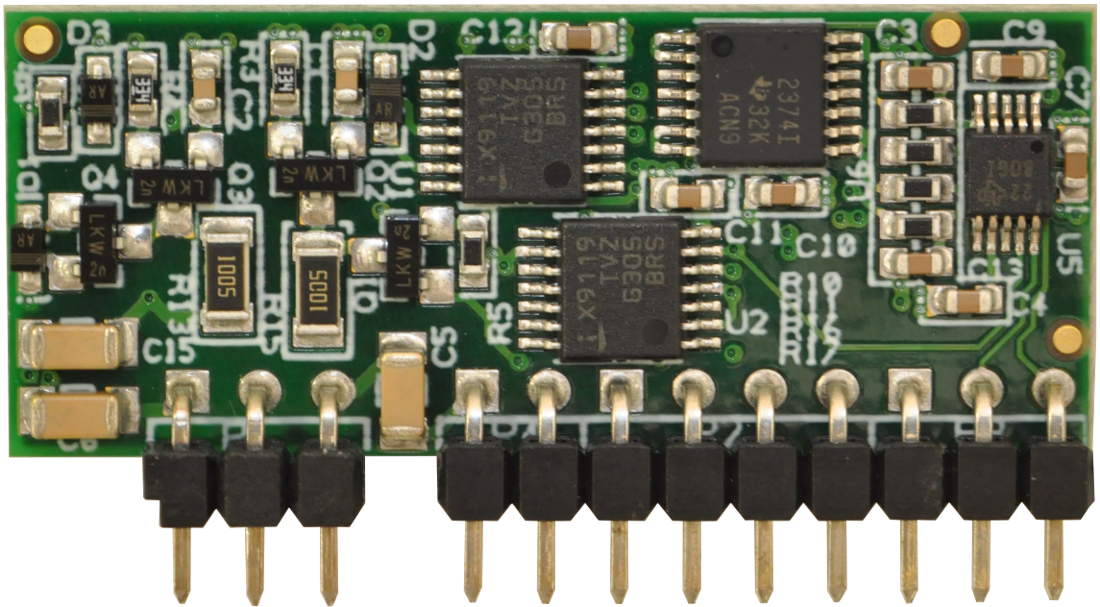}
\caption{Photograph of the bias supply card.}
\label{pic:HV_top}
\end{figure}

\subparagraph*{Implementation}
Fig.~\ref{pic:HV_top} shows a picture of one APD bias supply card, containing outputs and monitoring circuitry for two APDs.

\begin{figure}[ht]
\centering
\includegraphics[width=.85\columnwidth]{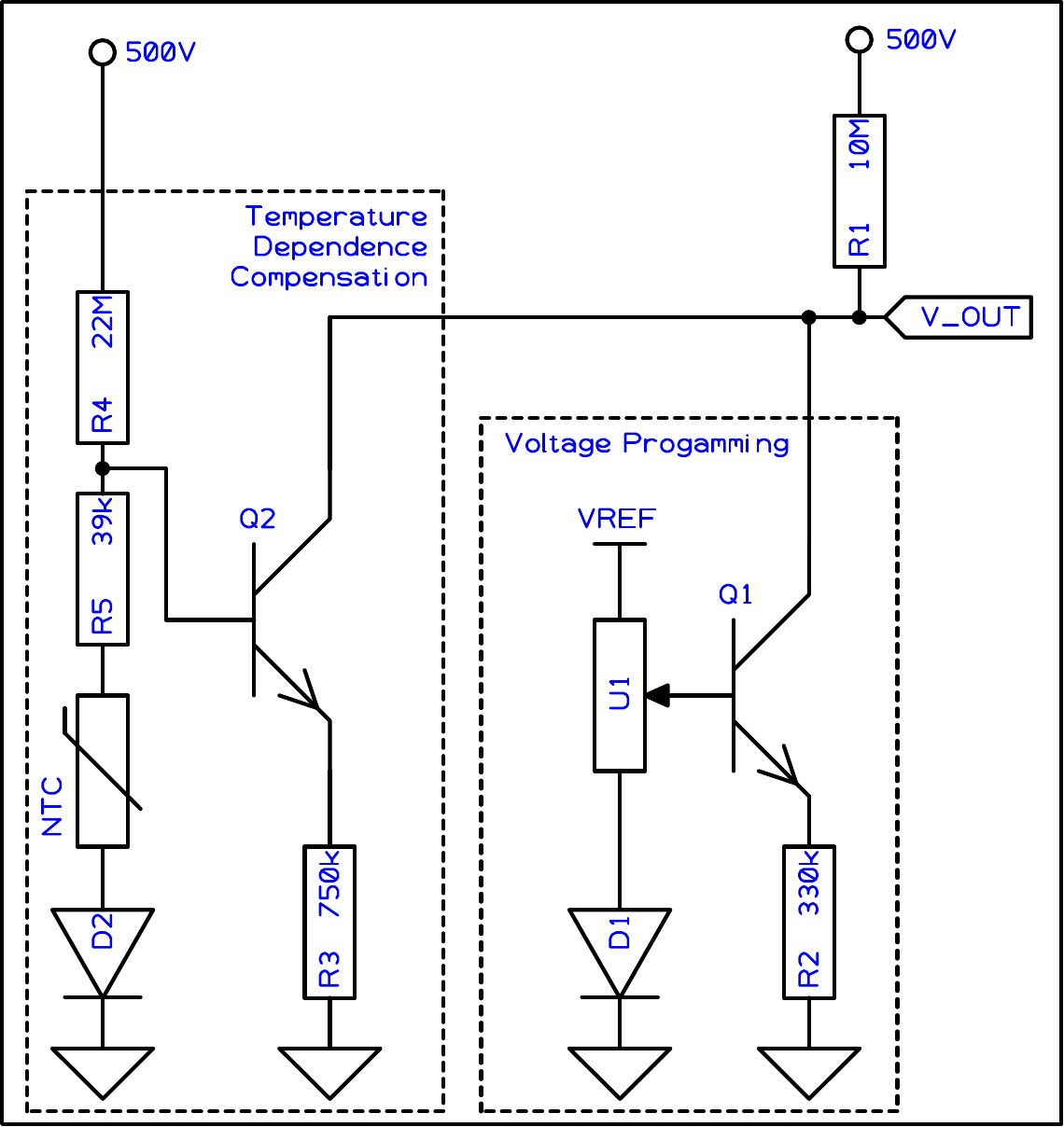}
\caption{Bias supply: voltage regulation circuit.}
\label{pic:HV_sch_reg}
\end{figure}
Each bias supply card is provided with a high voltage of \SI{500}{\volt}, which is used to generate the individual bias voltages. A simplified schematic of the circuit is shown in Fig.~\ref{pic:HV_sch_reg}. U1, Q1, R2, and D1 form a programmable current sink, where U1 is a 10 bit digital potentiometer (Renesas X9119).\\
The programmed current $I_P$ flows into the collector of Q1, originating in the \SI{500}{\volt} supply. $I_P$ generates a voltage drop across R1, therefore $V_\text{OUT}=\SI{500}{\volt}-I_P\cdot R1$.
D1 is used to compensate the temperature dependent voltage drop on the BE diode of Q1.

The remaining parts (Q2, R2, D2, NTC, R5, R4) build up the gain-temperature compensation. The NTC was selected to match the ratio of the gain coefficients ($\frac{\alpha_T}{\alpha_V}$).
The current which is generated by this compensation circuit also flows through R1 and thus alters the output voltage.
%The this way generated current flows also through R1, altering the output voltage.

This part of the circuit has an output impedance of $\SI{10}{\mega\ohm}$. While the impedance of the bias circuit needs to be high to work in combination with the preamplifier, a too large value will also cause problems. Signal current and dark current cause a voltage drop according to the output impedance, which reduces the bias voltage of the APD. %it also should not exceed an upper limit.\\
%The latter is result of the detector rate dependent signal current flowing through the APD. This current will create a voltage drop on the bias voltage, according to its output impedance.\\
%This has to be taken care of 
This can degrade the detector resolution in particular, if the detector rate varies over time. At the \CBE this is clearly the case as the beam of the accelerator has a macro structure in the magnitude of seconds.
An output impedance of $\sim\SI{1}{\mega\ohm}$ was found to result in a negligible distortion of the gain \cite{Honisch_15_diss, Urban_18_diss}.

\begin{figure}[ht]
\centering
\includegraphics[width=.7\columnwidth]{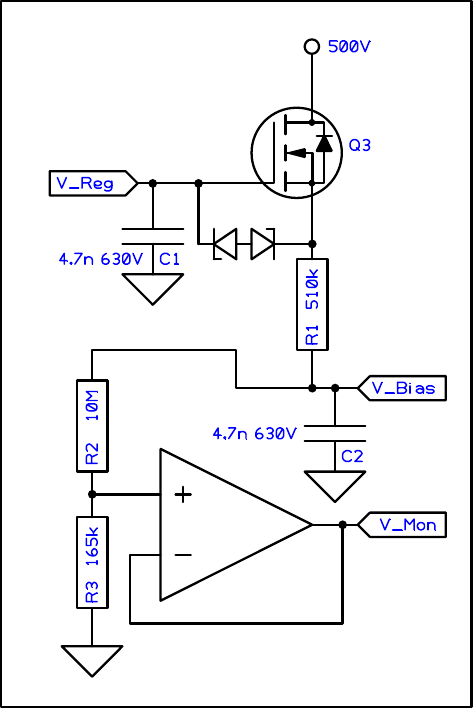}
\caption{Impedance conversion, filtering, and monitoring of the bias voltage.}
\label{pic:HV_sch_mon}
\end{figure}

Decreasing the value of R1 (in the schematic from Fig.~\ref{pic:HV_sch_reg}) would significantly increase the power dissipation of the supply module and is therefore disadvantageous. Instead, the impedance is decreased by a FET in source follower configuration. The corresponding circuit is shown in Fig.~\ref{pic:HV_sch_mon}.\\
C1 and C2 are part of low-pass filters to stabilize the bias voltage. If those capacitors were omitted, the noise on the bias voltage line would contribute to the noise of the preamplifier output. The biasing resistor with a value of $\SI{510}{\kilo\ohm}$ is located on the preamplifier directly at the input of the amplifier circuit.\\
R2 and R3 form a voltage divider, allowing to measure the bias voltage using a standard ADC. Connecting the divider at the lower end of R1 allows so see half of the bias voltage drop introduced by a high detector rate, which would not be possible if the divider was connected directly to the source of Q3.\\
A further improvement might be achieved by connecting the divider directly to the bias voltage at the APD.
However, in that configuration
%it should be investigated that no 
interference originating in the measurement circuit might couple into the extremely sensitive input of the charge sensitive preamplifier.\\
The divided voltage is measured by a 16-bit ADC (ADS1115). As the input bias current of the ADC might cause distortions on the high impedance of the voltage divider, an operational amplifier is inserted as buffer.\\
Further ADC inputs are connected to the voltage divider built up with the NTC and can be used to measure the temperature.
%10 bit poti (NV)
%16 bit ADC
% Analog APD-TK compensation
% Bachelor thesis: active compensation

\subparagraph*{Critical Review and Possible Improvements} While the bias supply card is successfully being operated at the calorimeter, there are still things that could be improved in future iterations.\\
One possible improvement is to connect the voltage divider for monitoring directly at the APD. This would allow to directly measure its bias voltage. In the current circuit, only half of the load current caused drop is seen in the measurement and the actual value of the APD bias voltage needs to be extrapolated.\\
Also the operational amplifier (see Fig.~\ref{pic:HV_sch_mon}) should be exchanged by a better suited version. The monitoring voltage should be identical to the unloaded output voltage of the voltage divider formed by R2 and R3. The input bias current of the operational amplifier is applied as a load to the voltage divider. \\
The input offset voltage of the operational amplifier also adds a deviation to the monitoring output. A constant offset voltage can be compensated by a calibration. However, more  severe are its temperature coefficient and the problem that rail-to-rail inputs have sometimes a steep change of the offset voltage at a certain common mode level (see Section 5.9.1.B of \cite{horowitz2015art}).

Another interesting option would be to replace the digital potentiometer (Fig.~\ref{pic:HV_sch_reg}, U1) by a DAC, as such are available with higher resolution than digital potentiometers. A microcontroller could be used to load the desired value on power-up.
Also, a more precise compensation of the gain-temperature coefficient could be implemented with a microcontroller. Here, a formula or a look-up table  may be used that matches the characteristics of the connected APD, instead of an NTC that resembles it "rather well". Furthermore, this compensation could simply be tuned for individual APD characteristics.
Using a look-up table to determine the optimal voltage was successfully tested with a prototype of the CB readout 
% was shown to work for the bias supply cards using the present digital potentiometers and a remote configuration
\cite{Pauli_14_bachelor}.\\
A charge sensitive preamplifier is an extremely sensitive device and therefore susceptible for electromagnetic interference (EMI). Digital electronics are a well known source of EMI. 
%While care needs to be taken when using a microcontroller in close proximity to the sensitive preamplifier, 
However, the authors think shielding the preamplifier from the microcontrollers EMI should be possible using proper layout techniques and proper filtering circuits.

%When building a stable bias supply without any analog temperature coefficient compensation, 
If the analog temperature coefficient compensation is not needed in an application, the whole part of the circuit should be removed (left half in Fig.~\ref{pic:HV_sch_reg}).
In this case, the temperature dependence of $V_{BE}$ of Q1 (Fig.~\ref{pic:HV_sch_reg}) seems to be the limiting factor for the output stability when the temperature is varied.

\begin{figure}[ht]
\centering
\includegraphics[width=.7\columnwidth]{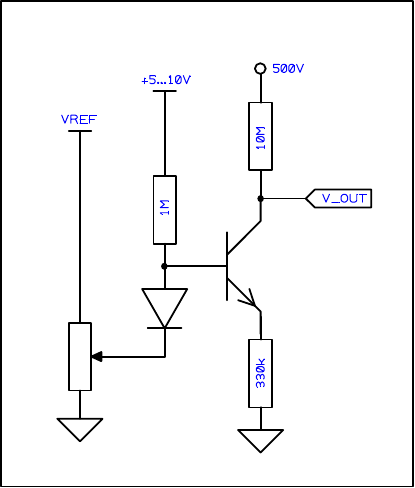}
\caption{Improved HV regulation circuit that cancels the temperature dependence of the base-emitter diode.}
\label{pic:HV_reg_imp}
\end{figure}

The forward voltage of D1 has approximately the same temperature coefficient as the base emitter diode of Q1. These two dependencies are supposed to cancel each other out, leaving the output voltage unaffected.
Unfortunately, this works only for a very limited range of potentiometer settings. The remaining temperature dependence is irrelevant for the readout of the CB. However, in an application that requires output voltages that are stable against temperature changes, the circuit should be modified:
%Unfortunately, the compensation via D1 (same schematic) does not work really well. For potentiometer settings that result in $V_\text{Q1,Base} \approx V_\text{REF}$, the diode D1 looses any effect.\\
The temperature dependence of $V_{BE}$ can be compensated with two emitter followers in series (PNP+NPN). As the first transistor does not need to have considerable current gain, it can even be replaced with a diode (see Fig.~\ref{pic:HV_reg_imp}).

\subsection{Signal Transmission}
To achieve immunity against pickup noise, the analog signals from the front-end are transmitted as differential signals via shielded cables (Samtec Twinax TTF-30100-01-01).\\
In order to maximize the immunity, the signal amplitude should be as high as possible. Two limitations apply here. First, the quiescent power dissipation of the line driver puts a limit on which peak voltage can be achieved. A maximum amplitude of \SI{4}{\volt} for each of the signal lines was chosen.\\
Second, the signal amplification is limited by the required dynamic range in units of MeV. Analyzing Monte Carlo simulations and existing data from measurements at ELSA, energy deposits of up to \SI{2}{\giga\electronvolt} were found \cite{Urban_18_diss, Honisch_15_diss}. As such high deposits are rare, the pile up of high deposits exceeding this energy can be neglected, even considering the slow decay time of the preamplifier signal ($\tau\approx\SI{58}{\micro\second}$).\\
To be able to handle unforeseen deviations in the system, a dynamic range of \SI{2.5}{\giga\electronvolt} was chosen.

\begin{figure}[ht]
\centering
\includegraphics[width=\columnwidth]{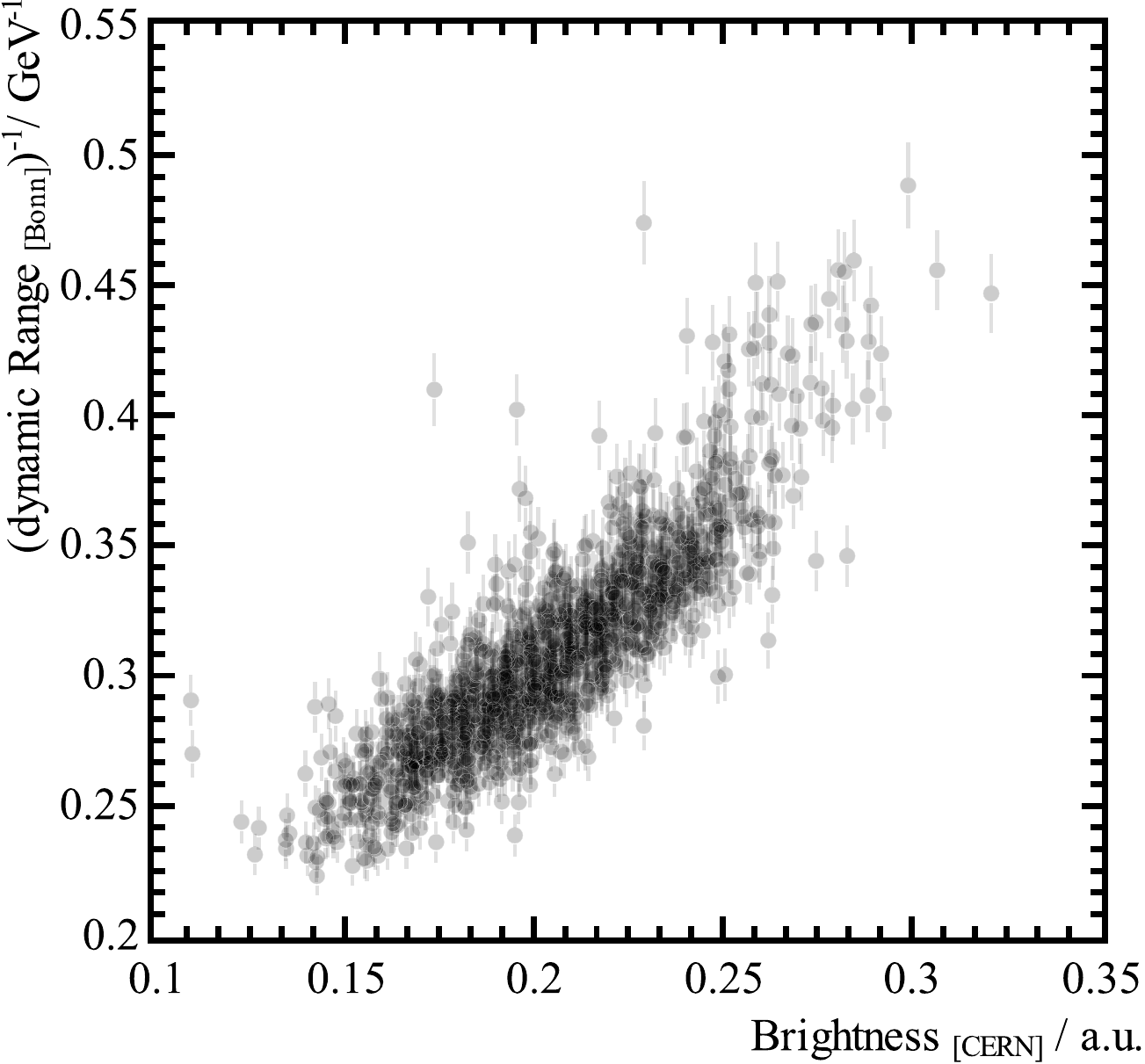}
\caption{Scintillation intensity of all detector crystals. Data from assembly of the calorimeter at CERN and from calibration during the upgrade in Bonn. \cite{Urban_18_diss}}
\label{pic:crystal_brightness}
\end{figure}

\subparagraph*{Variable Gain}
A disadvantageous  property of the CB's detector crystals is the large variation in scintillation brightness.\\
The light output varies by a factor of 2.6 among the individual crystals. This was seen in both, calibration data with the new readout and in existing data from the time when the the calorimeter was built up at CERN. Fig.~\ref{pic:crystal_brightness} shows both datasets in comparison \cite{Urban_18_diss}.

To compensate this variation and therefore to equalize the dynamic range on the transmission line, the single-ended-to-differential converter was built up with a programmable gain. The gain can be chosen using 4 bit solder jumpers. The remaining spread after calibration was found to be roughly 5\% \cite{Urban_18_diss}.

\subparagraph*{Signal Switch}
As introduced in the beginning of Section~\ref{sec:newConcept}, each detector module is read out by two APDs. The combination of both APDs reduces the noise level, compared to a single APD. Also, if one APD or the corresponding preamplifier malfunctions at some point during the operation of the calorimeter, it can just be switched off. In this backup mode, only the remaining APD is used at a slightly worse noise performance.

%The main benefits are a lower noise level when both signals are combined properly and a higher safeguarding against failure.\\
In order to not double the number of cables leaving the calorimeter, the two signals are combined to one signal inside the front-end.\\
A multiplexing circuit can be configured to forward the signal of APD 1, 2, or the average of both. The multiplexer is implemented \cite{Honisch_15_diss} using an operational amplifier (OPA2889) in the style of the video multiplexer, proposed in its datasheet \cite{datasheet:OPA2889}.

\begin{figure}[ht]
\centering
\includegraphics[width=.9\columnwidth, angle=180]{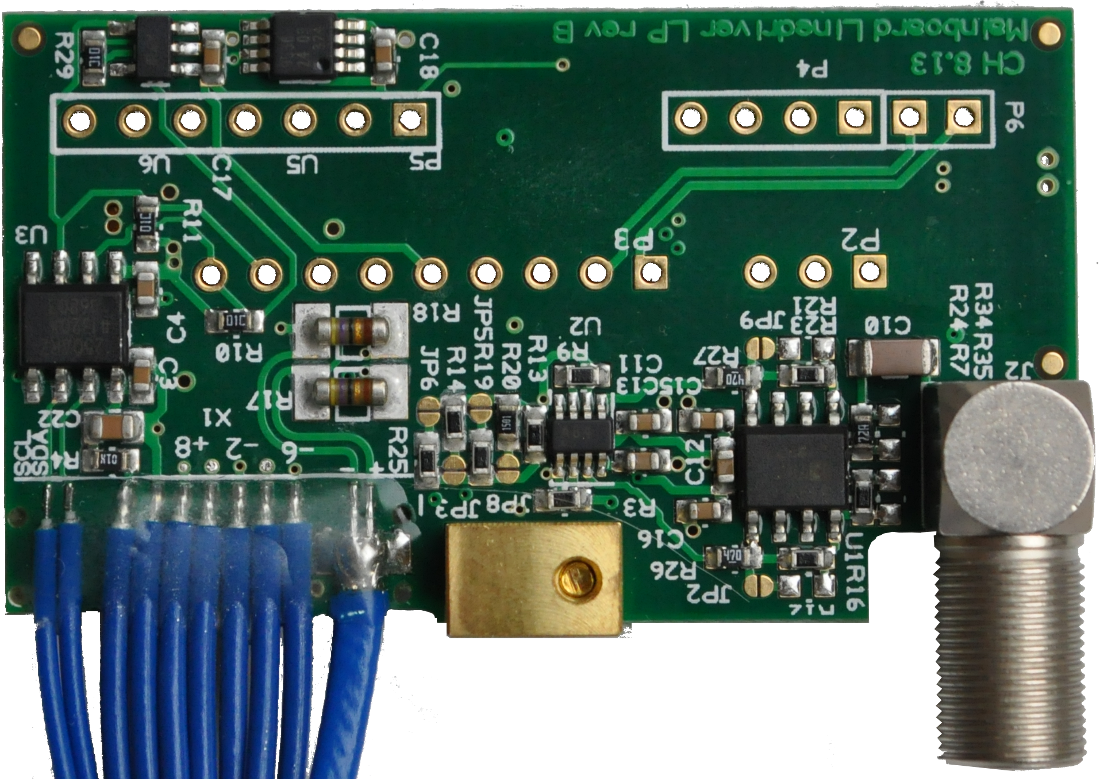}
\caption{Mainboard of the front-end housing line driver, multiplexer, galvanic isolation of the I2C slow control.}
\label{pic:mb}
\end{figure}

\subparagraph*{Implementation}

Fig.~\ref{pic:mb} shows the PCB that contains the line driver and the mutliplexer. It also contains a galvanic isolation for the I2C slow control, a temperature sensor, and voltage regulators. The bias supply card and the preamplifier are installed as daughter boards on this PCB.\\
The connection to the back-end is implemented using custom made cables (top right corner in Fig.~\ref{pic:mb}). The analog signal is transmitted via a twinax cable. The power supply is provided using low impedance cables $Z=\SI{20}{\ohm}$ (TCF-2620 by Samtec).\\
For the I2C signals SDA and SCL, \SI{75}{\ohm} cables are used (TCF-3875F by Samtec). The high impedance was chosen to minimize the parasitic capacitance introduced by the shield.\\
These cables are connected to a connector of type ECDP-08 with a customized pinout.
The \SI{500}{\volt} bias supply is connected through a Lemo 00 series connector.

\subsection{Further Components}
The main components of the front-end are the preamplifier, the bias supply, and the line driver. Additionally, there are a few periphery components: The I2C bus, which is galvanically isolated from the back-end via an ADuM1250. An I2C temperature sensor (MCP9802) allows to easily measure the temperature inside the front-end. A PCA9536 GPIO chip is used to configure the signal multiplexer. An EEPROM (24AA02E48) located on the HV supply card, contains a unique ID and therefore allows to identify each of these cards, as well as all assembled detector modules. Voltage regulators on the mainboard reduce the number of supply rails that need to be provided externally.

\subparagraph*{Critical Review and Possible Improvements}
While also this part of the front-end is successfully operating, a few things could be improved further.\\
The thin \SI{75}{\ohm} slow control cables turned out to be extremely sensitive to mechanical stress. Unfortunately, the conductor broke right at the strain-relief in a few cases. Choosing thicker cables seems to be the most obvious solution. Even if this implies the usage of \SI{50}{\ohm} cables with higher parasitic capacitance, the decreased maximum bus frequency seems acceptable.

During a beamtime, the multiplexer allowed the continued usage of detector units in which one of the APDs had failed. Also, the option to only enable one of both signals simplified the calibration procedure.\\
Unfortunately, the multiplexer circuit contains a bug. It uses a feature of the operational amplifier OPA2889, which is to shut down each of the two outputs individually. The datasheet states that the output goes into a high-impedance state and specifies an off isolation of \SI{70}{\deci\bel}. However, if the voltage between IN+ and IN- rises above roughly \SI{600}{\milli\volt} the current flow into the input increases exponentially, corresponding to a low input impedance.\\
In practice, turning off one channel does not work if high amplitudes are present. This occurred in the CB in a few units in which one APD had an abnormal dark current of \SI{10}{\micro\ampere} (see Section~\ref{sec:failure}).

To allow maintenance, connectors are required between front-end and back-end. 
The connector can be placed directly at each individual detector module. Prototypes in this configuration showed an insufficient light tightness. Ambient light entering the crystal and finally reaching the APD would cause increased noise, which can look like an increased dark current for continuous light sources or like mains hum for e.g. incandescent lamps.\\
As it was much easier to seal cables rather than connectors, it was decided to have a cable of \SI{30}{\centi\meter} mounted directly to the detector module, with a connector only on its other end.

Only when the detector was assembled, we realized how much this fixed cable complicates the assembling process and also the procedure of removing one single module for maintenance.
Retrospectively, we think having a connector directly on each module would be the better choice.

% with a connector on its end coming out of each crystal. This decision solved the required light tightness of each unit. Ambient light entering the crystal and finally reaching the APD would cause increased noise, which can look like an increased dark current for continuous light sources or like mains hum for e.g. incandescent lamps.\\
%On the downside, having no connector directly on the individual detector modules (i.e. a unique cable on each one) makes the assembly and maintenance of the detector a lot more difficult. 
%However, having cables fixed to the individual units complicates assembly and maintenance of the detector a lot.
%An increased effort in the design phase to get the connector light tight,
% for light tightness 
%seems to be worth the gain in flexibility.

The last design decision discussed in this section might affect the noise level in the trigger branch.
Another upgrade of the experiment is planned in which a TPC and a \SI{1}{\tesla} magnet are to be installed. The magnet available from the CB setup at CERN has tight restrictions on the space available to guide cables out. Therefore, very early in the design phase of the readout, it was decided to place the timing shaper into the back-end. Placing it into the front-end would result in doubling the number of signal cables that have to be routed through the magnet.\\
As it will be shown in Section~\ref{sec:timing-filter}, the frequency components in the preamplifier signal that are used in the timing branch contribute only a small fraction to the full amplitude. Therefore, the timing signal needs to be amplified after removing the slow components that make up for the major part of the amplitude.\\
A guideline in analog signal processing is to amplify the signal as much as possible and as soon as possible, to reduce the contributions to the noise level by elements
%in the rear 
at the back end
of the signal chain. In other words, no other than the unavoidable physical noise sources (e.g. dark current of the APD) should limit the SNR.
While the physical noise sources (see Sec.~\ref{ssec:apd}) dominate the noise level, for specific conditions (programmable gain in front-end is minimal) the contributions by the line driver in the timing branch are not entirely negligible.\\
Locating the timing shaper in the front-end, or maybe only the high-pass part of it, might resolve this and further improve the SNR for these cases.

%next: was noch auf dem mb ist: temp sens, isolation, v reg, mux.

% Custom developed cables, twinax, ...
% I2C: set voltage, monitor voltage / temp. Isolated.

%critical review: kabel anders herum. t-shaper in frontend wg. noise, hier verworfen wegen kabel, magnet.
%\FloatBarrier
\section{The Back-End and the Ultra-Fast Cluster Encoder}
\label{sec:backend}

\subsection{Timing Signal Shaper Module}
\label{sec:timing-filter}
The timing signal shaper is the key component to obtain hit information from the CB within the available latency of $\mathcal{O}\left(\SI{500}{\nano\second}\right)$. In the combination with threshold discriminators, two aspects of the shaping characteristics are important: First, the output signal needs to rise fast enough that a proper threshold is crossed early enough. Second, the signal should decay quickly, as deadtime results from the signal being above the threshold. Which decay time can be considered fast enough depends on both, the amplitudes and the corresponding rates of pulses.

%kompromiss zwischen SNR und peaking time
%CR with fixed PZC, 
The filters consist of a first order high-pass including a pole-zero compensation and a second order low-pass, implemented in multiple feedback topology.

The key parameter of this filters is the time constant used. It determines which part of the input signal's frequency spectrum is maintained and therefore determines the rise time of the output signal.

The smaller the chosen time constant, the faster a maximum is reached by the resulting output signal.\\

The minimally choosable time constant is limited by the relatively slow light emission of CsI(Tl).
%Unfortunately, the scintillation light is emitted relatively slowly, 
%which 
This results in a preamplifier signal, that has only small contributions from higher frequency components. Therefore, the filter's output signal amplitude also decreases when its time constant is tuned for faster rise times.\\
In combination
with the noise density spectrum of the preamplifier,
this results in a worse SNR the faster the rise time of the signal is shaped.

In order to maximize the SNR, the characteristics of the signal shaping filters have to be chosen such, that the output signal rises just barely fast enough to be processed in the trigger.

% schematic, tradeoff peakingtime vs snr, messung an zwei beispielen

\begin{figure}[ht]
\centering
\includegraphics[width=\columnwidth]{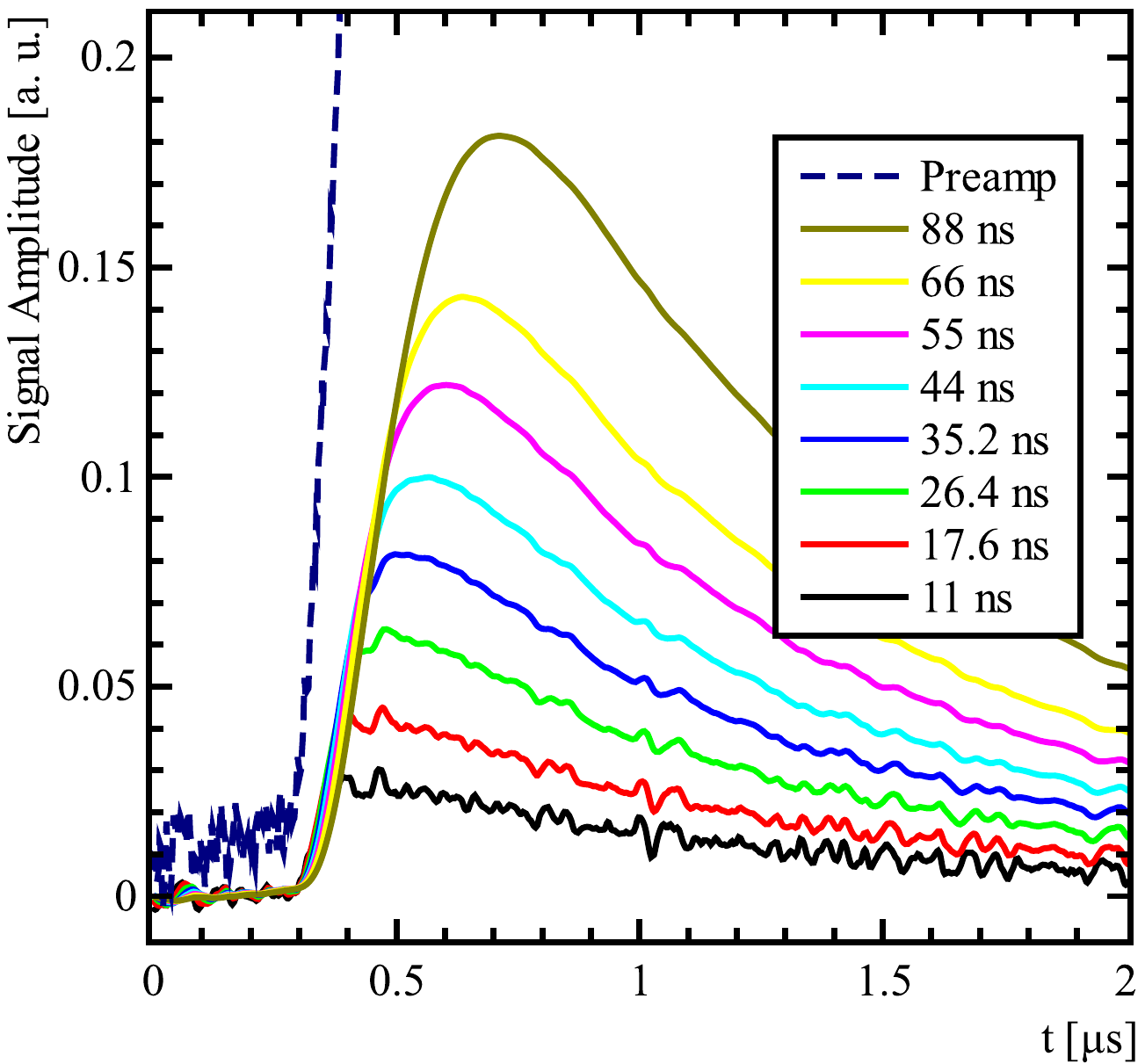}
\caption{Illustration of the trade-off between rise time and SNR.}
\label{pic:timing-signals-spice}
\end{figure}
Fig.~\ref{pic:timing-signals-spice} illustrates the trade-off between a fast rise time and a good SNR. The picture was created using a digitized scintillation signal from the CB front-end which was processed in a spice simulation implementing CR-RC$^2$ filters with different time constants. It is clearly visible how a smaller time constant results in a faster peaking time but also in a smaller amplitude and a higher noise level, which is best visible in the falling signal part.

For a quantitative analysis, the following pulse properties are determined: the amplitude and the peaking time,
% can be determined as well as 
and the noise level in standard deviations. At this point, only relative
% statements between different configurations are needed. 
comparisons between different configurations are considered.
Therefore all quantities can have arbitrary units.\\
The signal-to-noise ratio is one factor that limits the precision
% with which the 
of the determined pulse properties.
% can be determined. 
To achieve better precision, the scintillation signal was simulated as a superposition of two exponential decays. The time constants and weighting factors as well as the rise time of the scintillation signal were determined from the digitized pulse ($I_1=\SI{92}{\percent}, \tau_1=\SI{930}{\nano\second}, I_2=\SI{8}{\percent}, \tau_2=\SI{6.35}{\micro\second}$, rise time $t_{10-90}=\SI{36}{\nano\second}$).

\begin{figure}[ht]
\centering
\includegraphics[width=\columnwidth]{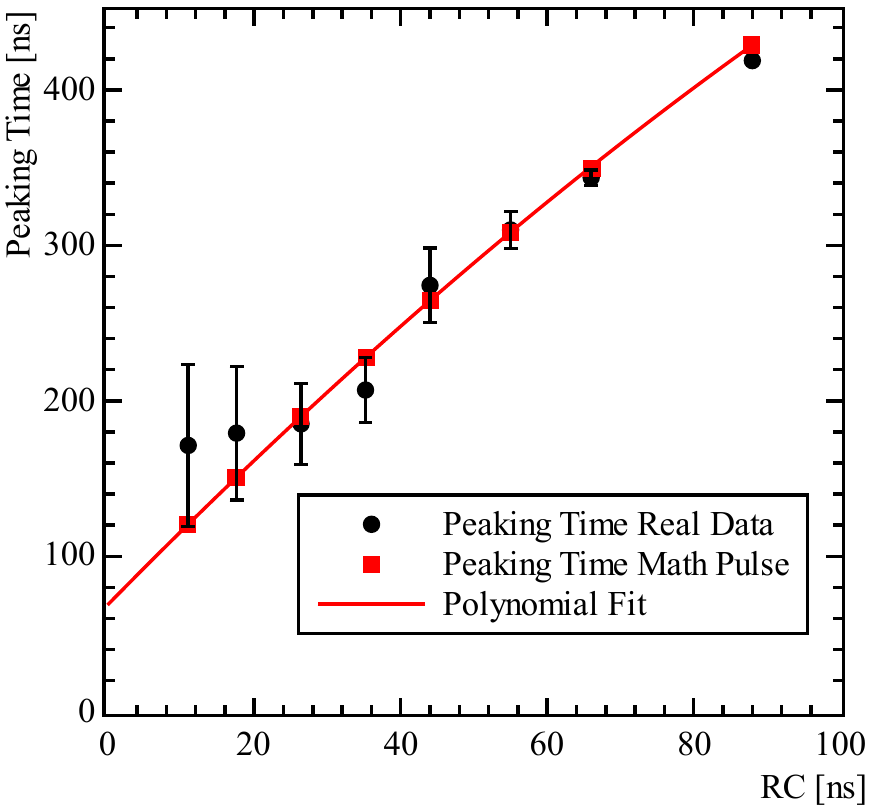}
\caption{Peaking time of a CsI(Tl) scintillation pulse after a CR-RC$^2$ shaper. Determined using a digitized signal and a signal generated with SPICE.}
\label{pic:math-modelling-peakingtime}
\end{figure}

Fig.~\ref{pic:math-modelling-peakingtime} shows the peaking time of the shaped signal using the digitized and the modeled pulse. It is important to remember that not only the digitized signal but also its noise 
%on the digitized pulse has the 
is identical
% time dependence 
for each simulated shaping time.
%case. 
Therefore also the noise in the output signals is correlated to some degree which introduces a systematic shift.\\
The noise level on the output of the shaper was determined at a time where no pulse is present and used for the error bars. This is assumed to overestimate the actual error, since also genuine scintillation pulses might be present at that time, with an amplitude barely exceeding the noise level.
Taking this into consideration, both data sets agree with each other.
%rauschen limitiert die genauigkeit mit der die parameter bestimmt werden können, daher:
%Szintillation als funktion I1 92%, tau1=930ns I2=8%, tau2=6.3µs

A polynomial fit is applied to the data to guide the eye. It is visible how in order to achieve a faster pulse (small peaking time) the time constant of the shaper needs to be reduced overproportionally: The data shows a trend towards a non-zero peaking time for the time constant approaching zero time. In other words, the effect of decreasing the time constant on the peaking time vanishes for small values. %This is the more pronounced, the faster the pulse is.

\begin{figure}[ht]
\centering
\includegraphics[width=\columnwidth]{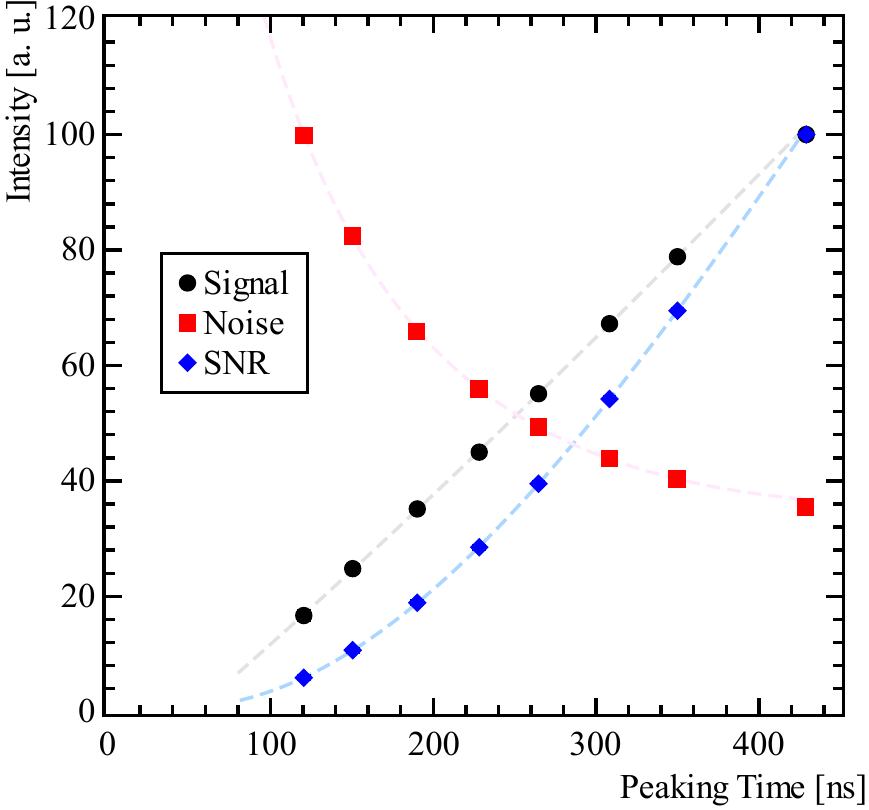}
\caption{Amplitudes of signal, noise, and SNR in dependence on the peaking time of the timing signal.}
\label{pic:s-n-snr-vs-risteime}
\end{figure}

Fig.~\ref{pic:s-n-snr-vs-risteime} summarizes the signal properties in dependence of the peaking time. The noise level rises for shorter peaking times, while the peak amplitude decreases. Both effects combined result in the SNR decreasing even faster with decreasing peaking time.\\
This underlines the importance of choosing the peaking time as slow as possible and just barely fast enough for the experiment trigger, in order to
% able to 
reach lowest possible trigger thresholds.\\

%Next: Messung: vergleich zweier konfigurationen

%\begin{figure}[ht]
%\centering
%\includegraphics[width=\columnwidth]{pic/Timing-Challenge_theo/VPvsPeakingCsI.pdf}
%\caption{Signal amplitude of a CsI(Tl) scintillation pulse after CR-RC$^2$ shaper in dependence of its peaking time.}
%\label{pic:}
%\end{figure}
%\begin{figure}[ht]
%\centering
%\includegraphics[width=\columnwidth]{pic/Timing-Challenge_theo/NoiseVsPeakingCsI.pdf}
%\caption{}
%\label{pic:}
%\end{figure}
%\begin{figure}[ht]
%\centering
%\includegraphics[width=\columnwidth]{pic/Timing-Challenge_theo/SNRvsPeakingCsI.pdf}
%\caption{}
%\label{pic:}
%\end{figure}

\begin{figure}[ht]
\centering
\includegraphics[width=\columnwidth]{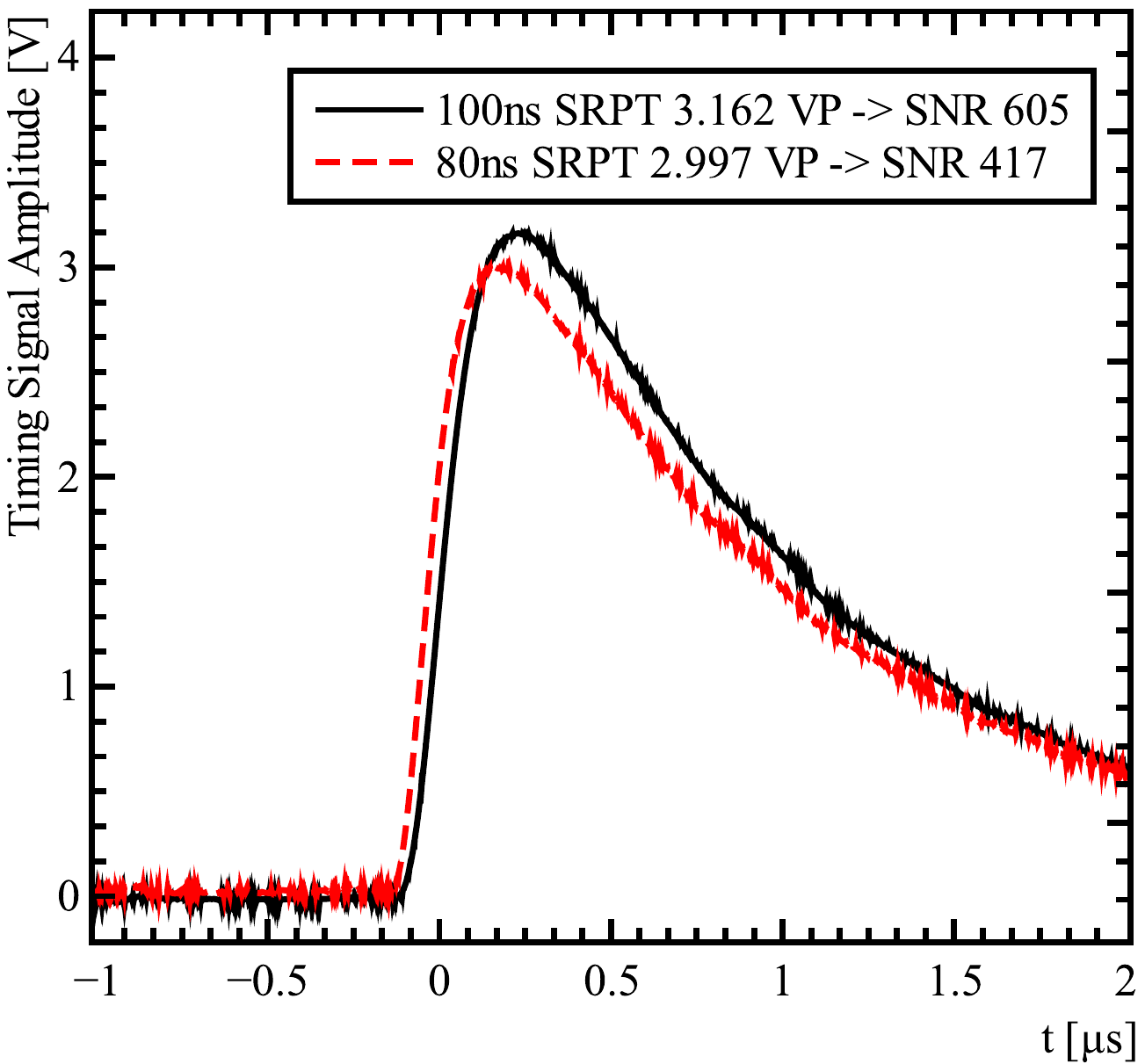}
\caption{Pulse shapes in the timing branch as used in prototypes (red) and the new readout (black). The SNR is higher for a longer peaking time.}
\label{pic:timing_vergr}
\end{figure}

To maximize the SNR, the final value of the shaper's time constant was chosen after all other constraints in the experiment readout were fixed.
Fig.~\ref{pic:timing_vergr} shows the signals in the timing branch as used in prototypes (red curve) and as finally implemented in the experiment upgrade (black). The increased SNR surpasses the gain expected by the simulations.
This might indicate the presence of a noise source, additionally to those covered in the spice simulation. For example, the linedriver's contribution is not fully modeled.
% An additional noise source for example in the line driver could explain this observation.

\begin{figure}[ht]
\centering
\includegraphics[width=0.9\columnwidth]{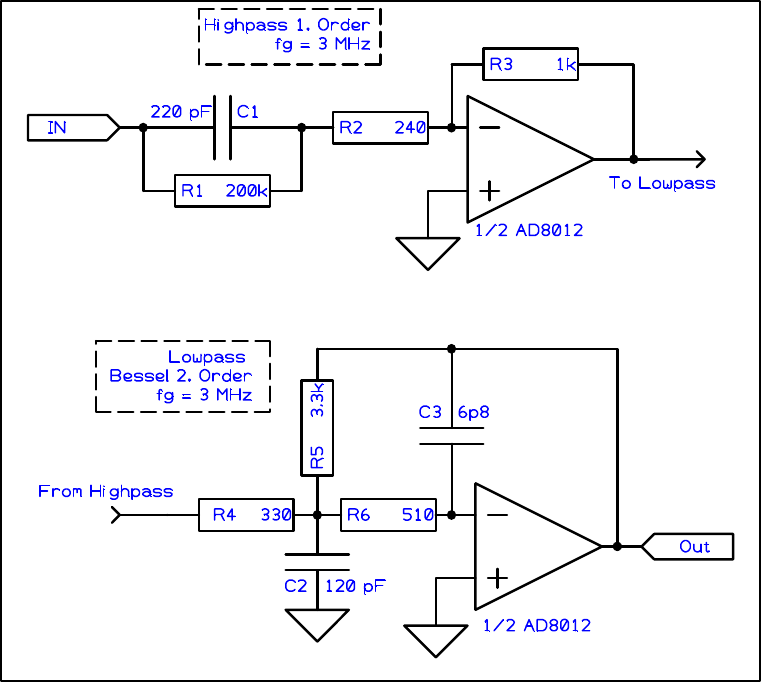}
\caption{Schematic of the pulse shaping filter of the timing branch.}
\label{pic:sch-t-shaper}
\end{figure}

\begin{figure}[ht]
\centering
\includegraphics[width=\columnwidth]{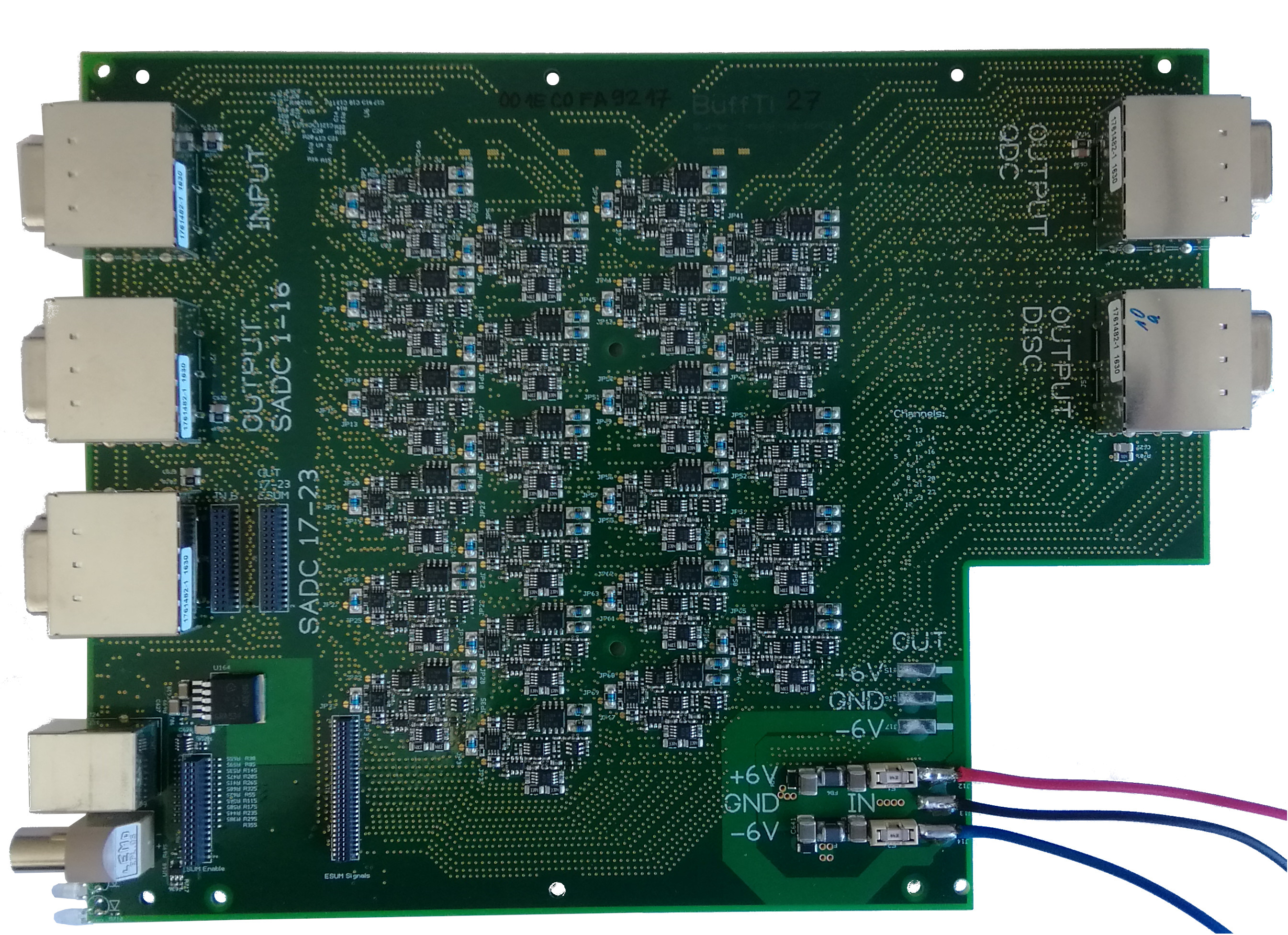}
\caption{Picture of the Module containing 23 channels of timing filter and preamp fanout for the energy branches.}
\label{pic:photo_buffti}
\end{figure}

Fig.~\ref{pic:sch-t-shaper} shows the schematic of the pulse shaping filter. Fig.~\ref{pic:photo_buffti} shows a picture of one module containing 23 of these filters. The module also contains a fanout of the timing signal for a fast energy sum (see Sec.~\ref{ssec:Esum}) and fanouts of the preamp signal for the energy branch of the readout.

\subsection{Energy Measurement}
The energy deposit is measured in a dedicated readout branch which has a signal shaping optimized for that purpose.
Actually, two branches exist that are optimized for energy measurement: The legacy readout utilizing QDCs and a new readout using modern SADCs \cite{Muellers_19_diss}.

\begin{figure}[ht]
\centering
\includegraphics[width=\columnwidth]{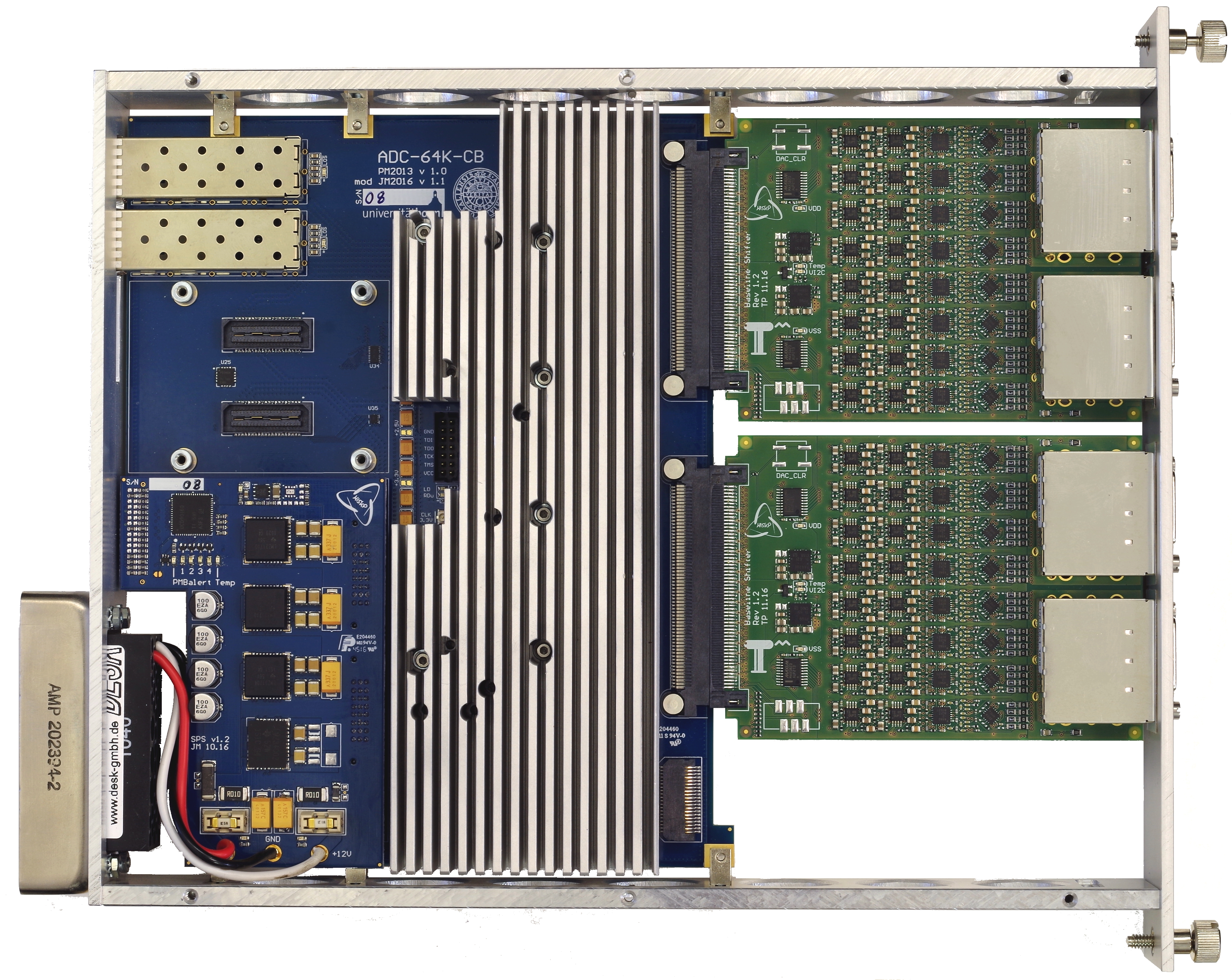}
\caption{64 channel SADC module for  the energy readout \cite{Muellers_19_diss}.}
\label{pic:sadc}
\end{figure}

The QDC branch is part of the old readout.
Due to its age, it has many disadvantages that are addressed by the modern SADC (see Fig.~\ref{pic:sadc}).
% readouts do not have.
The most important limitation arises from the processing time. About \SI{500}{\micro\second} dead time is needed to read out one event.
% occur when an event is being read out, which 
This is the slowest part in the experiment. Thus, it limits the achievable readout rate.\\
As the digitized QDC value corresponds to the integral of the input signal during an applied gate, this readout is susceptible to pile-up and offers neither a detection nor a correction of affected events.\\
Furthermore, the QDCs use an automatic  dual-range switching, which requires an additional step of calibration.\\

The new SADC readout \cite{Muellers_19_diss} uses 12-bit 80-MSPS ADC chips. A Kintex-7 FPGA decimates, buffers, and pre-processes the ADC data.
The feature extraction covers two moving averages with \SI{6.4}{\micro\s} and \SI{0.8}{\micro\s} which can be used as gated integrals.
%In addition, f
Further algorithms obtain timing information using constant fraction discrimination and a maximum finder. Also, a pile-up detector and per-event base line determination is implemented.\\
While the time information of the timing branch is more precise in most cases in the present implementation, the SADC time information is also available below the detection thresholds of the timing branch.

During normal operation, only the extracted features are transferred.
However, if pile-up is detected, the whole sample buffer of the affected channel is transferred for off-line pile-up correction.
%Additionally, raw data is transferred if pile-up was detected.
This mode allows a sustained readout rate of around $10\cdot 10^3\text{ events/s}$.%\SI{10}{\kilo\per\second}.

For debugging and development purposes, reading out the full sample buffers of all channels can be enabled. In this mode, a readout rate of around $2\cdot 10^3\text{ events/s}$ is achievable.
%The main benefits of this new readout are a higher maximum readout rate, pileup detection and recovery, and providing time stamps for signals which are below the discriminator threshold in the fast branch.
%sadc: keins dieser probleme, zusätzlich zeitinformation für kleine energie

Both readouts use signal shaping filters with a time constant of \SI{1}{\micro\second} which results in pulses of approximately \SI{8}{\micro\second} duration (see Fig.~\ref{pic:signals-pte}).\\
Both, the old and the new shaping filters have an adjustable pole-zero compensation and an option to trim the baseline.
In the new readout, these parameters are remotely configurable, while the QDC readout requires in-situ adjustments with screwdriver-operated potentiometers.

%The output pulse has a width of $\sim\SI{8}{\micro\second}$ (see Fig.~\ref{pic:signals-pte}).\\
%The legacy energy readout utilizes 12 bit dual range QDCs. This readout was maintained for the first few production beamtimes and is being replaced with a SADC readout \cite{Muellers_diss}.

%The new shaping filters maintain most features of the existing ones (shaping time, adjustable pole-zero compensation, baseline trim), which are remote configurable.

%Another benefit compared to the QDCs is the lack of need for range alignment.

The resolution of the new readout achieves the same performance or even surpasses the QDC resolution
% achieved with QDCs
when pile-up events are discarded \cite{Muellers_19_diss}.
The data quality is sufficient over the whole range, which makes a dual-range operation unnecessary \cite{Muellers_19_diss}.

It should also be noted that the new electronics provide a vastly increased channel density. The old readout used signal shapers which were built up as 9U NIM modules housing 8 channels each. The new readout uses 6U NIM module which contain 64 shaper and SADC channels each. This means that the energy readout of the whole calorimeter now fits into two NIM crates.
%todo: 

% seit wann sind sadcs mit dabei?
%shaper: tau 1us - 6us puls.
%alt: poti für pzc, gain, pedestal
%neu: alles elektronisch

%qdc: 12 bit dual range (switching), needs aligmnent

%sadc: 12 bit 80msps, dabei seit ersten teststrahlzeiten nahc umbau
%vorteile: höhere ausleserate möglich, pile up detection, time stamp für niedrigere energien als disc, auflösung ähnlich/besser

%SADC / QDC, beide mit shaper
\subsection{High Density Discriminator}
\label{ssec:HDDisc}
A custom module was developed to fulfill several tasks that are performed in the timing branch. This VME module contains two time-over-threshold discriminators per channel, which are used for pulse detection and
a slew rate measurement thereof.
%to measure the slew rate of each pulse. 
This measurement is performed in a Spartan-6 FPGA and is used for an online time-walk correction. Afterwards the same FPGA processes the obtained information in the first clustering stage. 
Also, a TDC is implemented in the firmware which 
digitizes the times of
%measures signals from 
all 184 comparators (two per detector channel).

\begin{figure}[ht]
\centering
\includegraphics[angle=90,width=\columnwidth]{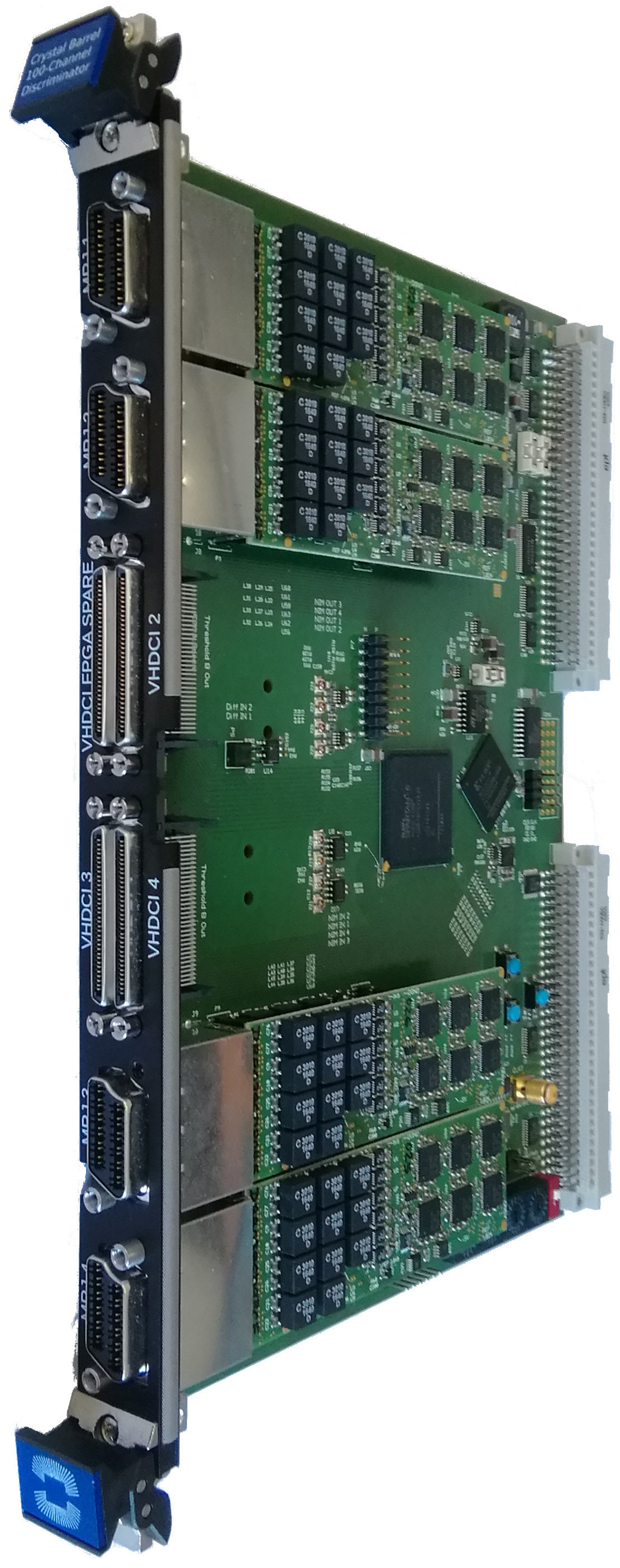}
\caption{High density discriminator which also implements online timewalk correction, TDC, and clustering of the CB timing signals.}
\label{pic:cbd}
\end{figure}

%In parallel, the firmware implements a TDC which is connected to each discriminator.\\
One module can process 92 signals from the CB. This means that the signals from the entire calorimeter are processed by a total of 16 modules which fit into one VME crate.\\
The discriminator module developed is also used to read out other detectors in the experiment. In those cases, a different firmware with a high resolution carry chain TDC is used \cite{github:jtdc, Grutzeck_16_bachelor}.\\
Currently the data is being read out via VME. A gigabit ethernet interface is under development (see Sec.~\ref{ssec:gb_if}). The module is shown in Fig.~\ref{pic:cbd}.

%custom vme module, 92 (+8) channels, 16 Modules for whole calorimeter, alles in einem vme crate
%TDC, partial cluster finder

%gbe under development

\subsection{Cluster Encoder}
\label{ssec:clusterfinder}
This section describes the concept of the new cluster encoder, its implementation, and its performance under realistic conditions. 
\begin{figure}[ht]
\centering
\includegraphics[width=0.55\columnwidth]{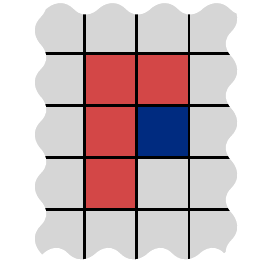}
\caption{Hit pattern used for cluster encoding. A crystal (blue) is treated as a top-left corner of a cluster if it detected a hit and no
hits were registered in the adjacent crystals, marked in red.
%ne of the crystals top-left of it (red) have detected a hit. 
Orientation as implemented, the $\theta$ coordinate increases from top to bottom, the $\phi$ coordinate from left to right.}
\label{pic:muster}
\end{figure}
To include any hit information of the calorimeter, its signals need to be processed within the available latency. Considering the comparably slow rise time of the timing signals and the limit on the total latency, a budget of $\sim\SI{100}{\nano\second}$ exclusively for clustering seems reasonable.\\
This goal was achieved using pattern matching. The implementation is based on the assumption that any cluster in the calorimeter has exactly one top left corner, which can be identified by merely processing the state of adjacent detector units.
% in the close neighborhood.

The pattern used is shown in Fig.~\ref{pic:muster}.  A crystal (blue) is treated as a top-left corner of a cluster if it has detected a hit and none of the crystals top-left of it (red) have detected a hit.

\begin{figure}[ht]
\centering
\includegraphics[width=\columnwidth]{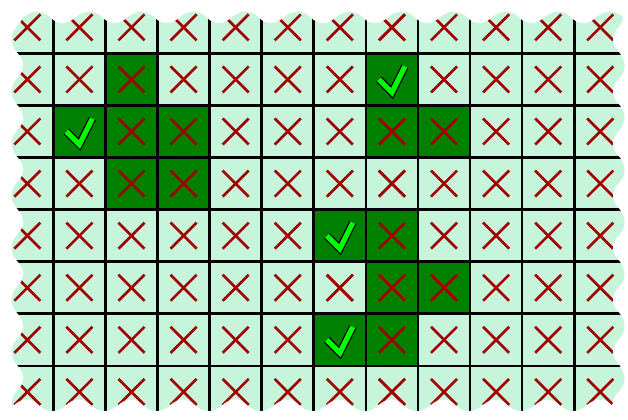}
\caption{Hypothetical hit pattern in the detector. Dark green: Energy deposit detected, bright green: no energy deposit, green check mark: pattern fulfilled, red cross: pattern not fulfilled.}
\label{pic:muster-bsp}
\end{figure}

As an example, Fig.~\ref{pic:muster-bsp} shows how this algorithm is applied to a hypothetical hit pattern. Dark (light) green boxes correspond to detector units which have (not) registered an energy entry.
A green check mark (red cross) represents a position where the pattern from Fig.~\ref{pic:muster} is (not) fulfilled.

For any position in the calorimeter, only very few signals need to be processed. This allows for massive parallelization. The algorithm is implemented in FPGAs and, in fact, the check is performed for all positions simultaneously.
%at the same time.

The output of this algorithm is a yes/no information for each of the 1320 detector units. To use this information in a multiplicity trigger, the number of positive results needs to be counted. The implementation of this processing stage in FPGAs utilizes massive parallelization of adders in a tree-like structure.\\
The 1320 signals corresponding to each individual detector unit are compressed in the first step by 320 instances of an adder to 320 signals. In the next step, the number of signals is reduced to 256. 
Afterwards, by adding two results from the previous stage, the number of signals is divided by 2, until a single value is reached.
%each step until reaching a single signal. 
At 6 positions in this processing chain, flip flops store the intermediate result to keep the logic delay low enough to be able to operate the logic at \SI{200}{\mega\hertz}. Further clock cycles are needed to handle numerical overflow and to account for the propagation delay between modules. In total, the full summation is executed in 9 clock cycles, i.e. \SI{45}{\nano\second}.
The cluster count is processed as a 5 bit signal. The most significant bit indicates an overflow. Thus, the number of clusters detected can range from 0 to 15, plus 16 or more clusters.

%The final output is truncated to 3 bit, allowing the indication of up to $N\geq7$  clusters found.

The final result is forwarded to the central trigger module of the experiment which currently accepts three signals from the CB clusterfinder: $N\geq1$, $N\geq2$, and $N\geq3$.
The information is updated every \SI{5}{\nano\second}. Therefore, the cluster finder can be considered free running.

\begin{figure}[ht]
\centering
\includegraphics[width=0.4\columnwidth, angle=0]{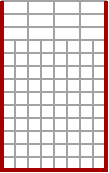}
\caption{Subsection of the calorimeter that is connected to one discriminator module. Red lines indicate the connection to another section.}
\label{pic:CBtopoDisc}
\end{figure}

One discriminator module processes the signals of up to 92 detector modules. A corresponding subsection of the CB is shown in Fig.~\ref{pic:CBtopoDisc}. To correctly process signals at the boundaries of these sections, modules need to exchange hit information of the corner detector units.

\begin{figure}[ht]
\centering
\includegraphics[width=\columnwidth, angle=0]{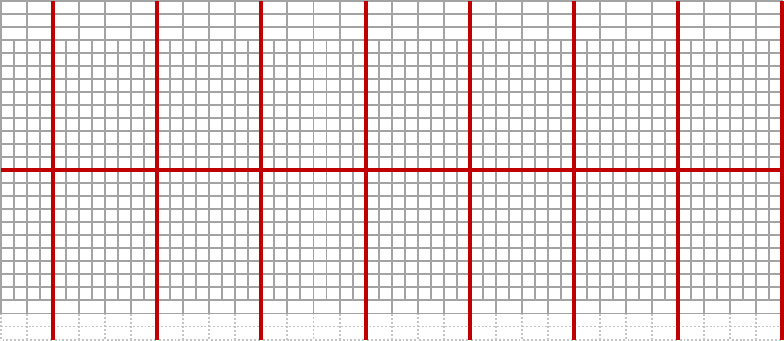}
\caption{Representation of the calorimeter. The right end is connected to the left end. One section highlighted in red corresponds to the section displayed in Fig.~\ref{pic:CBtopoDisc}. Gray, dashed boxes (two rows on the bottom) represent modules on the upstream end of the detector which are not installed in the current setup.}
\label{pic:CBtopo}
\end{figure}

%The number of hit status bits that needs to be exchanged does not depend on the orientation in which the cluster detection pattern is implemented, as long as the possibility should maintained to re-install the three currently missing rings.\\
The topology in which the hit status bits need to be transferred depends on the orientation of the cluster detection pattern.
In the implemented orientation (shown in Fig.~\ref{pic:muster}), one module in the top row needs to transmit 8 bit to the module below, 13 bit to the right-hand module, and one bit to the module right below. The topology of the whole calorimeter is represented in Fig.~\ref{pic:CBtopo}.

As this information is required in the cluster encoder, it needs to be transmitted not only at high speed but also at low latency.\\
For this purpose, an add-on backplane was designed, which is plugged into the P2 RTM ports on the back side of the VME crate. The order of modules was chosen to minimize the longest track length between modules. Signals are transmitted differentially.
The module has enough interconnecting signals to allow processing the full calorimeter. I. e., the currently removed detector modules can be re-installed and processed by the clusterfinder.
%This is done via differential signals which are transmitted on a custom made backplane. Each discriminator needs to transmit 20 hit signals to other discriminators, and the 

\subsubsection*{Simulation of the Algorithm}
Monte Carlo simulations were performed before the development of the clusterfinder was started.\\
The analysis compared the number of clusters found with the existing algorithm with the new method.
The existing algorithm precisely determines the number of topologically connected groups of detector modules with energy deposit, while the new algorithm performs an estimation based on the hit pattern in the local neighborhood of each module.

\begin{figure}[ht]
\centering
\includegraphics[width=\columnwidth, angle=0]{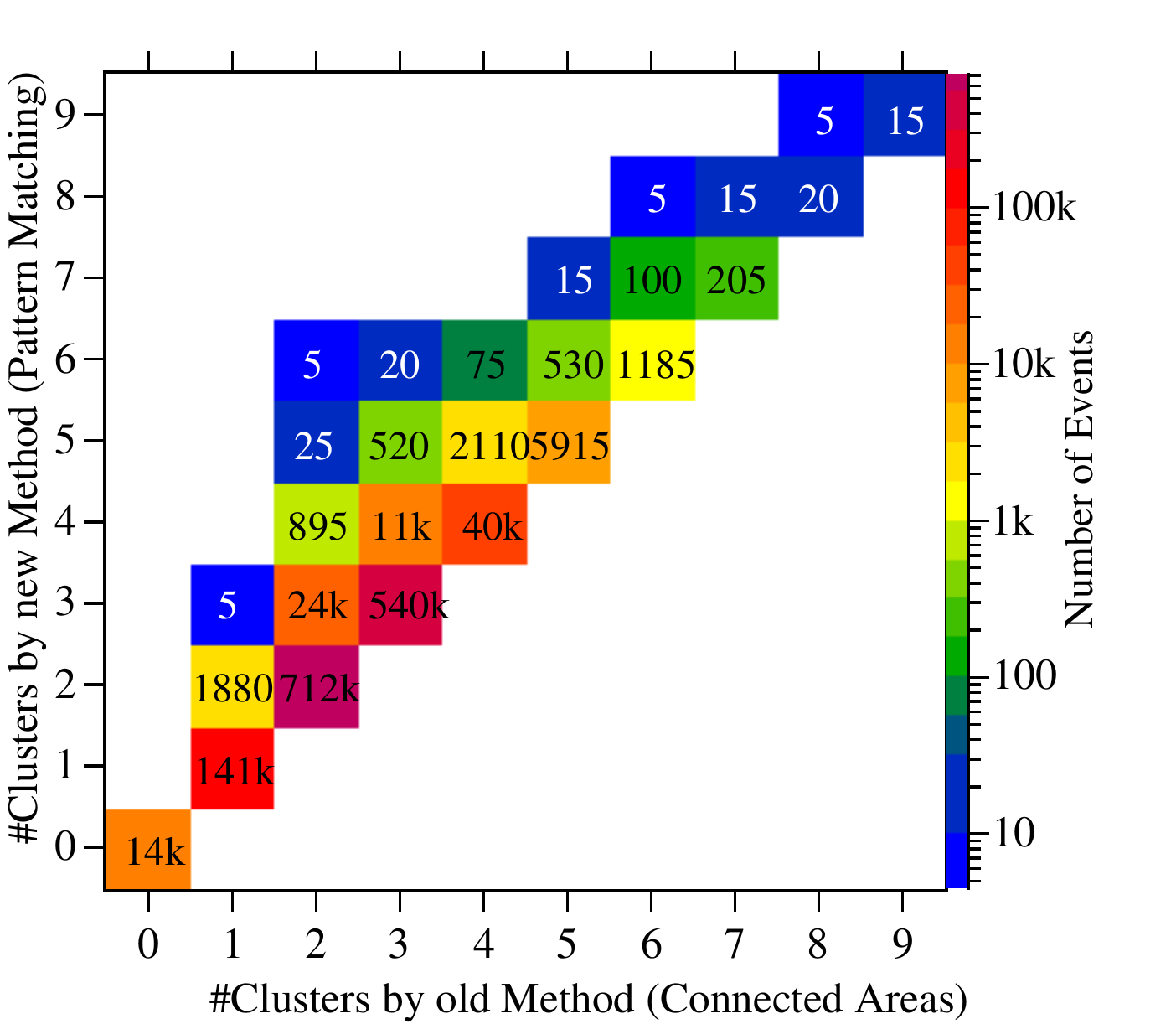}
\caption{Number of clusters found with new and old clustering method. Entries above (below) the diagonal correspond to an overestimation (underestimation) by the new algorithm.}
\label{pic:cf_sim}

\end{figure}

As an example, Fig.~\ref{pic:cf_sim} shows the result for the reaction products of $p\pi^0$. Further cases can be found in ref. \citep{Honisch_15_diss}.

It was found that the new algorithm never underestimates the number of clusters, and overestimates only in a small fraction of all cases.\\
This means that the sensitivity does not decrease compared to the old cluster encoder. The sensitivity of the new encoder might actually be better. Neither algorithm truly evaluates the number of detected particles. As clusters might overlap, the number of connected areas (determined by the old algorithm) can be lower than the number of impinging particles.\\
In some of these cases the clusters
% shape 
might be
% have concave features which can be 
shaped in a way that the new algorithm detects two clusters, although it is one connected area.

%Evaluating quantitatively if the sensitivity increases and how the selectivity changes has not been done yet since it is a complex undertaking.\\
%, however, is a lot more involved and has not yet been done.\\
%A proper analysis of the sensitivity should not consider the number of clusters, but the number of particles found in the detector. As their clusters can overlap, the number of clusters potentially underestimates the number of particles.\\

Predicting the selectivity of the trigger is even more complex. To understand how well undesired reactions can be suppressed, their signature needs to be known exactly, which is an information that can be obtained from Monte Carlo simulations. However, to predict the degree at which these unwanted reactions contaminate the useful data, many factors have to be well known. Among these are not only the exact beam characteristics, target dimensions, and cross sections but also parameters like composition and dimension of support structures and magnetic field strength in regions that are usually not of interest.\\
A quantitative prediction of the selectivity was therefore not attempted. Instead, the parameters of the trigger (i.e. threshold levels) were tuned during the commissioning beam time until a high quality of the recorded data was achieved. Values like live time, detected pions per unit time, and signal-to-background ratio were used during the optimization.

\begin{figure}[ht]
\centering
\includegraphics[width=\columnwidth]{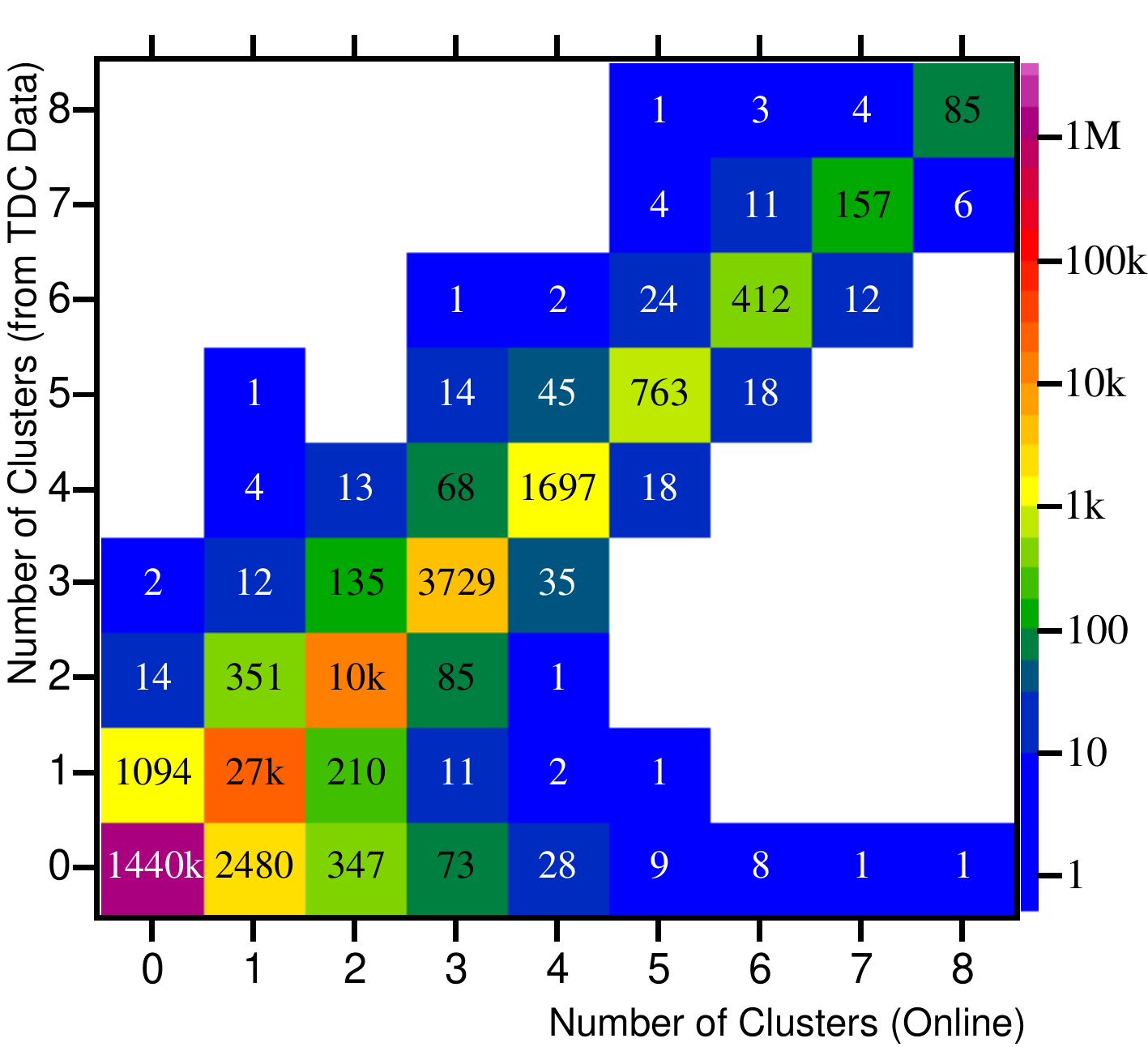}
\caption{Commissioning of the cluster finder electronics: Number of clusters expected according to TDC data vs number of clusters found by electronics.}
\label{pic:triggerCom}
\end{figure}

One of the tests done during the commissioning of the setup was intended to verify that the digital signal processing is working correctly. In other words: The output of the cluster finder setup should correspond to its input.\\
The cluster finder was tested within the full installation during a beamtime. To achieve a data set which is unbiased by the CB, the fiber detector inside the CB was used as trigger source for the DAQ.\\
The input data of the CB cluster finder corresponds to its TDC data. The clustering algorithm was implemented in software to predict the output of the cluster finder electronics.\\
To allow for a comparison of output and predicted output, the cluster finder firmware includes a TDC that samples the number of clusters currently found.\\
Fig. \ref{pic:triggerCom} shows the comparison of both results. The majority of events has an equal number for both methods.\\
A high number of events has zero clusters for both methods. This is a result of the fiber detector chosen as trigger source. For example electrons or positrons with a comparably low energy can be detected in the fiber detector, while not reaching the CB.\\
The entries with unequal numbers of clusters can be traced back to an imperfection of the analysis: The time cuts applied to both data sets were not perfectly equal. Accidental hits at the borders of the time window have a chance to be seen only by one of both methods.

% einer der tests: werden die cluster vom online clusterfinder genau so erkannt, wie es laut tdc daten passieren müsste?

%Trigger= innendetektor (no bias, daher aber auch viele einträge bei 0, 0)
% online cluser: aus sampler in clusterfinder ermittelt
% tdc cluster: algorithmus in software auf tdc daten angewendet.

% die zeitfenster je event sind für die beiden dattensätze nicht identisch. daher ergibt pile-up einträge neben der diagonalen.

\subsubsection{Time dependence of clusters}
An important problem that needs to be addressed is the time dependence of the number of clusters during one event.
%As the time resolution is limited, not all channels with an energy entry are detected at the same time. 
The time at which a hit appears is shifted by time walk and by the time resolution of the signal. The walk is compensated for (introduced in Section~\ref{ssec:fast_branch}), so only the random distribution caused by the time resolution remains.
Therefore, detector units belonging to one cluster can appear in the hit matrix in random order. This can lead to a high overestimation of the number of clusters for a short time and therefore
decrease the
%result in a poor 
selectivity of trigger conditions that require multiple clusters in the detector.

\begin{figure}[ht]
\centering
\includegraphics[width=\columnwidth, angle=0]{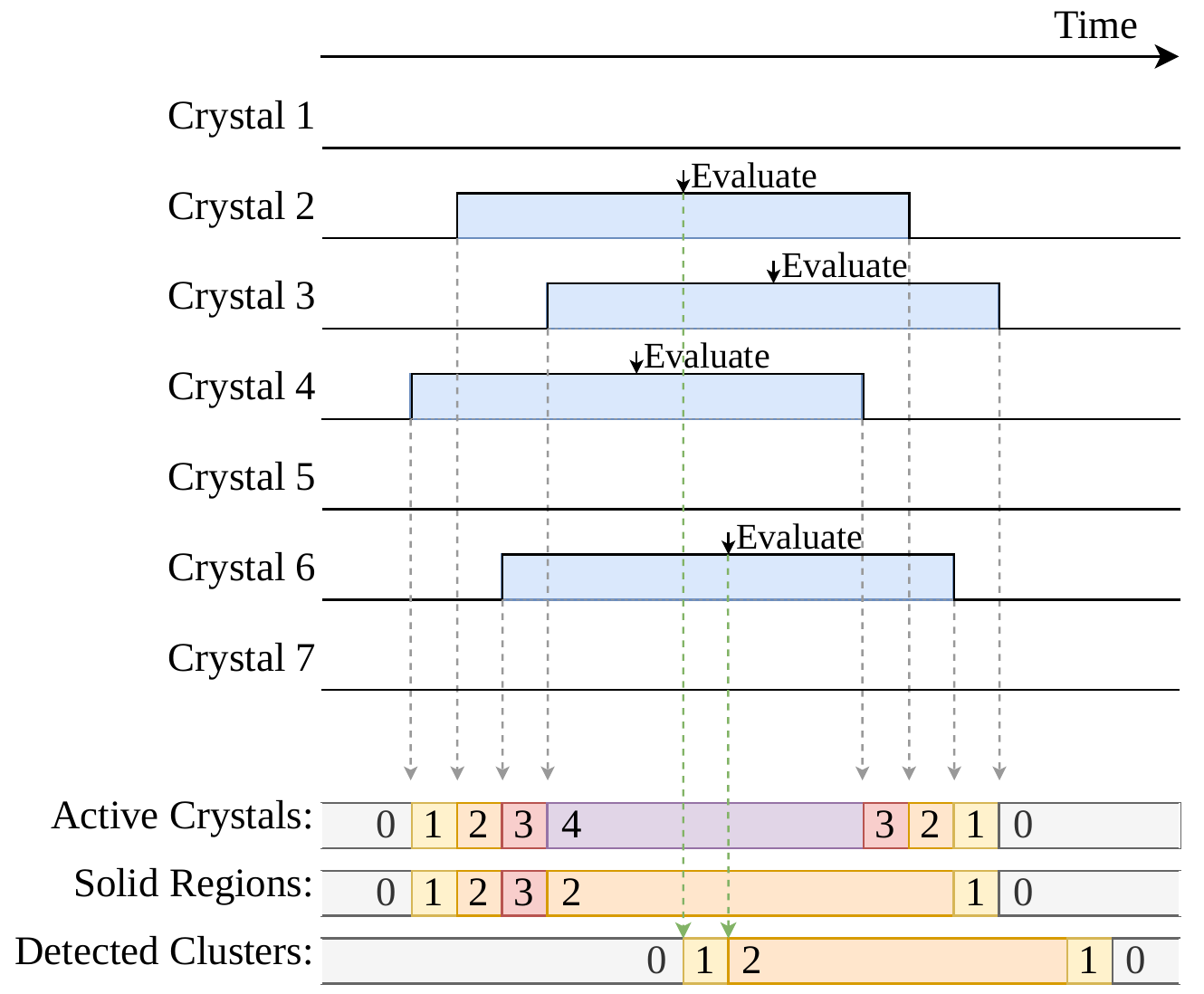}
\caption{Overestimation of the number of clusters. Illustration on a simplified 1-dimensional detector consisting of 7 crystals. Solid regions are formed by neighboring crystals. A crystal is counted as a cluster if the folowing pattern is fulfilled: The crystal has a hit itself and the previous crystal did not have a hit.}
\label{pic:cf_over_1D}
\end{figure}

The concept of this problem is illustrated in Fig.~\ref{pic:cf_over_1D}. The example shown applies for a simplified detector that has a 1-dimensional detector matrix consisting of a row of seven detector modules. The pattern of the cluster finder needs to be modified for this case: A crystal is counted as a cluster if itself has a hit while the previous module does not.\\
In the example, at first crystal 4 detects a hit, then successively the crystals 2, 6, and 3. Crystals 2, 3, and 4 are neighboring to each other and form one cluster. Because crystal 3 detects the hit only after crystal 2 and 4, this cluster forms 2 solid regions until the gap is closed.\\
If the cluster pattern was evaluated at the leading edge of each pulse, overestimation would occur. If the cluster pattern is only evaluated with a certain delay, the overestimation can be avoided.\\
The delay and the pulse duration have to be chosen properly to securely avoid overestimation and to also not lose true coincidences.\\

The actual magnitude of the overestimation depends on several factors like the discriminator threshold and the cluster energy: the smaller a cluster in the hit pattern is, the less likely it is that overestimation will occur.

Three ways are presented here, which show this effect for specific cases. It should be noted that the data analyzed here contains cosmic muons hitting the calorimeter. Therefore, clusters can have a very elongated shape, unlike hit patterns from electromagnetic showers.
This leads to a more pronounced overestimation and the results shown here should be taken with a grain of salt. However, the data is well suited to illustrate the problem.\\

\begin{figure}[ht]
\centering
\includegraphics[width=\columnwidth]{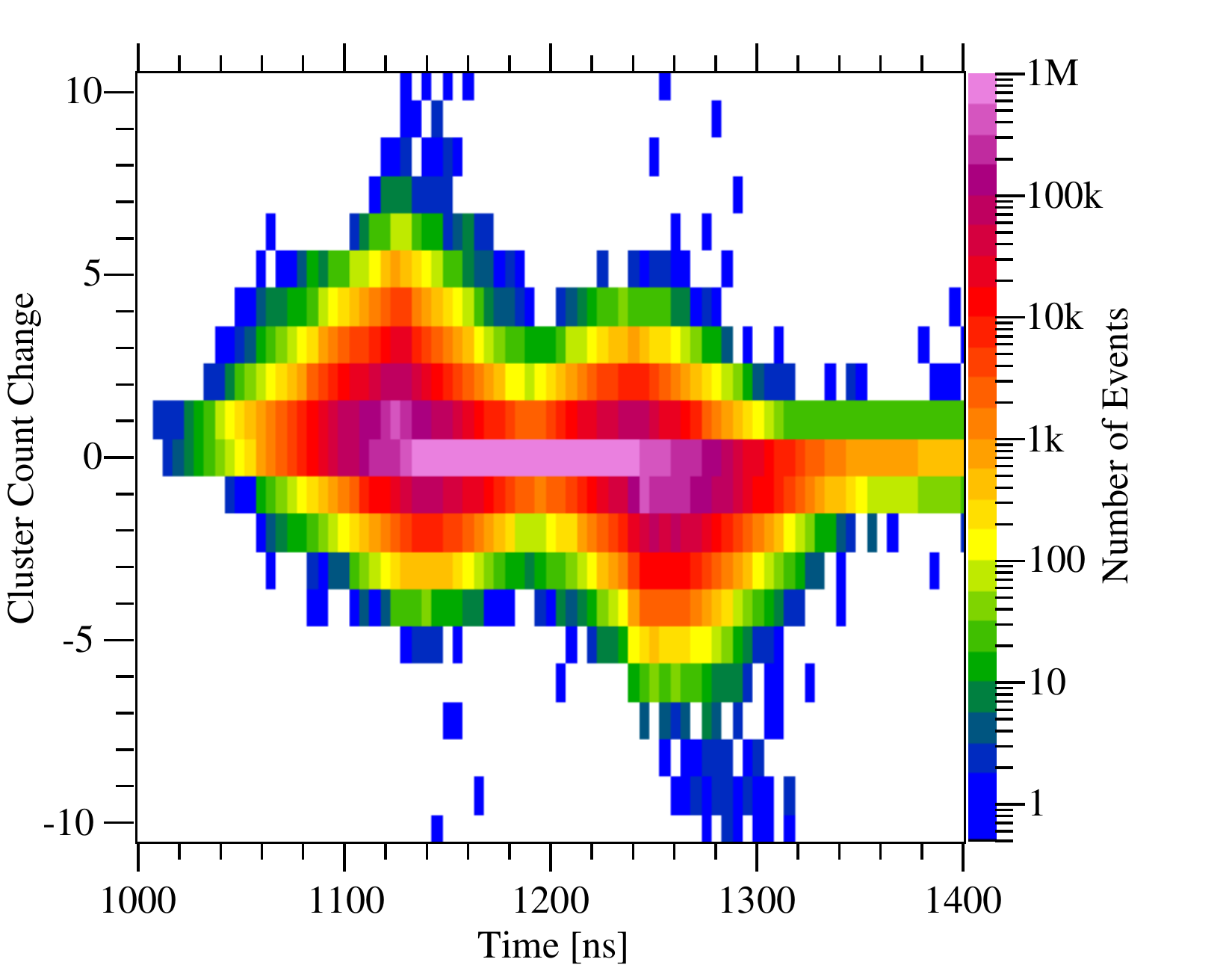}
\caption{Change in the number of clusters in the hit matrix \cite{Klassen_20_diss}.}
\label{pic:dcluster}
\end{figure}

The first illustration uses the number of detected clusters $N_C$ in dependence of time. The hit pattern is recorded in a \SI{5}{\nano\second} stepping. For each frame, the number of clusters is evaluated using the pattern detection method.
To improve the visibility of the effect, the change of the number of clusters is plotted in Fig.~\ref{pic:dcluster}. Frames with $N_C=0$ and $\Delta N_C=0$ are not plotted.

At very early times, no change in $N_C$ is visible. At roughly $t=\SI{1050}{\nano\second}$ clusters start to appear but also to disappear. Here, the full cluster appears gradually in the hit matrix and gaps cause an over-estimation. At about $t=\SI{1200}{\nano\second}$, the rate of changes decreases which means that $N_C$ is stable. The hit signals have a pulse duration of \SI{120}{\nano\second}. Therefore the cluster gradually disappears in the hit pattern, which again causes an overestimation of $N_C$.

Disappearing clusters before and appearing clusters after the reference time both indicate over-counting at these times.
As clusters might disappear in the same event at different times, this depiction does not allow to conclude by which number the cluster over-counting occurs.

The same raw data is used to show the effect of over-estimation in a different way. In this case, only events are plotted that have exactly 2 clusters within the prompt peak. Before these are determined, the time dependence was removed by a projection. In other words: In the resulting hit matrix, a detector unit is marked as hit if it was hit at any time in this event.  To limit any effects of event pile-up, a cut of \SI{265}{\nano\second} around the prompt peak was applied before analyzing the data.

\begin{figure}[ht]
\centering
\includegraphics[width=\columnwidth]{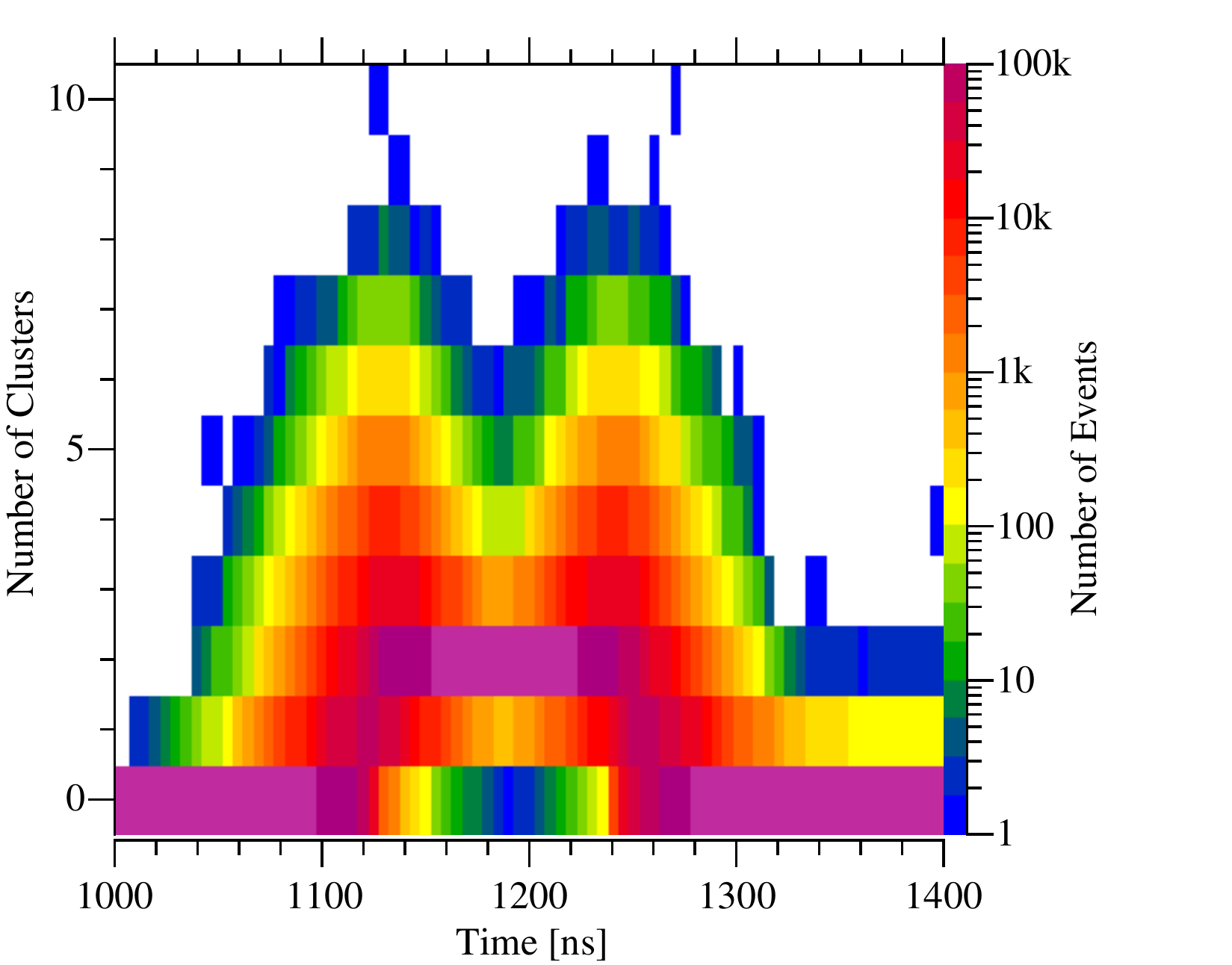}
\caption{Time dependence of clusters found when 2 clusters are found in the prompt peak \cite{Klassen_20_diss}.}
\label{pic:cluster_over}
\end{figure}

All remaining events are analyzed time-dependently, i.e. the number of clusters is determined at each time frame.
The result is shown in Fig.~\ref{pic:cluster_over}. While the data should only contain events with 2 clusters, more than these are detected before and after the prompt peak.

\begin{figure}[ht]
\centering
\includegraphics[width=\columnwidth]{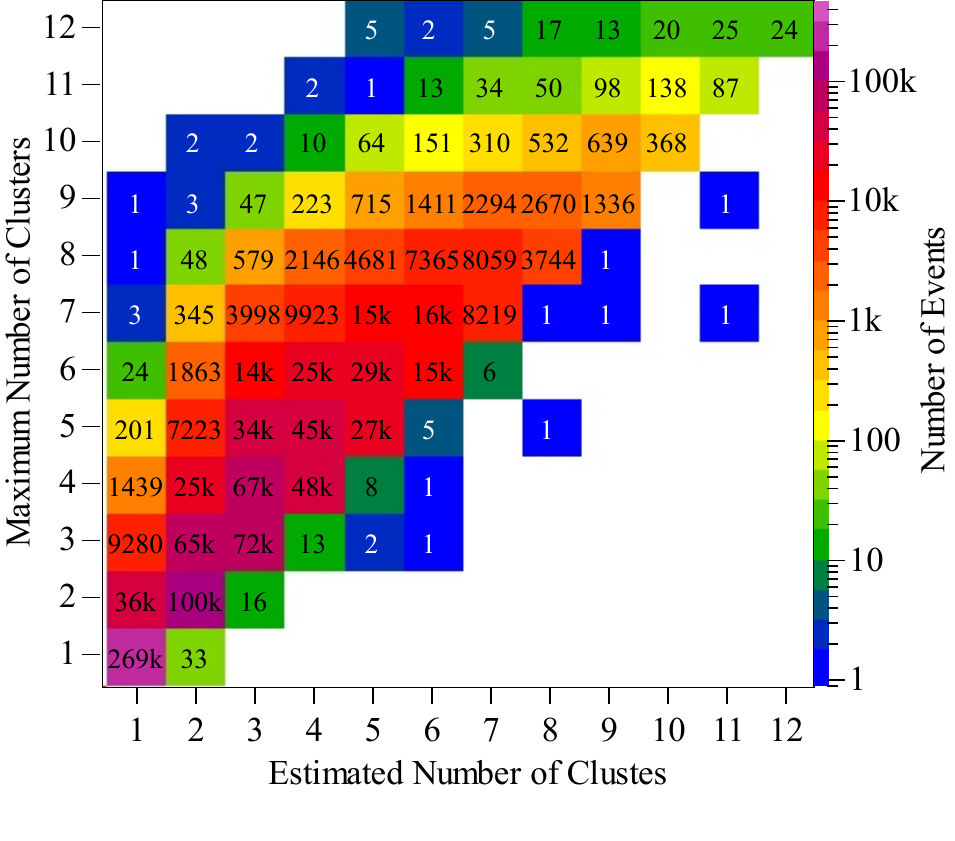}
\caption{
%Over-estimation of the number of clusters due to gradually appearance of clusters.
Maximum number of clusters vs estimated true number of clusters. Entries above the diagonal correspond to an erroneously high number. \cite{Klassen_20_diss}}
\label{pic:state_over}
\end{figure}

The third illustration is intended to provide a more quantitative idea of the degree of over-counting. It generalizes the previous analysis and all numbers of clusters in the projected frame.\\
For each event, the 
frame in the prompt peak with the highest 
%maximum number of 
cluster count $\text{max}\left(N_{C}\right)$ is determined.
This value is plotted against the number of clusters in the prompt peak projection.
% in the same time frame but evaluated for each time bin individually.
%For each event, the maximum number of clusters was determined as well as an estimate for the true number of clusters.\\
%The maximum number in one event was determined by evaluating each time slice in the hit pattern individually and selecting the maximum of all results in the studied event.\\
%To obtain the estimate for the true number of clusters, the time dependence was removed by projection. In other words: In the resulting hit matrix, a detector unit is marked as hit if it was hit at any time in this event. The resulting matrix is analyzed with the same pattern algorithm.
%To limit any effects of event pile-up, a cut of \SI{265}{\nano\second} around the prompt peak was applied before analyzing the data in any of both ways.
Fig.~\ref{pic:state_over} shows the result.

%\{bild ergänzen: overshoot bei projektion=2}

%overshoot qualitativ gezeigt
%in kleinem zeitfenster: projektion / zeitabhängigkeit entfernen
% # clusters in projektion vs max(#clusters) in allen einzelnen samples.
%cosmics, cf1 trigger, daher nur qualitativ. lösung ist aber einfach

To solve the problem of over-counting, a cluster check is only performed
% tag is only enabled 
a certain time after the hit was detected. This time is matched to the time it takes for the cluster to fully appear in the hit matrix.

\begin{figure}[ht]
\centering
\includegraphics[width=\columnwidth]{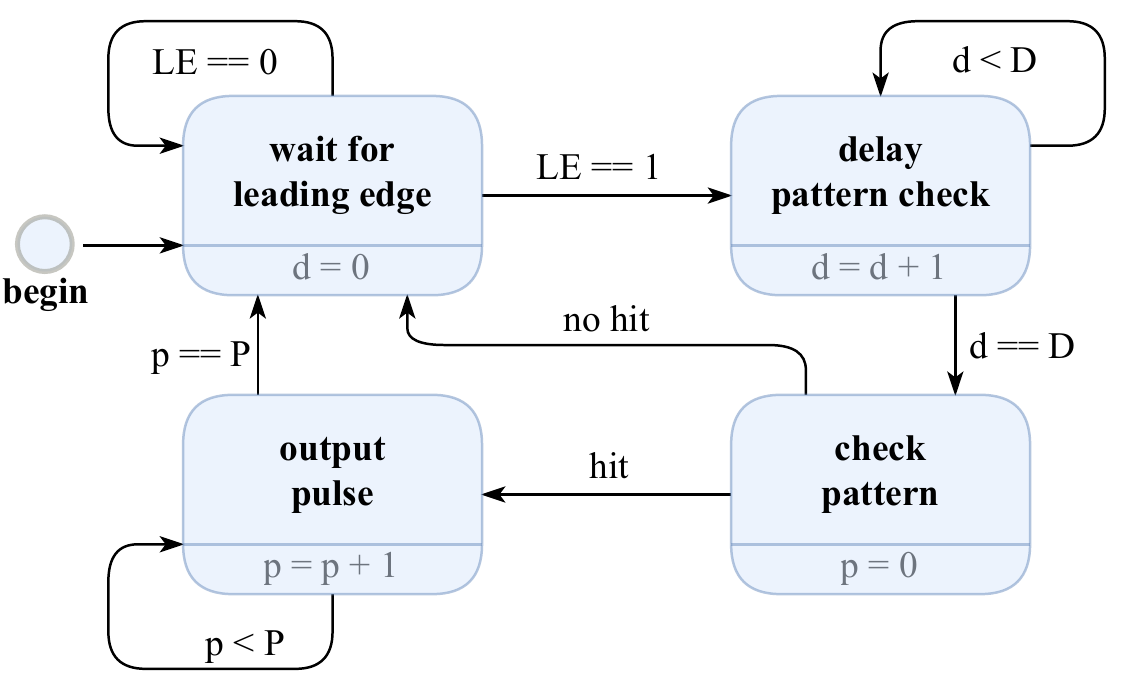}
\caption{State machine that prevents cluster count overestimation during build-up. $d$: delay counter, $D$: delay setting, $p$: pulse duration counter, $P$: pulse duration setting \cite{Klassen_20_diss}.}
\label{pic:state_cf}
\end{figure}

The algorithm is implemented as a state machine which
%The approach chosen to solve this problem 
is depicted in Fig.~\ref{pic:state_cf}. The shown state machine is implemented for each individual detector module.
When one hit is found, the algorithm waits for a specified number of clock cycles $D$. Only then, the hit pattern is checked and when fulfilled, the cluster output is enabled.\\
The delay has to be chosen such that all neighboring channels had enough time to be registered in the hit matrix. As the pattern also requires the checked channel to be still active, this algorithm also prevents false detection of clusters during the falling edge of the timing signal. At this edge, electronic noise can cause extra entries in the hit matrix.\\
The parameters were optimized during a beamtime using an oscilloscope. The values were chosen as small as possible while still effectively preventing a short term overestimation.  In the final configuration, pulses of \SI{120}{\nano\second} duration enter the cluster encoder. \SI{60}{\nano\second} after one cell has seen a signal, it checks whether the cluster pattern is locally fulfilled and, if that is the case,  generates a \SI{130}{\nano\second} output pulse.

\subsection{Calibration Light Pulser}
\label{ssec:led-pulser}

The light pulser generates flashes which are fed into the detector modules using light fibers.
% The system fulfills different purposes
%\begin{itemize}
%\item Measurement of the APD Gain,
%\item Generating calibration data to align the two ranges of the QDC,
%\item General testing and debugging of the calorimeter.
%\end{itemize}
The two main purposes of the system are to measure the APD gain and to generate calibration data to align the two QDC ranges.\\
Additionally, the system is helpful to generate signals to test and debug the calorimeter.\\
These applications yield different requirements on the light pulser system.

The APD gain is measured by comparing the signal amplitudes at nominal bias voltage (i.e. $M=50$) and the signal amplitude at zero bias voltage (i.e. $M=1$). 
%The temperature sensitivity of the APD can be eliminated using such a set of measurements, as the main contribution to the APD's temperature coefficient is the temperature dependence of the gain mechanism.\\
The measurement at $M=1$ suffers from a lower signal amplitude and also a higher noise level, due to the increased junction capacitance of the APD.
Therefore, a high intensity flash is needed to achieve a good SNR.\\
Stability over time is required to a degree that allows to perform one measurement at $M=1$ and one at $M=50$. A good long term stability allows for a long delay between reference measurements ($M=1$).

To align the high and low range of the QDC, pulses need to be digitized in both ranges.\\
For this purpose, either the relation between intensity setting and resulting flash intensity needs to be known very precisely, or a set of intensities needs to be chosen that can be digitized in both ranges.\\
The latter requires a fine granularity of the intensity settings for two reasons: The scaling and offset of both ranges vary from channel to channel in the QDC and the energy equivalent of the pulser system varies from detector module to detector module.\\
To be able to use one set of intensity settings in measurements in the timescale of months or years, 
%To not have to select intensities for measurements over and over, 
a decent long term stability is necessary.\\
As both measurements, gain monitoring and range alignment, have to be performed during production beam times, a high pulse rate is desirable in order to not waste valuable beam time.\\
To test and debug the readout, the light pulser has a few advantages over pulses from cosmic muons. The intensity can be set on demand and the time of the pulse is known.\\

\begin{figure}[ht]
\centering
\includegraphics[width=0.8\columnwidth]{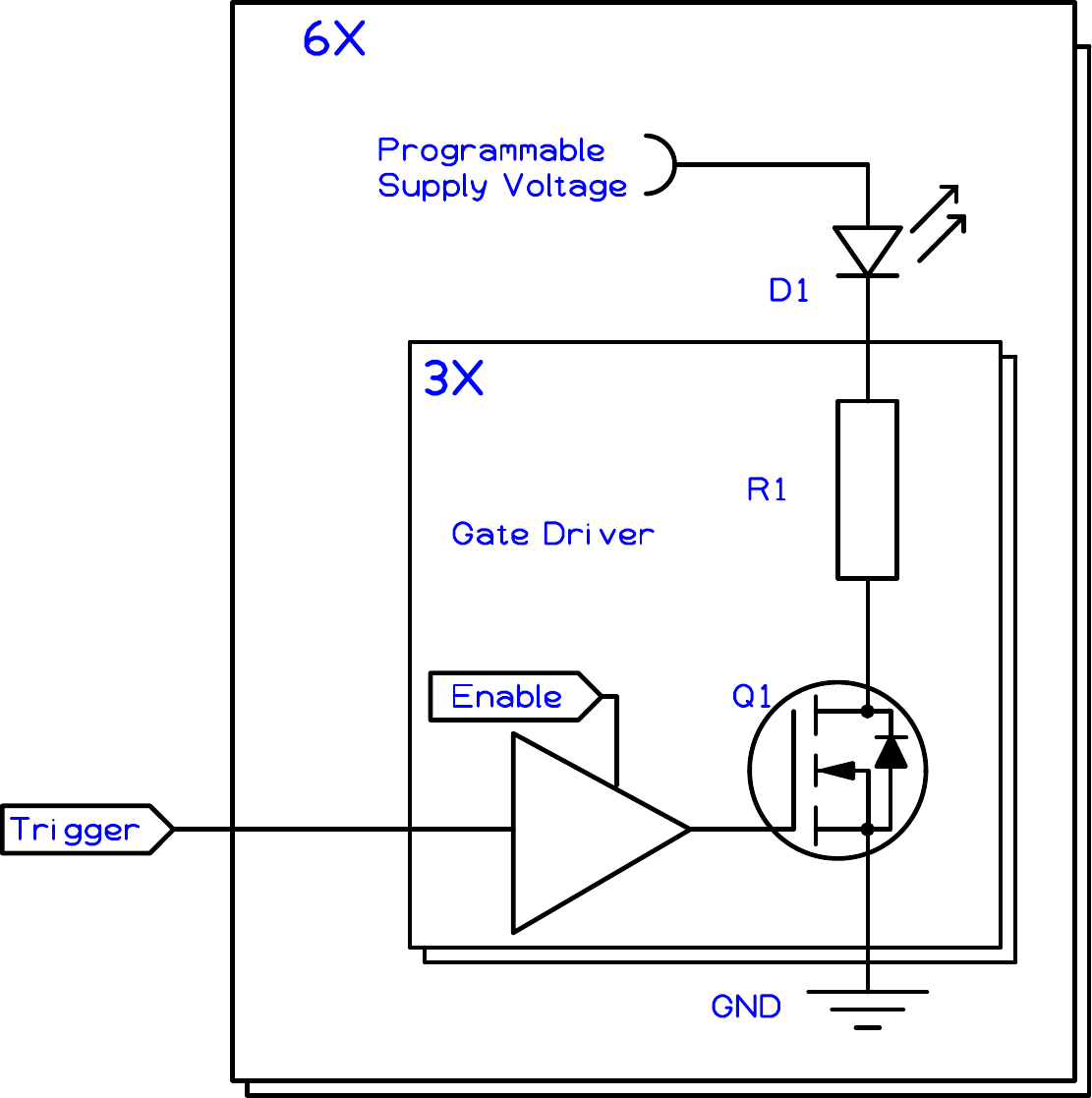}
\caption{Schematic of the essential parts of the light pulser system.}
\label{pic:LightpulserSch}
\end{figure}

Fig.~\ref{pic:LightpulserSch} shows a schematic of the core of the light pulser system. It consists of 6 identical units. Each unit uses a high power LED (LZ1-10G100) which illuminates an individual bunch of light fibers. 

Light pulses are generated by turning on and off the LEDs with power MOSFETs. The light intensity is roughly linearly related to the current flowing through the LED. The magnitude of the current results from the supply voltage, the series resistor (R1), and the voltage drop over the LED (D1).
To be able to control the intensity in fine steps, the supply voltage source is programmable (implemented with a 12 bit DAC).
To extend the accessible intensity range, each LED has three of those units with different series resistors ($R_1= \SI{3.3}{\ohm}$, \SI{60}{\ohm}, and \SI{280}{\ohm} respectively).

%The intensity can be controlled in two ways: A 12 bit DAC can be used to adjust the intensity in a fine stepping while three different series resistors allow to select a coarse range.

The highest available intensity generates pulse amplitudes equivalent to an energy deposit of $\sim\SI{15}{\giga\electronvolt}$ (with APD gain $M=50$). During a measurement with $M=1$ (APD gain off), this corresponds to amplitudes of $\sim\SI{300}{\mega\electronvolt}$ in the energy readout.

The light pulser was tested and characterized for stability using pulse rates up to \SI{5}{\kilo\hertz}. Under realistic conditions, the pulse amplitude was found to be stable within less than \SI{0.1}{\percent} \cite{Urban_18_diss}.

%DAC / verschiedene widerstände

%online gain calibration, qdc: range matching, general testing

%features: dyn range ~0 - 15 GeV_eq
%~3000 stufen + 3 ranges
%pulse rate virtually unlimited, practical limit: DAQ

\begin{figure}[ht]
\centering
\includegraphics[width=\columnwidth]{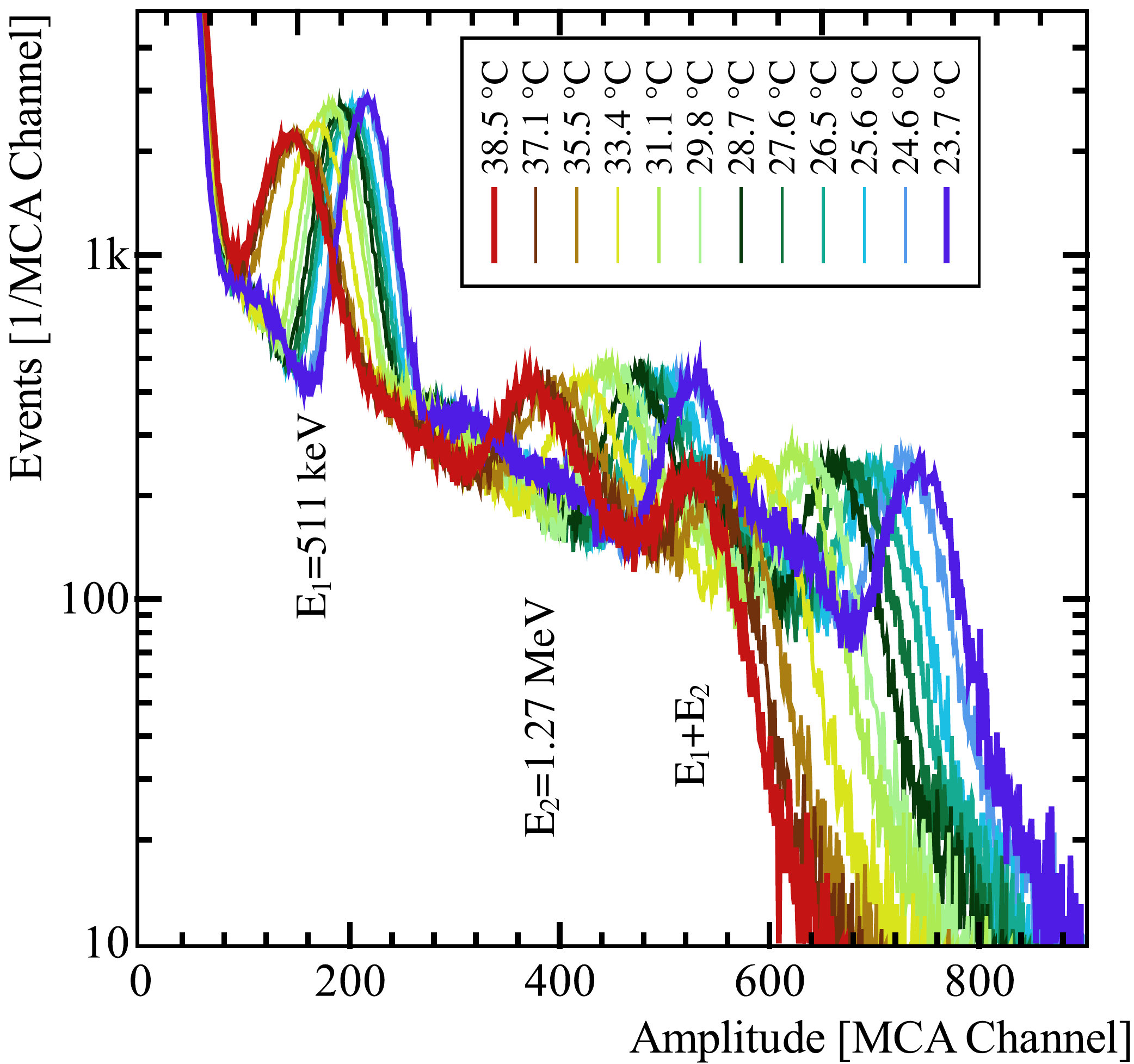}
\caption{$^{22}$Na spectra at different temperatures, measured without the compensation circuit in the front-end. The decreasing gain scales the measured spectrum to smaller MCA values. \cite{Urban_18_diss}}
\label{pic:na_temp_var}
\end{figure}

\begin{figure}[ht]
\centering
\includegraphics[width=\columnwidth]{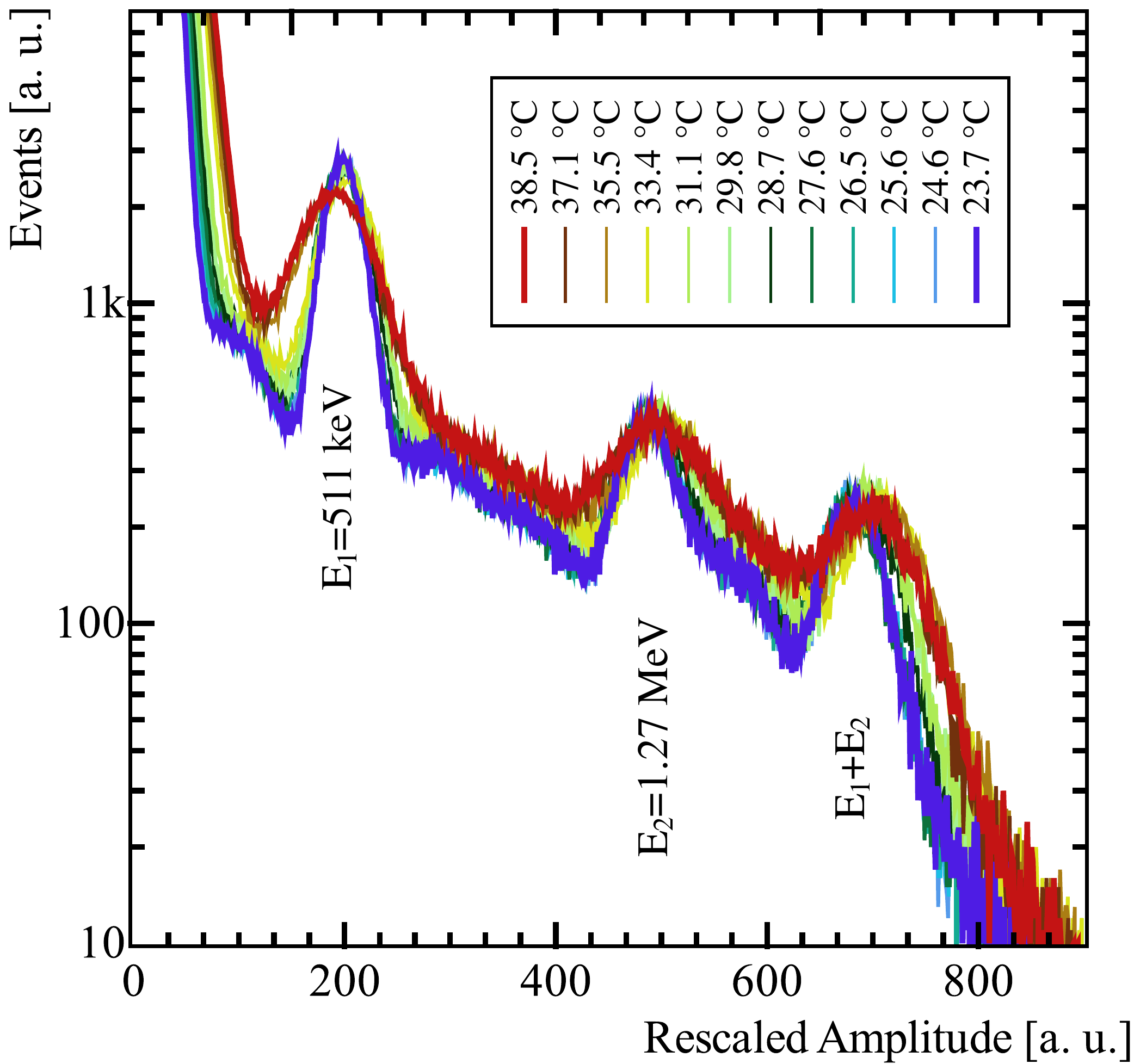}
\caption{Same measurement as shown in Fig.~\ref{pic:na_temp_var}, spectra rescaled according to the gain measured with the light pulser. \cite{Urban_18_diss}}
\label{pic:na_temp_var_comp}
\end{figure}

%co60 plot
%\begin{figure}[ht]
%\centering
%\includegraphics[width=\columnwidth]{pic/Photo/Lichtpulser_Fasern/Lichtpulser_Fasern.png}
%\caption{}
%\label{pic:fasern}
%\end{figure}
\paragraph*{Application of Gain Measurement}
As introduced in Section~\ref{ssec:apd} the APD's gain suffers from a temperature dependence.
The impact of a temperature change can be seen in Fig.~\ref{pic:na_temp_var}. A $^{22}$Na source is used to calibrate the output of one detector unit.
One can see how the positions of the photopeaks shift towards smaller MCA channels, corresponding to smaller amplitudes.
The unit used for this test was modified: The temperature compensation circuit introduced in Section~\ref{ssec:hv_front} was disabled and the APD was operated with a fixed bias voltage.
Using the gain information obtained from light pulser measurements, the spectra can be rescaled to compensate for the gain change.
The result can be seen in Fig.~\ref{pic:na_temp_var_comp}. The photopeak positions overlap very well, considering the wide temperature range covered.
%It is visible that s
Spectra recorded at higher temperatures exhibit wider peaks.
We assume this is caused by
%This is assumed to result from 
the increased dark current at higher temperatures which 
%impacts
increases the noise level.

\begin{figure}[ht]
\centering
\includegraphics[width=\columnwidth]{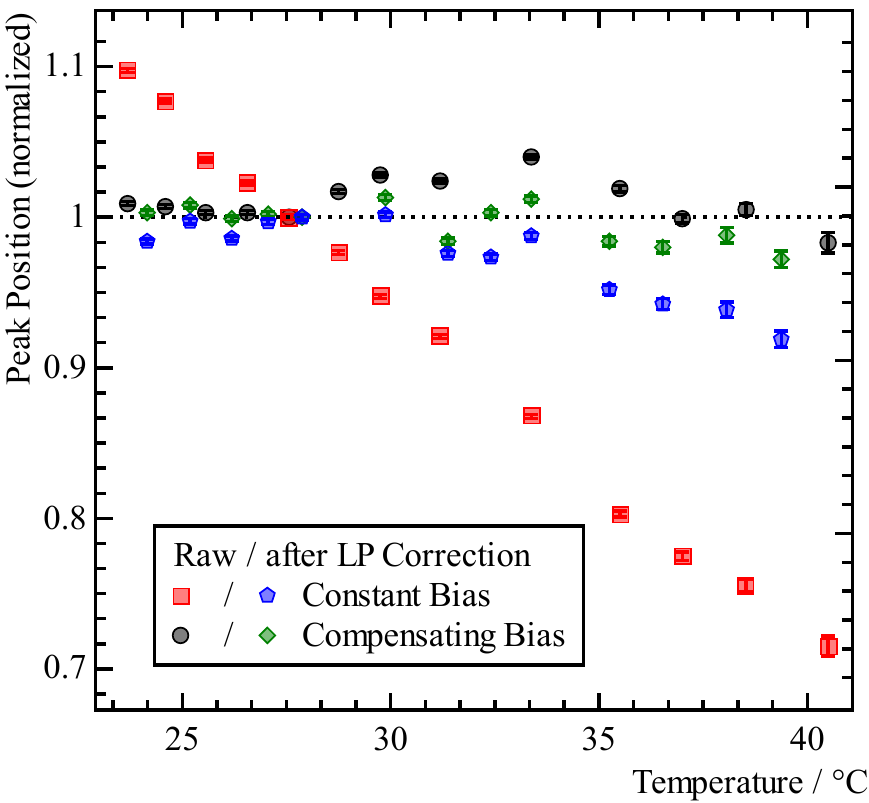}
\caption{Position of the photopeak in dependence of the temperature for four scenarios: fixed bias (red), fixed bias + rescaling (black), compensating bias (blue), and compensating bias + rescaling (green). \cite{Urban_18_diss}}
\label{pic:peak_vs_T}
\end{figure}

For a quantitative discussion, the peak positions are plotted against the temperature in Fig.~\ref{pic:peak_vs_T} for four scenarios. The strongest dependence is found when a constant bias voltage is used and the gain's temperature coefficient is not accounted for (red data points).\\
If the spectra are rescaled using the light pulser, a low variation over the full temperture range is achieved (black data points).\\
The bias supply with automatic voltage tuning successfully compensates the influence of the temperature over a considerable range (blue data points).
These spectra can addidtionally
% The such acquired spectra can also 
be rescaled using the gain measured with the light pulser. An improvement is visible in particular for high temperatures (green data points).

For all data points, only statistical errors from the fit are shown, which seem to underestimate the total error. A known error source of not quantitatively known magnitude results from the fit range.
% of the data used in the fit. 
A gaussian was used for the fit and the spectrum was truncated to a region around each peak.
Still, the data shows that over a large temperature range
%\begin{itemize}
%\item
either or both methods combined improve the gain stability.
%\item 
Better results are obtained when including the LED calibration.
%\end{itemize}
A quantitative analysis showed \cite{Urban_18_diss} that the temperature needs to stay within $27...\SI{28}{\celsius}$ to have a gain variation of at most $\pm\SI{0.1}{\percent}$. This will not significantly affect the energy resolution, which is $\SI{1.8}{\percent}$ at the highest occurring photon energies and worse at lower energies (see Sec.~\ref{ssec:E_res}).

\begin{figure}[ht]
\centering
\includegraphics[width=\columnwidth]{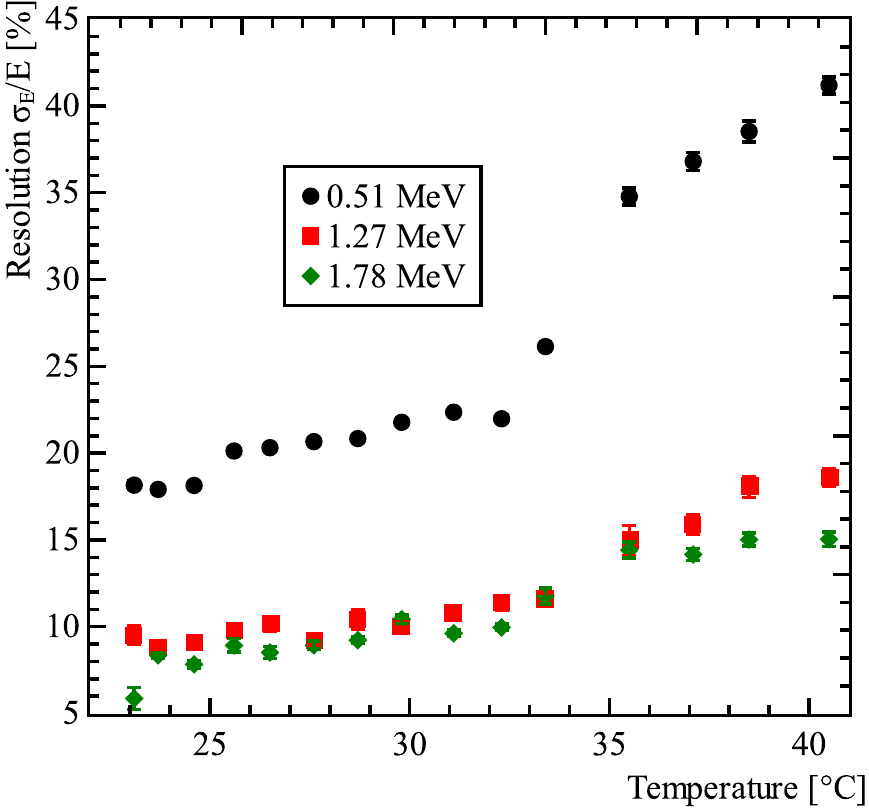}
\caption{Energy resolution in dependence of the temperature at low energies.}
\label{pic:sigmaE_vs_T_LE}
\end{figure}
Fig. \ref{pic:sigmaE_vs_T_LE} shows the energy resolution extracted from the $^{22}$Na spectra. The resolution degrades at high temperatures.

%We assume that a LED calibration decreases the stability in the case that temperature variations are very small.
%How much the temperature has to vary before the improvement starts to outweigh the error on the gain measurement itself cannot be determined from the data shown in in Fig.~\ref{pic:peak_vs_T}.
%If precise knowledge needed, the raw spectra should be analyzed using more sophisticated methods or further measurements need to be performed.\\
%\subparagraph*{Critical Review and Possible Improvements}
\section{Characterization of the Detector Performance}
This section presents the performance achieved with the new readout, in particular the energy and time resolution. Results are discussed from both, prototype tests and data from production beam times.
\subsection{Prototype Tests at Tagged Photon Beams}
To study the performance of a calorimeter, measurements at a tagged photon beam aimed directly at the detector, are a very conclusive test case. For the case of the Crystal Barrel, the incident photon energy range is comparable to those hitting the detector in production beamtimes. The energy resolution of the tagger is usually much better than that of the calorimeter, allowing 
the determination of %to determine 
the energy resolution. The event rate can be varied beyond realistic scenarios. The precise time information provided by the tagger allows the determination of the time resolution of the detector under test.

Measurements on a detector prototype consisting of a $3\times 3$ array of the CsI(Tl) scintillation crystals with the new readout, were taken at ELSA ($E_{e^-}=\SI{800}{\mega\electronvolt}$, $\SI{2.4}{\giga\electronvolt}$, and \SI{3.2}{\giga\electronvolt}) and at MAMI \cite{Jankowiak:2006yc} ($E_{e^-}=\SI{180}{\mega\electronvolt}$, \SI{450}{\mega\electronvolt}, and \SI{1.5}{\giga\electronvolt}).

\begin{figure}[ht]
\centering
\includegraphics[width=\columnwidth]{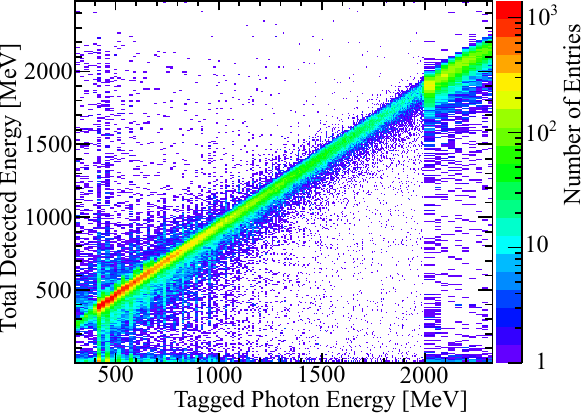}
\caption{Detected energy vs. tagged photon energy. Data from measurement at ELSA, $E_{e^-}=\SI{2.4}{\giga\electronvolt}$. Binning changed for $E_\text{Tag}>\SI{2.0}{\giga\electronvolt}$ accounting for a change in the tagger readout.}
\label{pic:EvsE}
\end{figure}

Fig.~\ref{pic:EvsE} shows the sum of the detected energy deposits in the $3\times 3$ detector array versus the tagged photon energy. The distribution is dominated by a linear correlation. Pile-up events were not removed from the data, which can result in much higher energy deposits than tagged (above diagonal). As photons might be absorbed in the collimator, zero energy entries are possible as well as random background (below diagonal).\\
The binning of the tagged photon energy corresponds to the granularity of the tagging detectors, which causes the sudden change at $\approx\SI{2}{\giga\electronvolt}$ in this data set.

\subsection{Energy Resolution}
\label{ssec:E_res}
\begin{figure}[ht]
\centering
\includegraphics[width=\columnwidth]{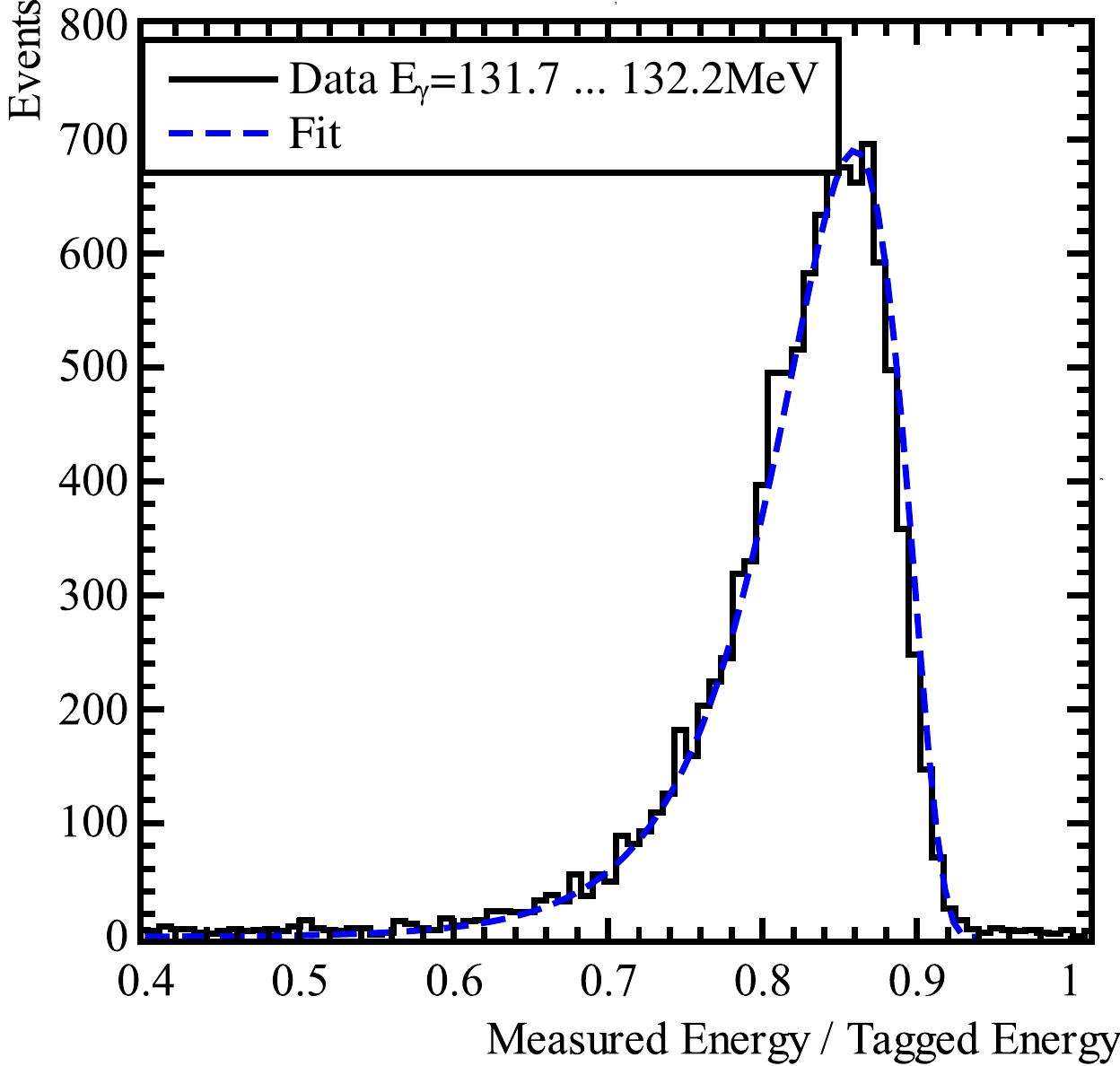}
\caption{Distribution of the measured energy at the photon energy bin $(E_\gamma=131.7...\SI{132.1}{\mega\electronvolt})$ \cite{Urban_18_diss}.}
\label{pic:eblock_at_E}
\end{figure}

Fig.~\ref{pic:eblock_at_E} shows the distribution of the detected energy for a fixed photon energy \citep{Urban_18_diss}. Such data was analyzed to get an estimation of the energy resolution of the calorimeter. Instead of extracting the FWHM numerically from the raw data, a function was fit to the data in order to reduce the impact of statistical fluctuations.\\
To account for the asymmetric shape of the distribution, the Novosibirsk function \citep{IKEDA2000401} was used for the fit.
This function has one parameter $\tau$ characterizing its asymmetry. For $\tau=0$, the function is identical to a Gaussian.
It has a parameter $\sigma$ characterizing the width of the function and, like for the Gaussian function, the relation $\text{FWHM}=2\sqrt{2 \ln 2}\cdot \sigma$ is valid, independently of $\tau$.

\begin{figure}[ht]
\centering
\includegraphics[width=\columnwidth]{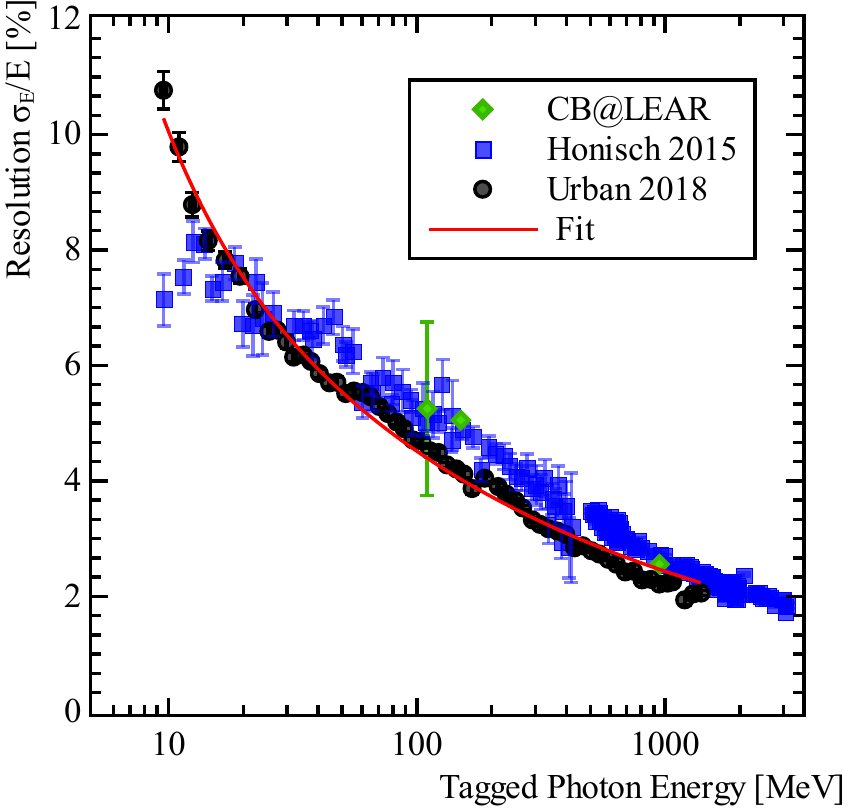}
\caption{Estimated energy resolution of the calorimeter. \cite{Aker:1992ny, Honisch_15_diss, Urban_18_diss}}
\label{pic:sigma_e_vs_e}
\end{figure}

%Y axis units are actually # per 0.758150113722517%
Fig.~\ref{pic:sigma_e_vs_e} shows the determined values for $\frac{\sigma_E}{E}$ from various measurements. These values are the currently best available estimate for the energy resolution of the Crystal Barrel calorimeter with the new readout.\\
The true energy resolution of the \CB might differ for several reasons:
%Compared to the actual energy resolution of the \CB, the results from the prototype might show systematic differences due to differences in the geometry. 
%The data was acquired using only a very small subset of all calorimeter crystals, which might be better or worse than the average.
%The following properties of the measurements might have an impact on the determined values:
%\begin{itemize}
%\item 
First, the geometry of the prototype detector differs slightly. The scintillator crystals in the prototype have all an identical geometry while the geometry in the full calorimeter depends on the polar angle $\theta$.

Also, as the scintillators vary in size along the polar angle $\theta$, the energy resolution might in fact be $\theta$-dependent.\\
Furthermore, only a small subset of all scintillators was used in the tests. The scintillators vary a lot in scintillation intensity (see Fig.~\ref{pic:crystal_brightness}), which may have an effect on the energy resolution, especially in the low energy range. While two different prototypes yielded similar results, this is still a too small subset to claim validity for the whole calorimeter.\\
%\item 
Finally, another point is that lateral shower leakage might have an impact on the energy resolution. As the prototype consists of only 3 $\times$ 3 scintillators, lateral leakage is potentially more pronounced.

In the prototype measurements, the photon beam was aimed at the center of the detector setup. Higher shower losses in dead material might occur for photons hitting the scintillators close to their boundaries.

For the energy resolution of the Crystal Barrel calorimeter with the old readout, three datapoints are available \cite{Aker:1992ny}. While the error on two of those is unknown, the new data seems to be in agreement.

The results can be parameterized with the semi-phe\-no\-me\-no\-lo\-gi\-cal function
\begin{equation}
 \left(\frac{\sigma_E}{E}\right)^2=
 \left( \frac{\SI{2.39(5)}{\percent}}{\sqrt[4]{E/\SI{}{\giga\electronvolt}}} \right)^2
+
 \left(\frac{ \SI{0.46(8)}{\percent}}{\sqrt{E/\SI{}{\giga\electronvolt}}}\right)^2
+
 \left( \frac{\SI{0.048(7)}{\percent}}{E/\SI{}{\giga\electronvolt}} \right)^2
 \end{equation} 
The first term dominates the total error and is purely phenomenological. The numerical value is in agreement with the old parametrization \cite{Aker:1992ny}.\\
The second term has an impact below $\sim\SI{200}{\mega\electronvolt}$. Statistical fluctuations in the shower formation lead to a contribution $\propto 1/\sqrt{E}$ \cite{doi:10.1146/annurev.ns.32.120182.002003}. The third term contributes below $\sim\SI{30}{\mega\electronvolt}$. Electronic noise in the readout gives a contribution $\propto 1/E$ \cite{doi:10.1146/annurev.ns.32.120182.002003}. However, the utilization of the phenomenological term might distort such relations and the formula should just be used as parameterization.

%für niedrige Energie: nicht zu viele kristalle reinnehmen.

\begin{figure}[ht]
\centering
\includegraphics[width=\columnwidth]{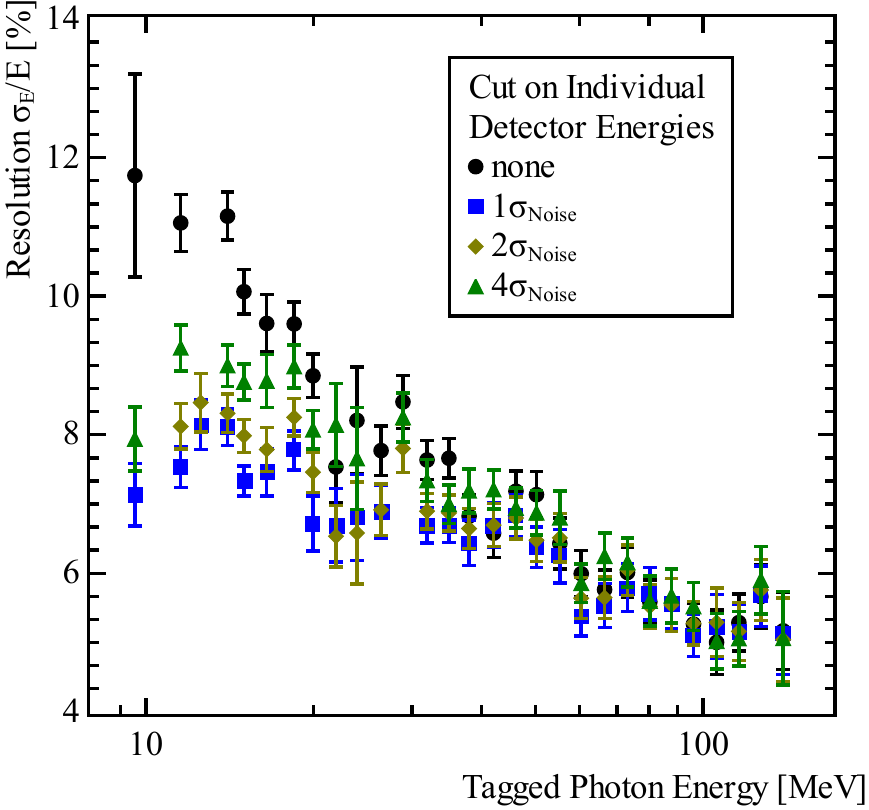}
\caption{Energy resolution at low photon energies.}
\label{pic:sigma_E_cut}
\end{figure}

To achieve the full performance at low photon energies, it is important to carefully consider the number of detector units used in the cluster reconstruction. For the values in Fig.~\ref{pic:sigma_e_vs_e}, the entries of all nine units were added up.\\
As each unit has a certain amount of electronic noise, each additionally considered unit increases the noise level in the sum. The resolution when only considering detector units with energy entries of 1, 2, or 4 times the unit's noise level is shown in Fig.~\ref{pic:sigma_E_cut}.\\
A major impact is visible for $E_\gamma < \SI{20}{\mega\electronvolt}$ while it seems to be negligible for $E_\gamma > \SI{30}{\mega\electronvolt}$. The data originates from a measurement using the $3 \times 3$ prototype \cite{Honisch_15_diss}. The situation might be more complex for the full calorimeter.

%ratenabhängigkeit. wichtig, dass der detektor unter realistischenbedingungnen funtkionerit.

The data corresponding to Fig.~\ref{pic:sigma_e_vs_e} was taken at a low beam intensity. An important property of a detector is its capability to handle high event rates. More practically speaking, the detector needs to still have an adequate resolution at production beam time conditions.

To fully characterize what beam time conditions are, it would actually be necessary to describe the hit rate as a function of energy, as this spectrum needs not to be identical to the spectrum of bremsstrahlung. Accepting the entailed loss of information, the intensity conditions are described here as hit rate above $E_\text{THR}=\SI{6.5}{\mega\electronvolt}$ in the central crystal.\\
For production beam time conditions (primary electron energy $E_{e^-}=\SI{3.2}{\giga\electronvolt}$, extracted beam current \SI{550}{\pico\ampere}, polarized-target phantom) the hit rate above \SI{6.5}{\mega\electronvolt} was measured to be in the order of $2000...\SI{3000}{hits/s}$ in the most forward scintillation modules \cite{Honisch_15_diss}.

\begin{figure}[ht]
\centering
\includegraphics[width=\columnwidth]{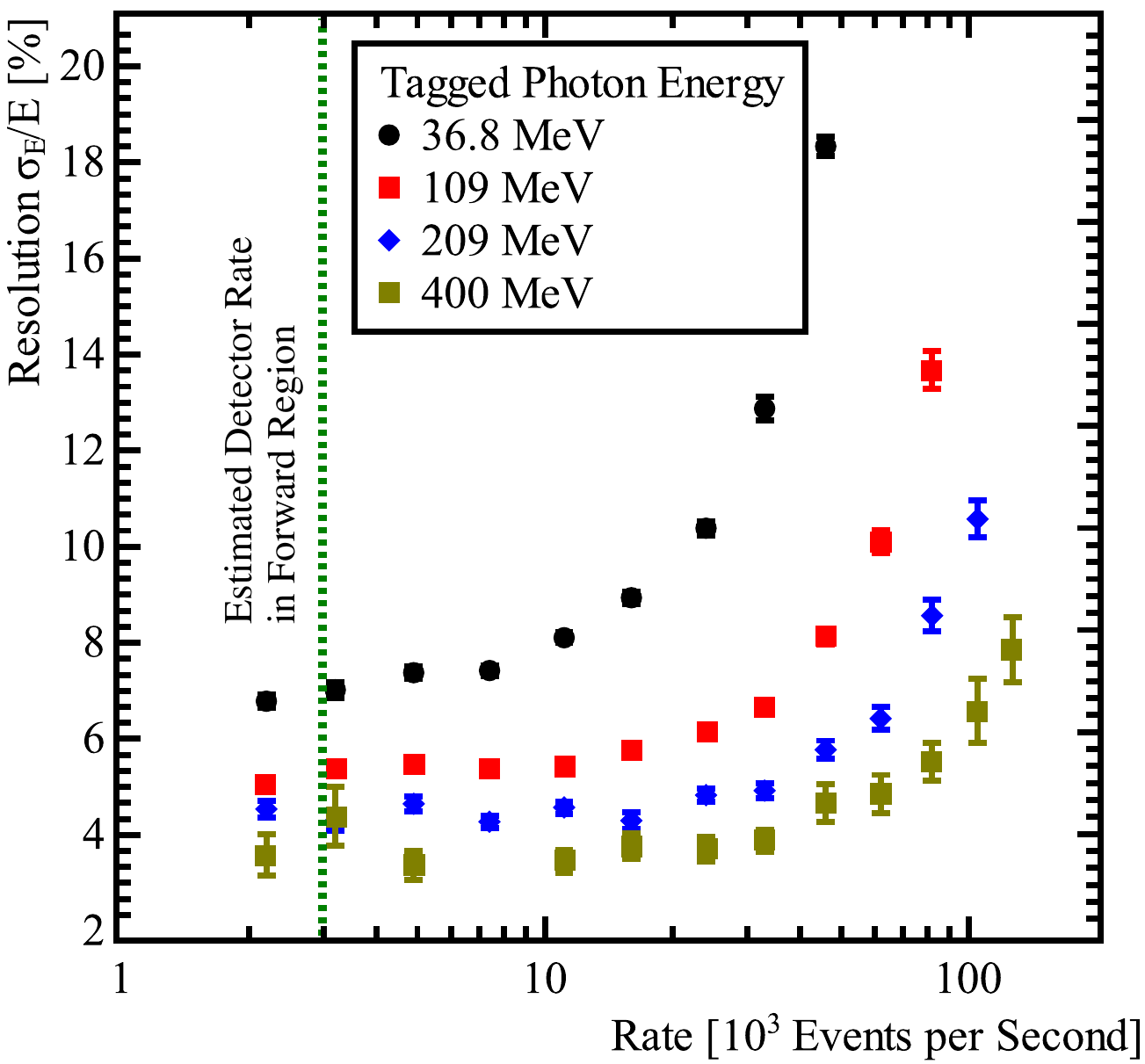}
\caption{Rate dependence of the energy resolution for different tagged energies \cite{Honisch_15_diss}.}
\label{pic:sigma_e_vs_rate}
\end{figure}

Fig.~\ref{pic:sigma_e_vs_rate} shows the estimated energy resolution at four photon energies in dependence of the detector rate. Only a small change is visible at small rates. The rate above which the resolution gets deteriorated depends on the energy in consideration. While the resolution of $\SI{400}{\mega\electronvolt}$ photons seems to be almost constant until a rate of $30\cdot10^3\;\SI{}{hits/s}$, an impact on the resolution of $\SI{37}{\mega\electronvolt}$ photons is already visible at $5\cdot10^3\;\SI{}{hits/s}$.\\
This study was performed with a QDC used for energy measurement which is sensitive to pile-up and baseline shift. The CBELSA/TAPS experiment is being upgraded with sampling ADCs which can compensate for both. The authors assume that this readout will be able to stabilize the energy resolution also towards higher rates.

\subsection{Fast Branch}
\label{ssec:fast_branch}
The most important goal of the upgrade was to provide a hit information that can be included in the first level of the experiment trigger.\\
The time resolution needs to be known to properly determine the coincidence time window in trigger conditions.\\
As time information improves the clustering during data analysis, the whole setup was equipped with TDCs.

This section presents the performance that was achieved with prototypes and the full setup.

Two factors play an important role on the achievable time resolution: the noise level and the signal shape. The effects will be discussed for leading edge discriminators as such are used in the readout of the \CB (see Sec.~\ref{ssec:HDDisc}).\\
This type of discriminator produces a digital time stamp when the signal exceeds a predefined threshold.
Electronic noise is a random variation on top of the scintillation signal. Therefore it can alter the time at which the threshold is crossed. The amount of distortion scales with the noise level and has a random distribution.\\
The signal shape also has an influence on this variation: the faster a signal rises (\SI{}{\volt\per\second}), the lower the introduced variation for a given noise level (\SI{}{\volt}).\\

The signal shape also leads to a deterministic error in combination with leading edge discriminators, which is known as time walk. A given set of signal shape, discriminator threshold, and signal amplitude lead to a certain delay of the digital signal. If the signal shape does not vary for individual pulses and all parameters are known, the error can be compensated for.

%For the case of the \CB's timing signals, the pulse merely scales with its energy while its shape remains constant. Therefore the time-walk can be removed during the off-line data analysis.

\subsubsection*{Online Time-Walk Compensation}
This is done for the  \CB timing signals in two ways. In the data analysis, the energy information of the slow branch is used to remove the time walk.\\
%Online: schwieriger
Unfortunately, this information is not available within the trigger decision time. Thus, other information has to be used. Constant-fraction discriminators (CFD) are a common way to remove time walk. However, those have the drawback that the time defining zero crossing occurs with a delay which is usually chosen to be close to the peaking time of the signal. While shorter delays can be used to generate a zero crossing, the lower latency comes at the cost of  a lower SNR in proximity of the zero crossing.

A dual threshold approach was chosen for the discriminators of the \CB. 
The time it takes the signal to rise from the first threshold to the second, allows the measurement of the slew rate of the signal and therefore an extrapolation the beginning of the signal $t_0$.

%%
%% bild einbauen in dem das verzögern illustriert wird.
%% cut off bei kürzerem delay-> kleinere latenz möglich.
%% evtl in eigenen abschnitt extrahieren?

\begin{figure}[ht]
\centering
\includegraphics[width=\columnwidth]{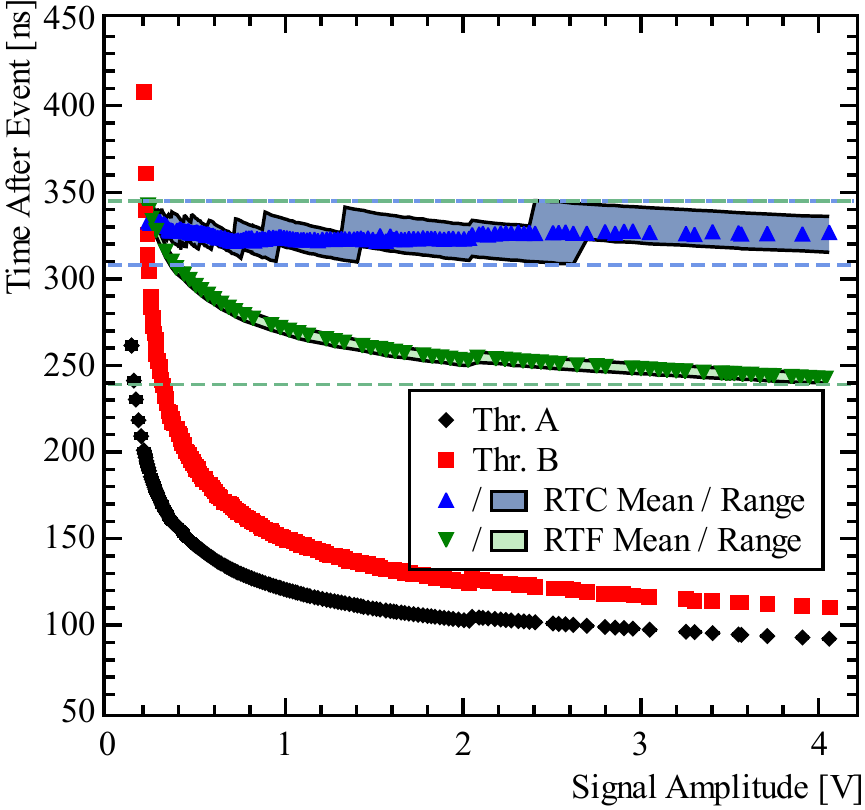}
\caption{Time walk of different discriminators. Black: Low threshold. Red: High threshold. Green: Truncated threshold. Blue: Truncated threshold + walk corrected. \cite{Klassen_20_diss}}
\label{pic:walk_correction}
\end{figure}

Fig.~\ref{pic:walk_correction} shows the latency in dependence of the signal amplitude for different stages in the signal processing. The data points in black (red) show the time walk of a lower (higher) threshold. These data points have the lowest latency, however the total time walk is the largest: The difference between earliest and latest time on each curve is larger than on any other.

The result of a first improvement is shown in green. The electronics discard time stamps originating from small amplitude pulses. This reduces the amount of walk, as most of it is found in the discarded low amplitude region. The scheme of the signal processing for this case can be implemented in an FPGA. When the lower threshold is exceeded, a timer is started. Once this timer reaches \SI{150}{\nano\second}, the status of the second comparator is evaluated. Only if also the second threshold was crossed, an output pulse is generated.\\
The choice of lower and upper threshold and delay results in a certain amplitude (or energy) cut-off.

\subsection*{Non-Linear Extrapolation}
A more sophisticated approach also makes use of the time difference between the crossing of the lower and the upper threshold. The shorter this time is, the higher the amplitude has to be, which corresponds to a low shift in time. This case occurs at the right end of Figure \ref{pic:walk_correction}.
An improvement compared to the previous case is achieved by making use of this measured inter-threshold delay. Depending on its value, a certain additional delay is applied to the lower threshold timing signal.\\
The result is shown in blue with a light-blue band. The data points represent the mean time of the output signal. The band illustrates the variation introduced by the signal processing in the FPGA. Steps in the band correspond to switching from one value in the look-up table to the next. Also the sampling frequency and the fact that the input signals are not correlated to the phase of the FPGA clock affect the shape of the band.

As it will become clear in Section~\ref{ssec:time_res}, neither the sampling quantization error nor the slight slope of the mean values at low amplitudes play a role. The accuracy of the signal's time information is dominated by the intrinsic time resolution (noise limited) and therefore there is no benefit from further improving the walk correction.

%The time walk is compensated for by delaying pulses with low time-walk (i.e. high energy) signals to match the timing of high time-walk (i.e. low energy) pulses. A look-up table is used to get the proper delay.\\
%The digital sampling and pulse generation introduce an quantization error which CFDs by principle do not have. However, it will be shown later \{kapitel einbauen} that this error has no significant contribution on the trigger signal. Also this approach offers a higher level of flexibility: To reduce the latency, pulses with a 

% einflüsse: noise -> random, pulse shape-> deterministic: walk
% walk kann korrigiert werden.
% unterschied online offline zeitauflösung: offline kann walk genauer korrigiert werden, da die energie bekannt ist.
% online: fenster = n* sigma(low E)/2 + n* sigma(high E)/2 + |t0(low E) - t0(high E)|

%Walk: funktion von Energie und Schwelle (aber kaum von E/S) -> skalierung
%Zeitaufl: komplexe abhänigkeiten, rauschen bricht skalierung (sigma * S ist unabhängig von E/S)

\subsubsection{Latency}
%When evaluating the time resolution of a detector, it would be nice to have only one number to 
To evaluate the performance of the timing branch, it would be desirable to have a single parameter specifying the time resolution.\\
Unfortunately, it is not that simple, since it is important to know that the achieved resolution in one system is affected by various parameters. Some of these mechanisms apply also to the latency. The dependencies are more simple in the case of the signal's latency. Therefore, this feature will be discussed before the time resolution.

\begin{figure}[ht]
\centering
\includegraphics[width=\columnwidth]{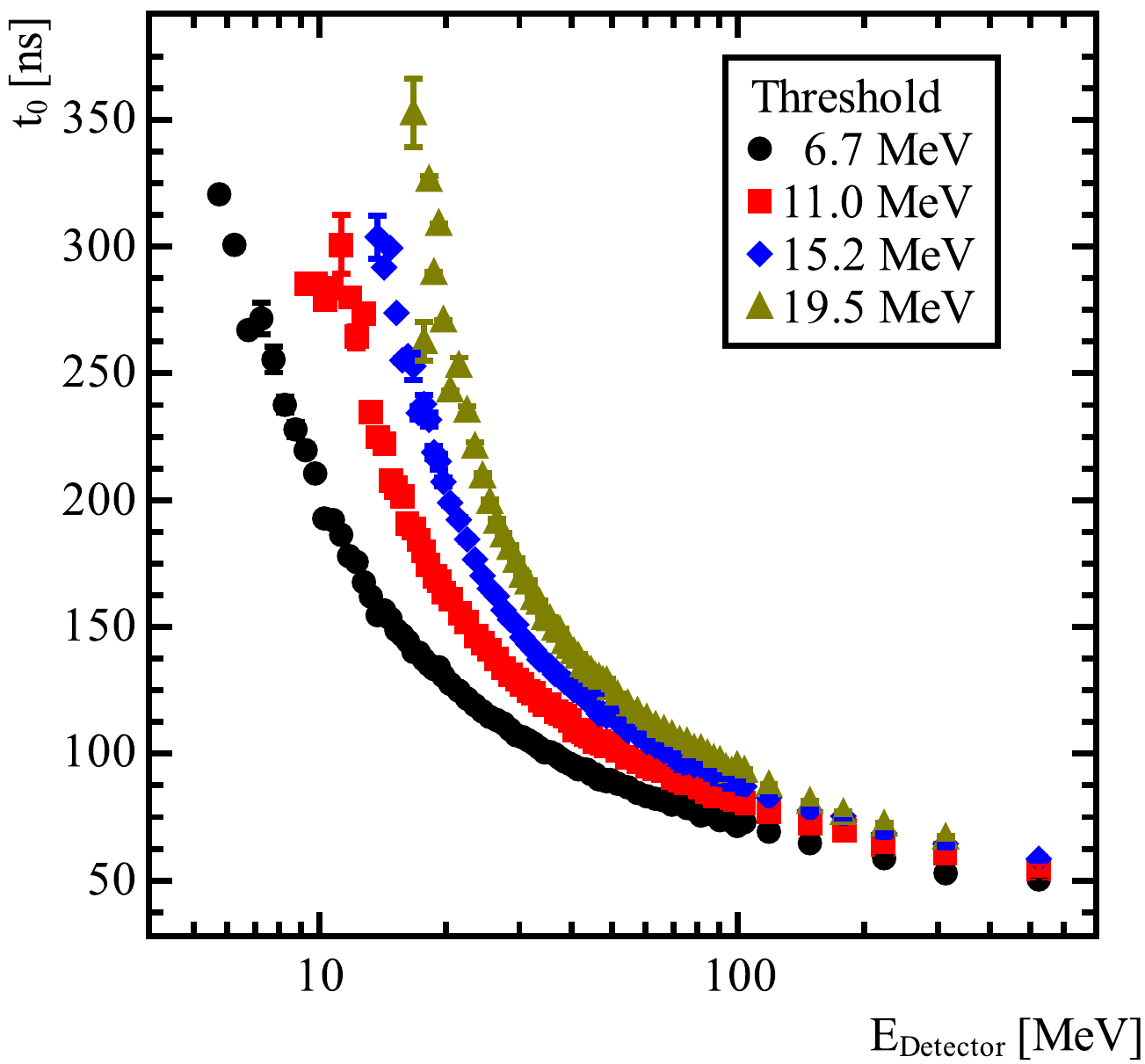}
\caption{The latency $t_0$ of the hit information of a crystal depends on the energy deposit $\text{E}_{\text{Detector}}$ and the discriminator threshold.}
\label{pic:lat_vs_E_T}
\end{figure}

The latency of the time information in the fast branch is shown in Fig.~\ref{pic:lat_vs_E_T}: It increases with an increased threshold and decreases with increased energy deposit. The dependence on two parameters can make it difficult to compare data from different measurements.

The diagram can be simplified by making use of one property of the signals: the pulse shape does not depend on its amplitude (i.e. the energy of the corresponding event), only its amplitude scales with it.
This means that a given fraction of the full amplitude is reached at the same time for any pulse. Thus, the curves for all threshold settings should be identical, if the x-axis is rescaled properly. Instead of plotting the amplitude (or energy) of the pulse directly, the value is normalized to the threshold of the corresponding measurement.
For example, if $V_\text{Pulse} / V_\text{Threshold} = 5$ (or, equivalently $E_\text{Pulse} / E_\text{Threshold} = 5$), then a pulse will always cross the threshold at 20\% of its full amplitude, independent of the actual values. Because of the constant pulse shape, this will happen at the same delay after the beginning of the pulse.

\begin{figure}[ht]
\centering
\includegraphics[width=\columnwidth]{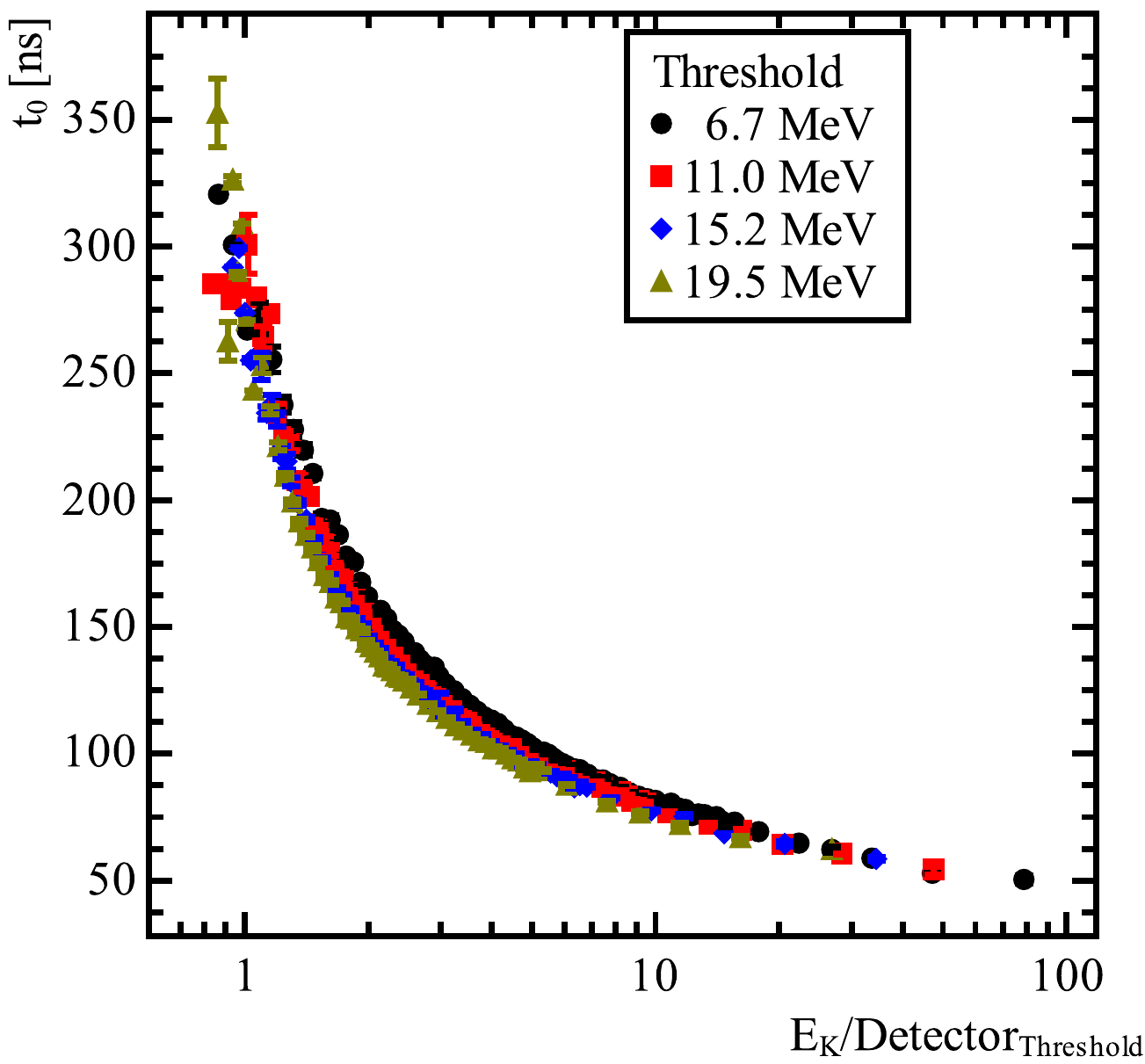}
\caption{Same data as in Fig.~\ref{pic:lat_vs_E_T}, $x$-axis rescaled.}
\label{pic:lat_vs_E_T_norm}
\end{figure}

The data rescaled this way is shown in Fig.~\ref{pic:lat_vs_E_T_norm}. A slight trend remains: Lower thresholds seem to have a slightly higher latency in this representation. However, an error in the threshold determination might result in such a systematic shift. Other sources of deviation might be the electronic noise, which is not constant in this representation, or a slight correlation between pulse shape and amplitude.

\subsection{Time Resolution}
\label{ssec:time_res}
\begin{figure}[ht]
\centering
\includegraphics[width=\columnwidth]{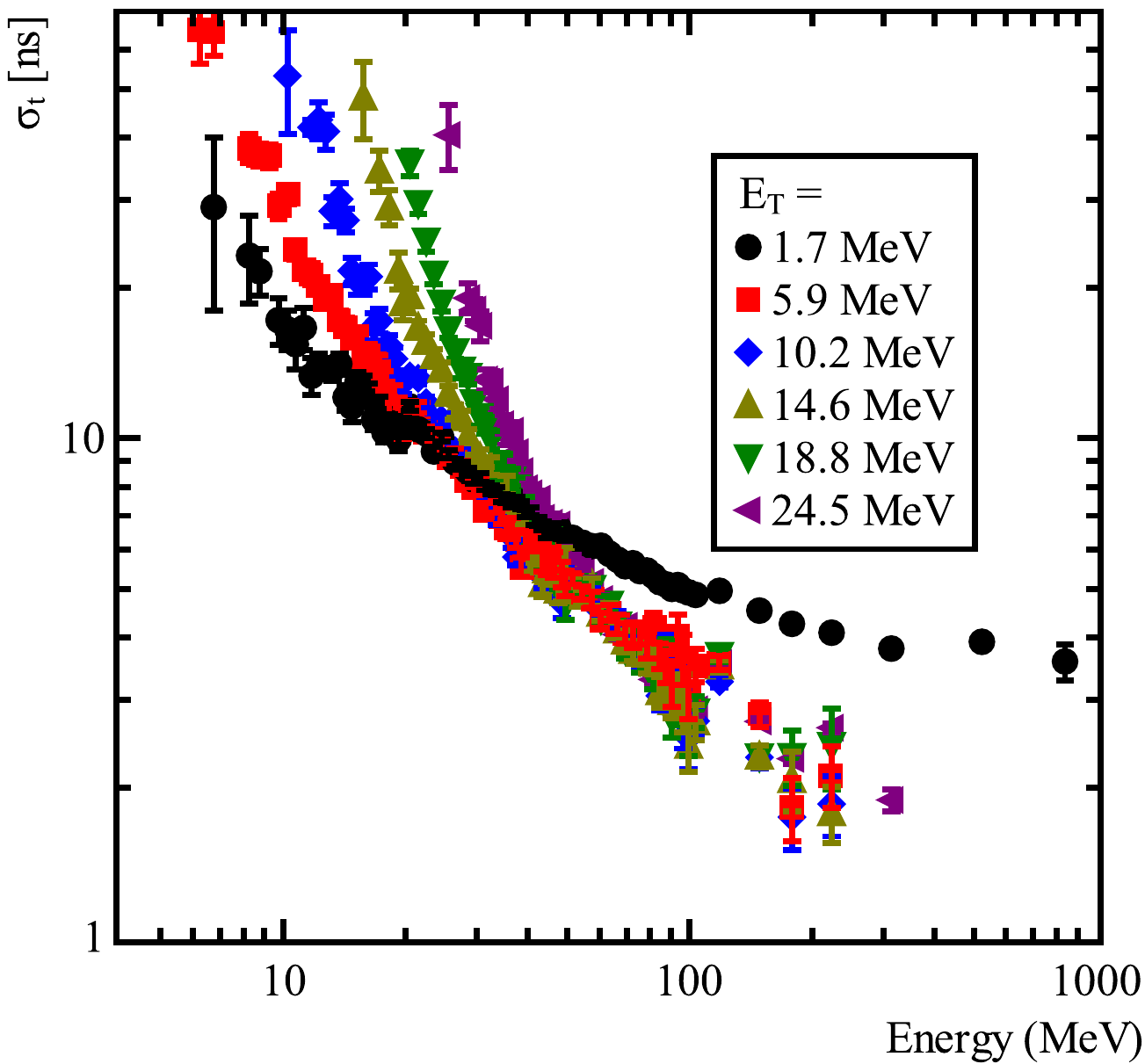}
\caption{Time resolution of one prototype detector, shown in dependence of the detected energy for different thresholds.}
\label{pic:st_vsE}
\end{figure}

The time resolution is one of the most important properties of a signal used for timing purposes. Unfortunately, the dependencies are even more complex here than for the case of latency, making it difficult to compare different prototypes or measurements.\\
The result of a measurement can be seen in Fig.~\ref{pic:st_vsE}. The visible trends are plausible: A higher energy corresponds to a steeper rising pulse. Electronic noise can cause the signal to cross the threshold a bit sooner or later. The steeper the pulse is, the smaller the resulting time variation will be.\\

The effect of the threshold is more complicated: Consider a detected energy of $E=\SI{20}{\MeV}$. Thresholds at (or very close to) this energy will have a bad time resolution. The threshold will only be crossed when the pulse reaches its maximum, which is flatter than any part of its leading edge. This means: At this energy, a lower threshold will be crossed by a steeper part of the pulse and therefore achieve better time resolution.

For very high energies, the opposite seems to be true, a lower threshold corresponds to a worse time resolution. Electronic noise means an additional random amplitude at the beginning of the pulse. This means that the signal has to rise a bit more or a bit less before crossing the threshold, compared to a theoretical noise-free signal. This effect should play a big role for very low thresholds. The data shown in Fig.~\ref{pic:st_vsE} supports this hypothesis.

\begin{figure}[ht]
\centering
\includegraphics[width=\columnwidth]{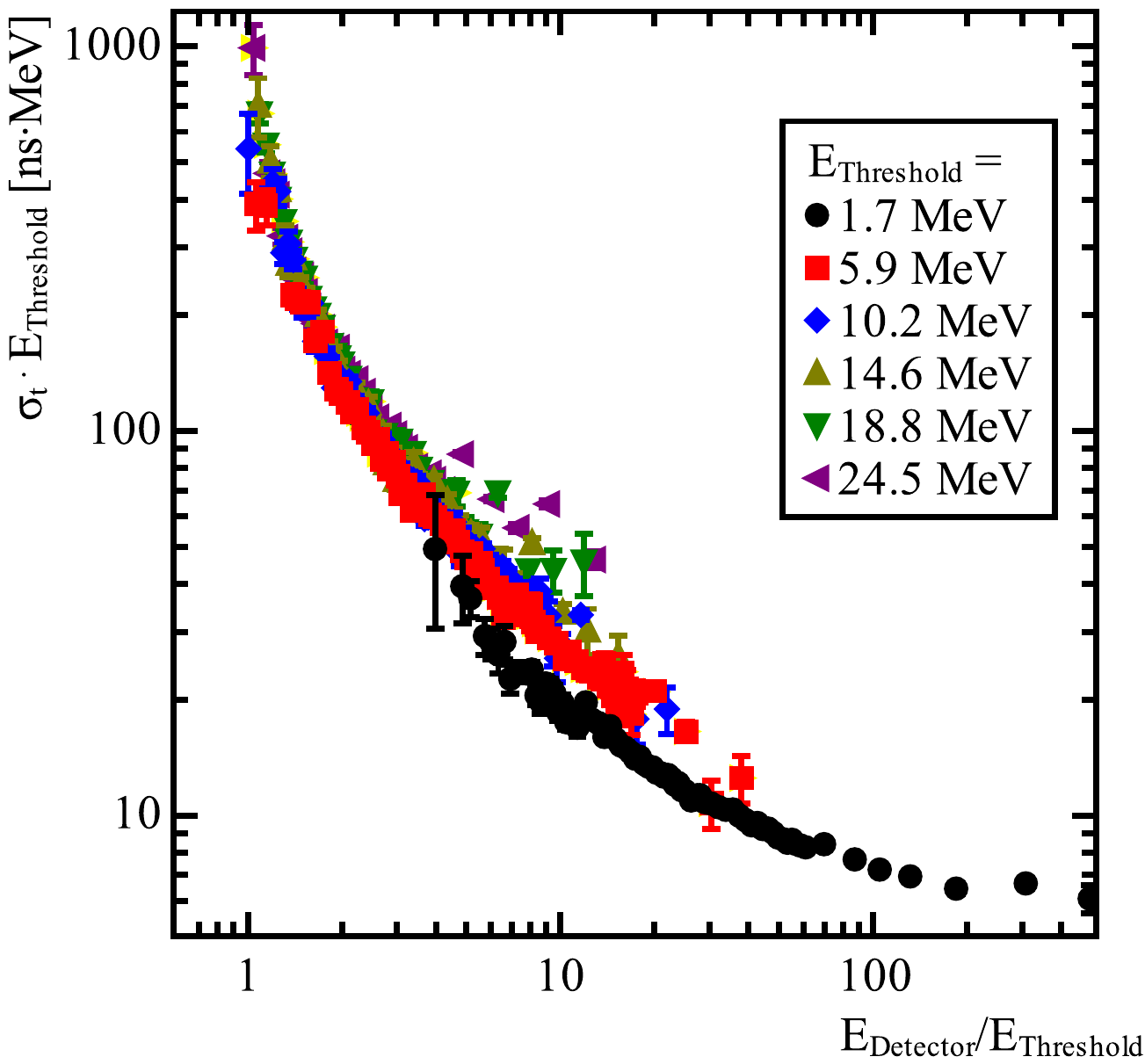}
\caption{Same data as in Fig.~\ref{pic:st_vsE}, $x$- and $y$-axis rescaled.}
\label{pic:st_vsE_norm}
\end{figure}

Plotting the data in the same way like the latency in Fig.~\ref{pic:lat_vs_E_T_norm} cannot resolve this complexity, as the noise level is constant and should not be scaled for different threshold settings.\\
However, there is another way to achieve a simpler dependence. Consider the case $E_\text{Pulse}\gg E_\text{Threshold}$. If now $E_\text{Pulse}$ is doubled, the steepness of the pulse while crossing the threshold should also roughly be doubled, resulting in half of the initial $\sigma_t$ (i.e. better time resolution). Therefore $\sigma_t \cdot E_\text{Threshold}$ should be constant (for a given $E_\text{Thr}/E_C$).
In Fig.~\ref{pic:st_vsE_norm}, this seems to be fulfilled particularly well for thresholds $E_\text{Threshold}\geq\SI{5.9}{\MeV}$ (red).

\begin{figure}[ht]
\centering
\includegraphics[width=\columnwidth]{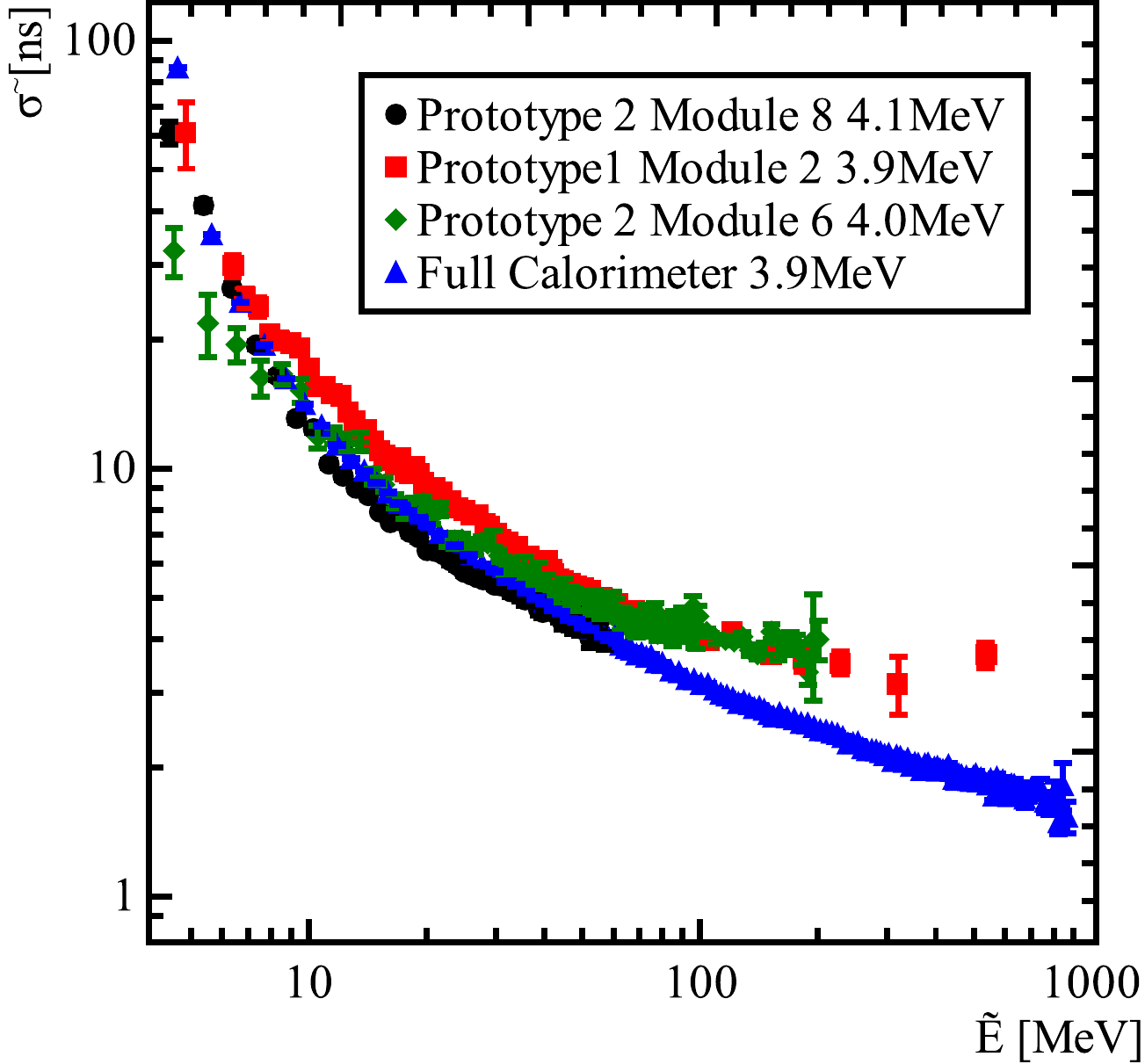}
\caption{Time resolution of different prototypes and final setup.}
\label{pic:st_vs_E_comp}
\end{figure}

This method is applied to data shown in Fig.~\ref{pic:st_vs_E_comp}. Data from the first and second prototype detector is shown, as well as results from the full calorimeter after the upgrade. The prototype data was acquired at test beam times, the full calorimeter data set during a production beam time.
All of these data sets have similar thresholds. To remove the remaining systematic shift, the same method as used in Fig.~\ref{pic:st_vsE_norm} was applied to the data. 

%kleine unterscheide in schwellen, methode oben um verbleibende unterschiede zu korrigieren, rescale für einfache einheiten.

In order to have more intuitive units, each data point $i$ of each data-set $j$ was rescaled to an equivalent threshold of $E_\text{Thr, eq}=\SI{4}{\MeV}$. So for each data-set $(\sigma_i, E_i)_j$ with a threshold $E_\text{Thr, j}$, data is plotted according to:
\begin{equation}
\tilde{\sigma}_{i,j}=\sigma_{i,j}\cdot\frac{E_\text{Thr, j}}{E_\text{Thr, eq}}
\end{equation}
and
\begin{equation}
\tilde{E}_{i,j}=E_{i,j}\cdot\frac{E_\text{Thr, eq}}{E_\text{Thr, j}}.
\end{equation}

%A few observations can be made:
%\begin{itemize}
%\item 
The results from the prototypes and the full setup diverge  at high energies ($\tilde{E}\gtrsim\SI{100}{\MeV}$) from each other. The prototype measurements are worse than those from the full calorimeter, trending towards a factor of 2. This was found to be caused by an error in the readout, where a faulty connection caused a bad resolution of all TDC channels.
While this problem affects data at all energies, its effects are not visible at low energies since the timing resolution there is much worse than the impaired TDC resolution.
%\item 
The average time resolution of the full calorimeter is similar to the performances achieved with the prototypes, which indicates a well working system.
%\end{itemize}
%abweichungen bei E>100MeV: fehler im tdc
%bei E klein: grün flacher, evtl da schwellen scheinen nicht komplett richtig bestimmt?
% in general: similar
% full cb: bedenke: alle kristalle, gemsichte auflösung, leichte laufzeitunterschiede, fpga tdcs, trotzdem gut.

%\begin{figure}[ht]
%\centering
%\includegraphics[width=\columnwidth]{pic/sigma_t_vs_E/sigma_t_vs_E.png}
%\caption{}
%\label{pic:sigma_t_vs_E}
%\end{figure}

%\subsection{Setup for APD Characterization}
%\label{ssec:apd-char}
%\{warum ist das noch hier? ist doch oben beschrieben... todo: nochmal oben lesen, klären ob das hier gelöscht werden kann}
%\FloatBarrier
\section{The First Year of Operation at ELSA}
%invariant mass, trigger rate, cosmic trigger rate
While data has already been taken that allows the extraction of polarization observables, this section focusses on results from the perspective of a detector developer, e.g. the achieved temperature stability and the occurrence of problems.

\subsection{Detector Sensitivity}
\begin{figure}[ht]
\centering
\includegraphics[width=\columnwidth]{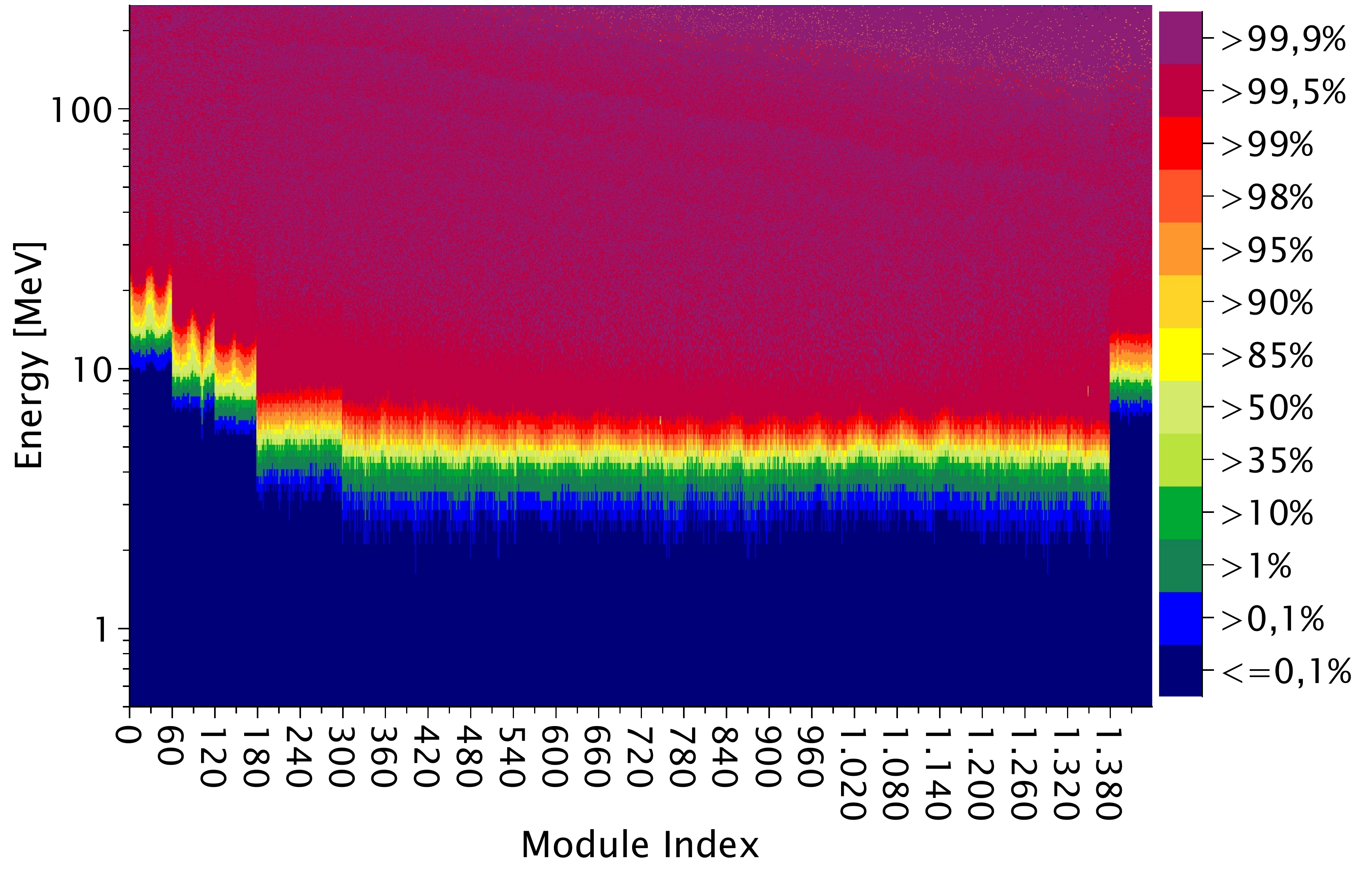}
\caption{Efficiency of the discriminator (lower threshold).}
\label{pic:thra2d}
\end{figure}

The detection efficiency was determined for a data set acquired during a beamtime.
The efficiency $\eta$ was calculated via
\begin{equation}
\eta_i\left(E\right) = \frac{H_{i}\left(E\right)}{N_{i}\left(E\right)},
\end{equation}
where $i$ is the index of the detector module, $E$ the energy measured in this module, $H_i(E)$ the number of events with a TDC hit in this module within the energy bin in consideration, and $N_{i}\left(E\right)$ the total number of events within the energy bin.
Fig. \ref{pic:thra2d} shows the detection efficiency of the lower threshold. The detection efficiency is high for all modules beyond the transition region.\\
The first few rings (low index) and the last ring (high index) have higher thresholds to reduce the impact of background reactions.

\subsection{Invariant Mass Spectrum}
\begin{figure}[ht]
\centering
\includegraphics[width=\columnwidth]{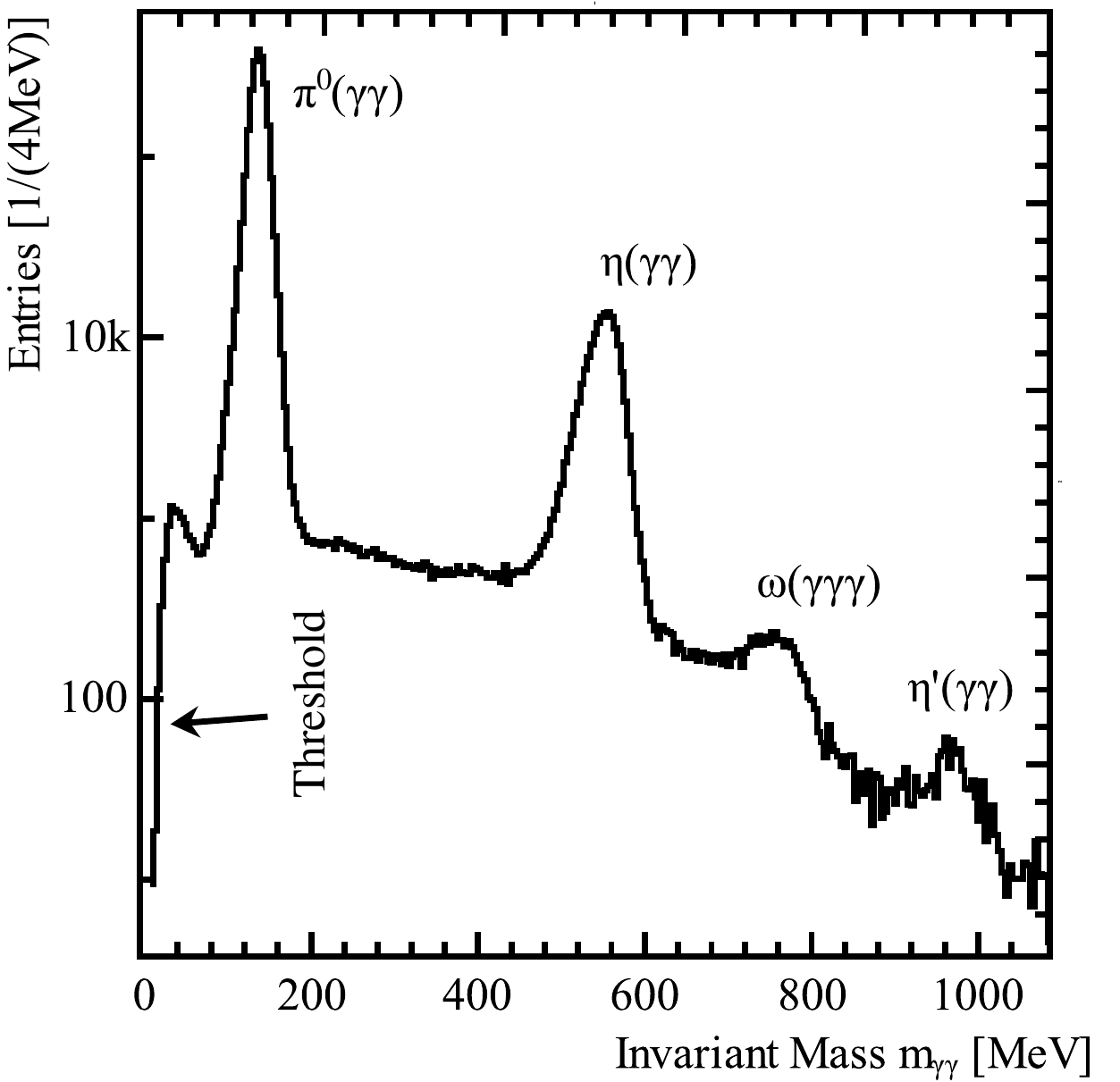}
\caption{Invariant mass spectrum of photon pairs. Data from first production beam time after the upgrade.}
\label{pic:mgg}
\end{figure}
A plot that indicates the proper operation of the full calorimeter is the invariant mass spectrum of detected photon pairs (see Fig.~\ref{pic:mgg}). Peaks corresponding to the $\pi^0$, $\eta$, and $\eta'$ decaying into two photons are visible at their masses. Another peak is visible at the mass of the $\omega$ meson which does not decay in two but in three photons ($\omega\rightarrow\pi^0\gamma\rightarrow\gamma\gamma\gamma$). The peak is assumed to result from events in which one of the three photons is below the detection threshold.

\subsection{Temperature Stability}
\label{ssec:tempStab}
\begin{figure}[ht]
\centering
\includegraphics[width=\columnwidth]{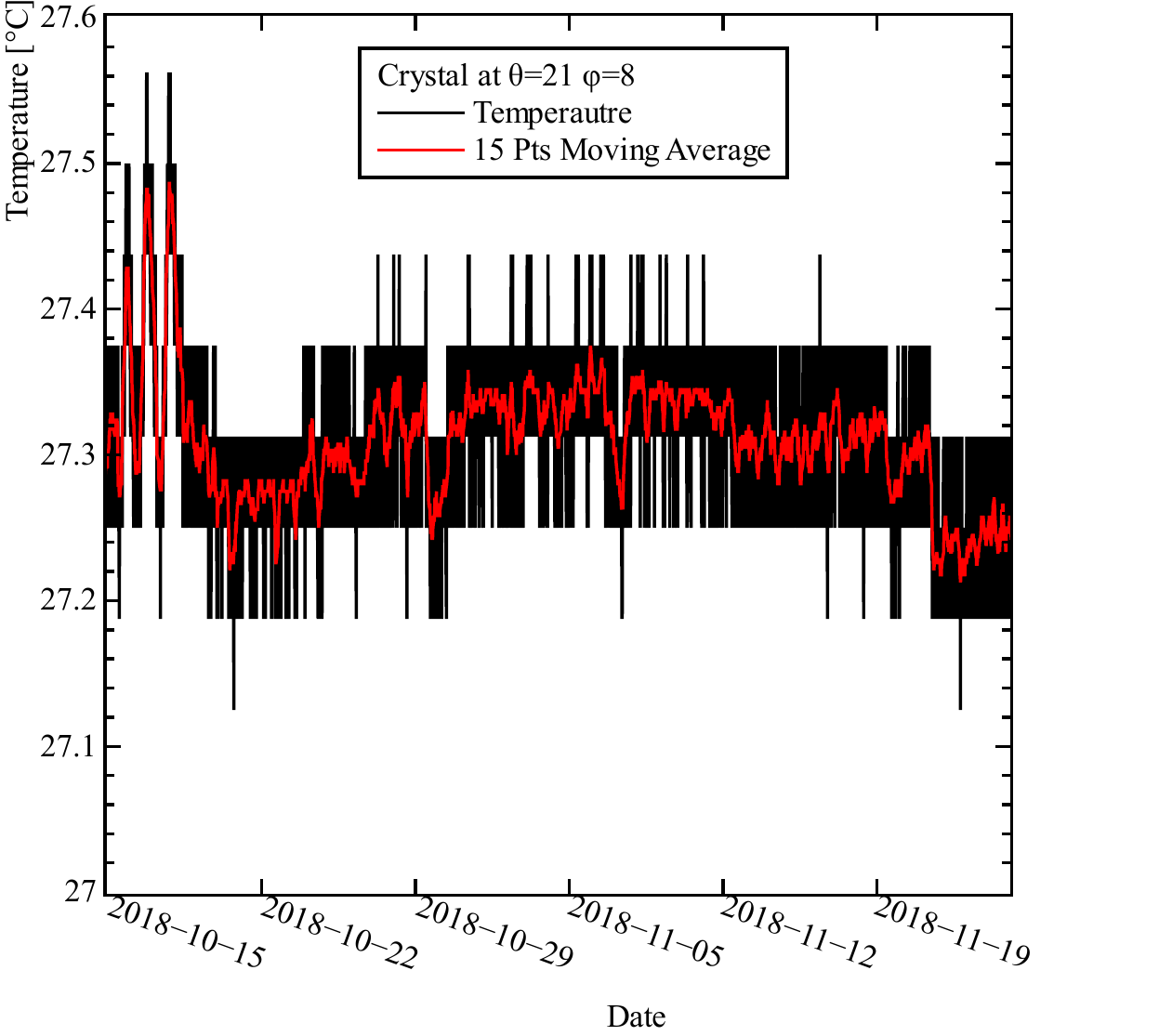}
\caption{Temperature in one representative detector module during a beamtime.}
\label{pic:temp_stab}
\end{figure}

Fig.~\ref{pic:temp_stab} shows the temperature inside one detector module during a production beamtime. During the 6 weeks shown, the temperature is stable within $\pm \SI{0.25}{\celsius}$. A strong variation with a 24 hour periodicity is visible at the beginning of the shown time frame. After settings were optimized in the temperature stabilization system, these variations do not appear anymore and an even better stability is achieved.

This stability is sufficient to maintain the full energy resolution while the gain calibration is not yet required.
More precisely: It was found \cite{Urban_18_diss} that the temperature needs to stay within $27...\SI{28}{\celsius}$ to have a gain variation of at most $\pm\SI{0.1}{\percent}$. This does not significantly decrease the energy resolution which is larger than $\SI{1.8}{\percent}$ at any occurring energy.

\subsection{Occurrence of problems}
\label{sec:failure}
\begin{figure}[ht]
\centering
\includegraphics[width=\columnwidth]{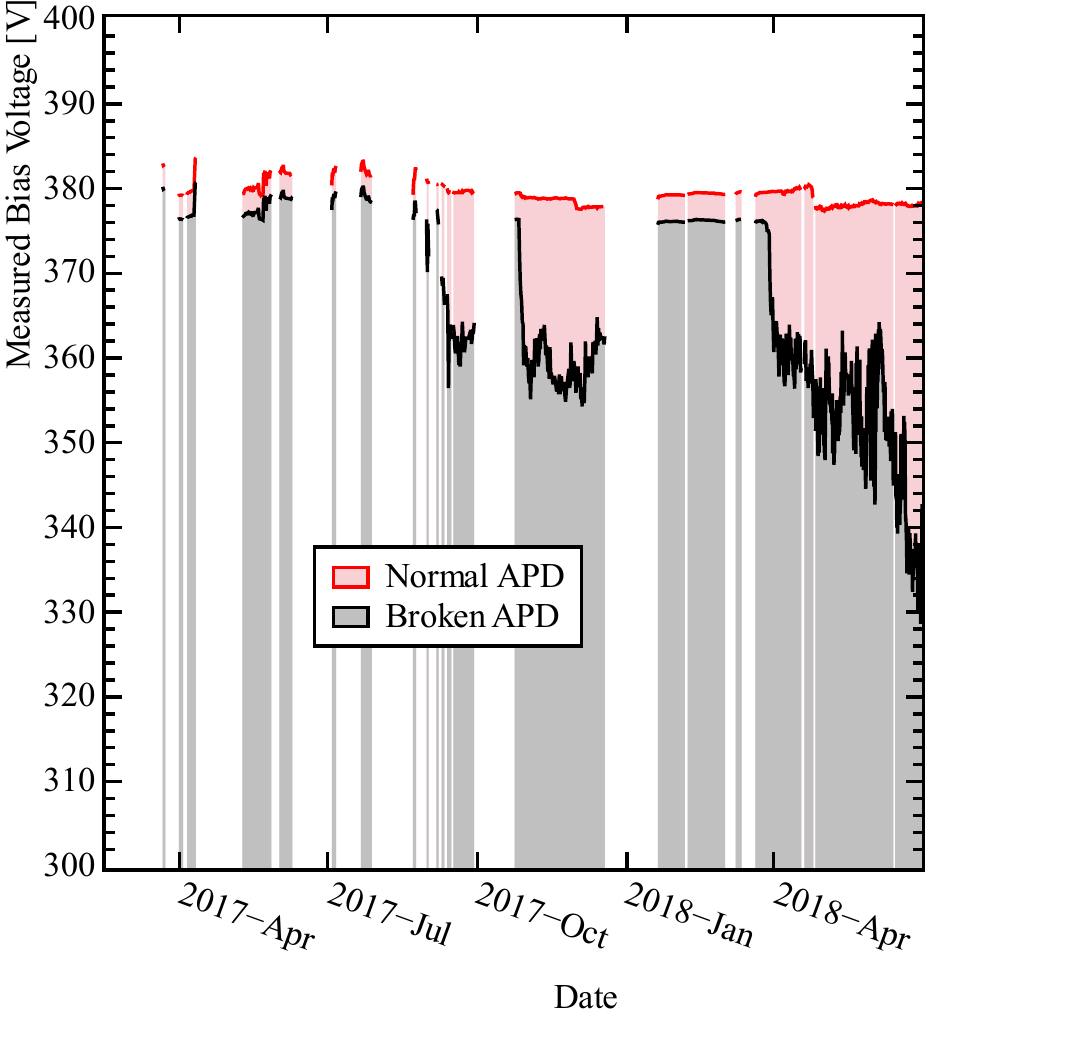}
\caption{APD bias voltage of a normal APD (red) and a broken one (black). Gaps in the data correspond to phases during which the bias voltage was turned off.}
\label{pic:brokenAPD}
\end{figure}

A few components broke after the calorimeter was installed in the experiment. The most frequent type of problem were broken APDs. This occurred for approximately 5 APDs, corresponding to \SI{0.2}{\percent}. Affected channels showed excessively increased noise levels, which could be traced back to an increased dark current.\\
The dark current increased so much that a voltage drop on the bias voltage could be observed, which is shown in Fig.~\ref{pic:brokenAPD}.
The voltage of an APD in normal operation is shown in red. Variations occur due to the build-in temperature compensation (see Sec.~\ref{ssec:hv_front}).\\
The bias voltage of a broken APD shows much larger variations (black curve).
%Much larger variations can be seen on the bias voltage of a broken APD (black curve).
At the end of the shown time frame, the voltage drops by \SI{50}{\volt}, which corresponds to a dark current of $I_D\approx\SI{100}{\micro\ampere}$. In normal operation, the dark current of this type of APD (X11048(X3), similar to S8664-1010 \cite{apddatasheet}) is typically $I_D\approx\SI{10}{\nano\ampere}$.

Localizing this error turned out to be difficult as affected APDs sometimes returned to normal operation after a power cycle, which can also be seen in Fig.~\ref{pic:brokenAPD}.
A few of the broken APDs were sent back to the manufacturer, who traced the problem back to stains on the inside of the ceramic frame. These were caused by a problem in the production process, which had already been resolved before the broken APDs were returned to the manufacturer.

\section{Summary}
The Crystal Barrel calorimeter consists of 1380 CsI:Tl scintillators. 1320 of these are currently installed and were equipped with an avalanche photodiode (APD) readout. The high signal-to-noise ratio allows to use the detector signals in the first level of the experiment trigger. This also required the installation of shaping amplifiers and discriminators to obtain hit information. An FPGA based cluster finder processes the signals fast enough to use multiplicity signals in the trigger, using a single level trigger scheme.

While a direct comparison is not possible, the energy resolution seems to be maintained: A prototype achieved a comparable resolution and, using the full calorimeter, the width of the pion peak in the $\gamma\gamma$ invariant mass spectrum has the same width.

The strong temperature coefficient of the APD's gain is counteracted with a temperature stabilization and an automatic analog bias voltage tuning. The performance achieved of both methods combined prevents the temperature dependence from degrading the energy resolution.

An LED based light pulser system can be used to measure each APD's gain online. This might be necessary if stronger temperature variations would occur.

In December 2017, the CBELSA/TAPS experiment resumed data taking with the new readout of the Crystal Barrel and its integration in the first level of the trigger. Analyses of the new data are ongoing for many different photoproduction reactions off protons and, due to the new trigger capabilities, off neutrons. Preliminary results for the polarization observables $T, P$ and $H$ exeed previously measured CBELSA/TAPS data regarding statistical precision as well as angular and energy coverage \cite{Afzal:2020geq}.  

\section{Outlook: Upcoming Improvements}
%possible (but unnecessary) improvements were discussed. here: really benefitial improvements.
\subsection{Energy Sum}
\label{ssec:Esum}
The energy of beam photons in the CBELSA/TAPS experiment is determined with a tagging system. It measures the momentum of electrons after producing bremsstrahlung and can detect electrons up to $\SI{87.1}{\percent}\cdot E_0$, where $E_0$ is the primary energy of the incoming electron beam.

Electrons with a higher energy are not detected, but the corresponding photons still contribute to the low energy part of the primary photon beam ($E_\gamma < \SI{413}{\mega\electronvolt}$ for $E_0=\SI{3.2}{\giga\electronvolt}$) and eventually cause events in the detectors.\\
While such type of event will be filtered out during data analysis, it still triggers the readout of the DAQ and increases the deadtime. \\
Removing these events already on trigger level would therefore result in a higher number of useful events, recorded in the same time under the same conditions.

The energy sum upgrade is an approach to increase the selectivity of the trigger regarding this aspect. The signals from all channels of the CB calorimeter are summed up with analog circuits. When properly calibrated, the output should allow the measurement of the energy in the whole calorimeter.

As this information is already needed at trigger time, the signals from the timing branch are used in the sum.

\begin{figure}[ht]
\centering
\includegraphics[width=\columnwidth]{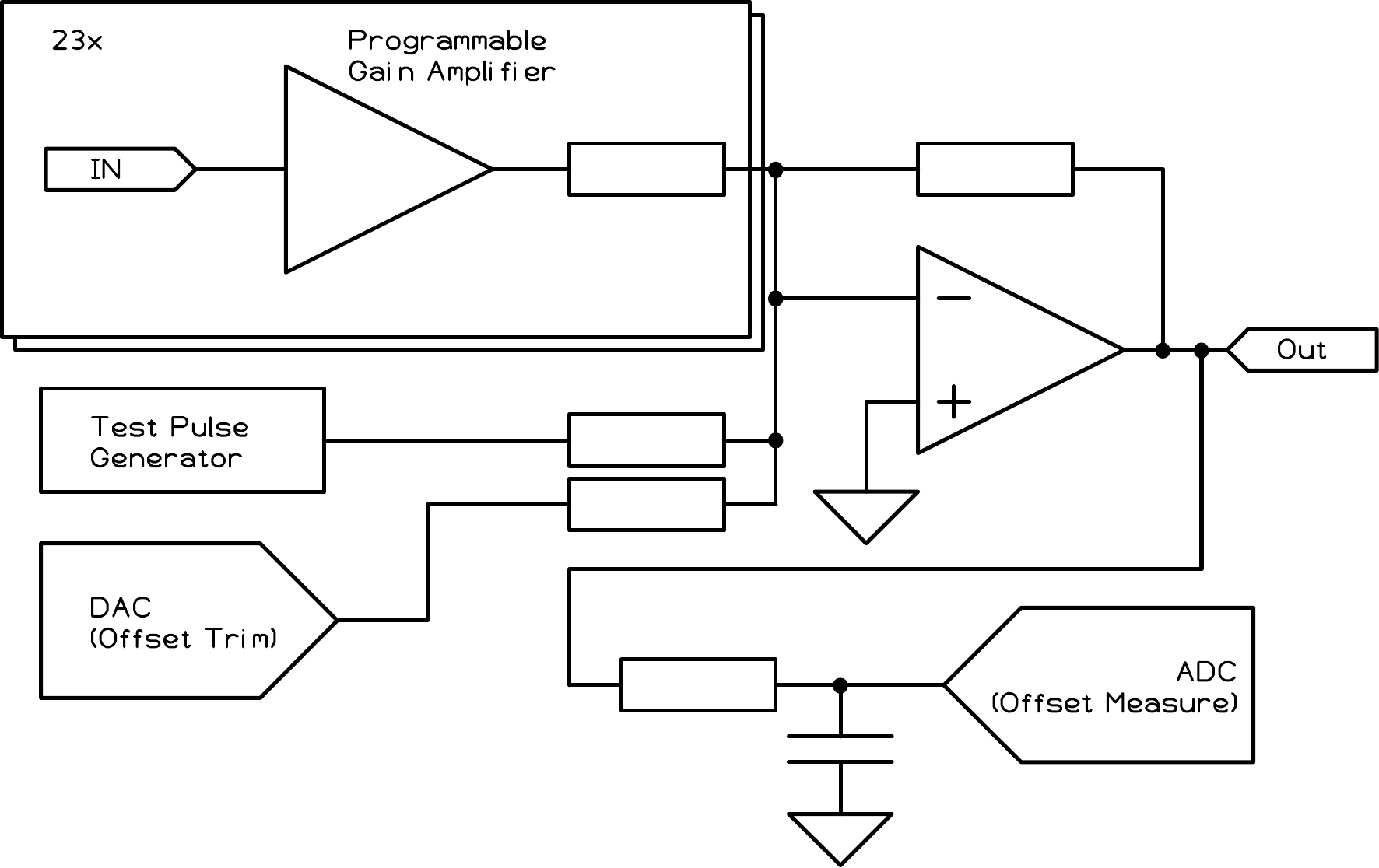}
\caption{Schematic of one fast energy summation module.}
\label{pic:esum_sch}
\end{figure}

\begin{figure}[ht]
\centering
\includegraphics[width=\columnwidth]{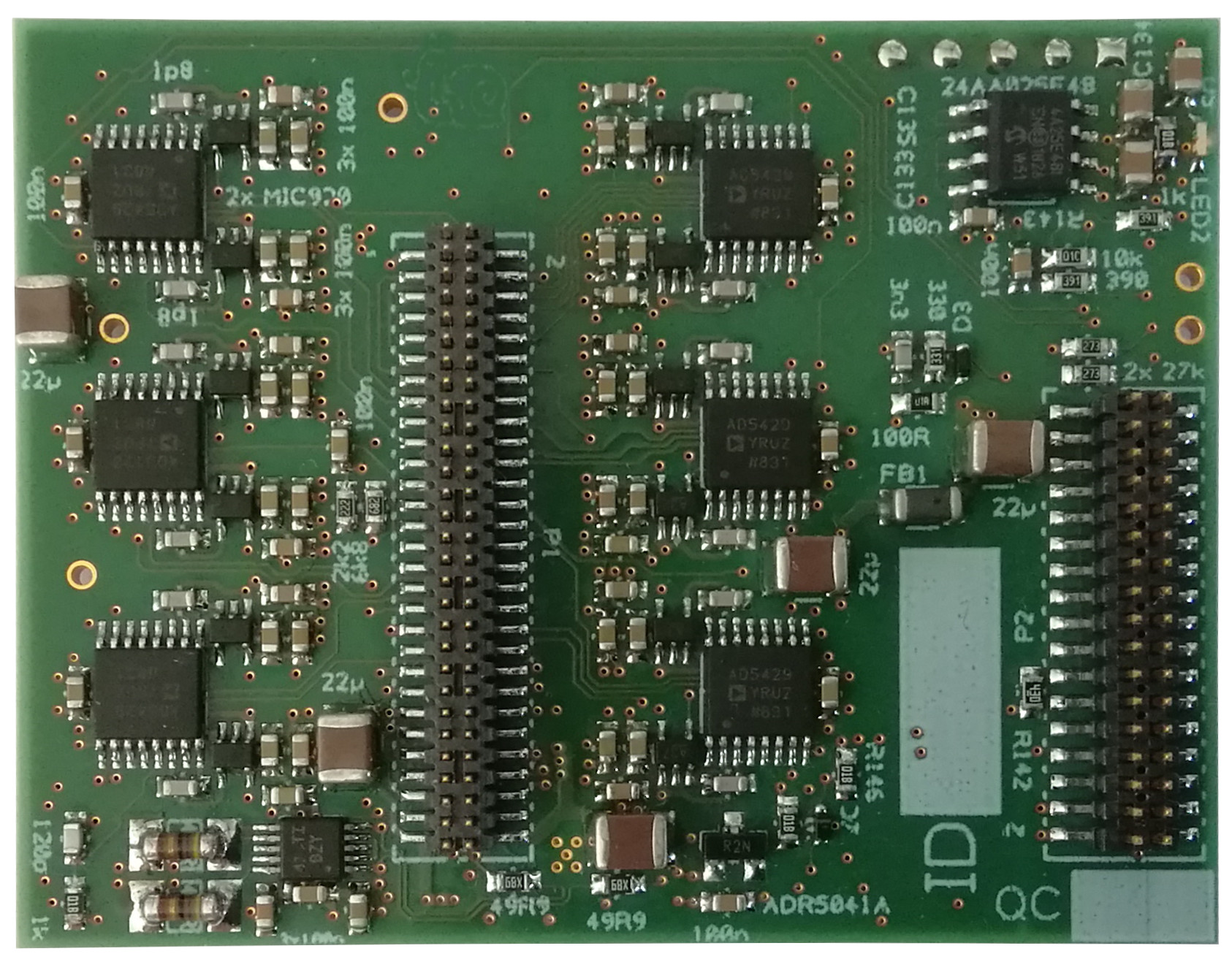}
\caption{Photograph of one fast energy summation module.}
\label{pic:esum_pic}
\end{figure}

The schematic of one summation circuit is shown in Fig.~\ref{pic:esum_sch}. It processes the signals of one subsection of the calorimeter, consisting of 23 detector modules.\\
Each channel has one variable gain amplifier using a multiplying DAC (AD5429). This allows a calibration to have the signals of all detector modules appearing with the same weight at the output of the sum.\\
The signals are summed using an ordinary operational amplifier summation circuit. An ADC can be used to measure the offset voltage of the output, and a DAC to remove it. A test pulse generator is implemented for debugging purposes. Fig.~\ref{pic:esum_pic} shows a picture of one PCB of the same circuit.

% viele events below tagger range -> can't be analyzed
% remvoe on trigger level to reduce deadtime
%sum of energy deposited in all detectors, start with cb

%implementation: VGA, Baseline shift, summing amplifiers

\subsection{Gigabit Interface}
\label{ssec:gb_if}
The most significant limitation of the Crystal Barrel discriminator modules is their readout. All modules on one VME bus have to share the bandwidth in the order of \SI{10}{\mega B \per\second}. The cluster finder VME crate houses 18 modules.\\
While this is currently not a limiting factor during production beam times, it will be once other components of the experiment are upgraded and the intensity of the photon beam is increased.\\
An extension module that can increase the data transfer rate significantly is in development. Each discriminator will be equipped with an individual board which houses a gigabit ethernet interface.\\
%\section*{References}

We thank the technical staff at ELSA and at all the participating institutions for their invaluable contributions to the success of the experiment. Funded by the Deutsche Forschungsgemeinschaft (DFG, German Research Foundation) Project-ID 196253076 TRR 110 and Project-ID INST 217/688-1 FUGG, the U.S. Department of Energy, Office of Science, Office of Nuclear Physics, under Contract No. DE-FG02-92ER40735, and the Schweizerische Nationalfonds (200020-175807, 156983, 132799, 121781, 117601).

\bibliography{CHpaper}
\end{document}